\documentclass[twocolumn]{aastex631}

\usepackage{graphicx}
\usepackage{amsmath}
\usepackage{cleveref}
\usepackage{float}
\usepackage{pgffor} 
\usepackage{tabularx}
\usepackage{array}
\usepackage{xcolor}
\newcolumntype{Y}{>{\raggedright\arraybackslash}X}

\usepackage{makecell}
\begin{document}
%\title{LAMOST Low Resolution Survey Wavelength Calibration:  Precision Verification and Correction}

\title{A General Framework for Radial Velocity Calibration in Low-Resolution Spectroscopic Surveys: Correcting Wavelength-Dependent and Global Systematics with Application to LAMOST DR9}

%Precision Radial Velocities from Low-Resolution Spectra: A Generalizable Calibration Framework for Segment-Level and Global Systematics

%Towards km/s-Precision Radial Velocities from R ~ 1800 Spectra: A Unified Correction Framework for Segment-Level and Fiber-Level Biases

\author[0009-0005-7743-6229]{Jinming Zhang}
\affiliation{Institute for Frontiers in Astronomy and Astrophysics, Beijing Normal University, Beijing, 102206, China}
\affiliation{School of Physics and Astronomy, Beijing Normal University, Beijing, 100875, China}

\author[0000-0003-2471-2363]{Haibo Yuan}
\affiliation{Institute for Frontiers in Astronomy and Astrophysics, Beijing Normal University, Beijing, 102206, China}
\affiliation{School of Physics and Astronomy, Beijing Normal University, Beijing, 100875, China}

\author[0000-0003-0220-7112]{Zhijia Tian}
\affiliation{Department of Astronomy, Key Laboratory of Astroparticle Physics of Yunnan Province, Yunnan University, Kunming 650200, China}

\correspondingauthor{Haibo Yuan}
\email{yuanhb@bnu.edu.cn}

\begin{abstract}
Radial velocity (RV) is crucial for stellar kinematics and Galactic archaeology. The Large Sky Area Multi-Object Fiber Spectroscopic Telescope (LAMOST) has obtained over ten million low-resolution spectra ($R \sim 1800$), yielding RVs for millions of stars, but these suffer from (1) wavelength-dependent inconsistencies (relative shifts between spectral segments) and (2) global zero-point offsets (uniform shifts of entire spectra). In this work, we comprehensively characterize and correct both. Each spectrum is firstly divided into eight $\sim$500~\AA\ segments. We organize the data at the spectrograph and fiber levels, measure segment-wise RV offsets relative to the full spectrum at each level, then fit these offsets with low-order polynomials to correct wavelength-dependent systematics. We then correct zero-points hierarchically: at the spectrograph level by minimizing a joint $\chi^2$ constrained by repeat observations and cross-matches with APOGEE and Gaia RVS; and at the fiber level by averaging seasonal offsets. After correction, RV precision improves significantly: for cross-night repeats, the standard deviation of RV differences at high signal-to-noise ratios drops by a factor of two from $\sim 3.6$ to $\sim 1.8$~km\,s$^{-1}$, implying a single-measurement precision of $\sim 1.3$~km\,s$^{-1}$. External checks with APOGEE and Gaia show dispersions drop from $\sim 4.0$ to $\sim 2.0$~km\,s$^{-1}$. The precision approaches, though slightly below, the theoretical limit at $R \sim 1800$. We release a value-added RV catalog with corrected velocities for $\sim5.7$ million spectra, providing a homogeneous and systematically corrected dataset. The framework established in this work is also applicable to RV calibration in other large-scale spectroscopic surveys.
\end{abstract}

%\begin{multicols}{2}
\section{Introduction} \label{sec:intro}
Large-scale spectroscopic surveys play an increasingly critical role in modern astronomy. They provide indispensable data for investigating the formation and evolution of stars and galaxies, with radial velocity (RV) being one of the most fundamental physical quantities. In recent years, international spectroscopic surveys such as the Radial Velocity Experiment (RAVE; \citealt{rave2006}), which provided high-quality radial velocities for large stellar samples with typical precision of $\sim$2 kms$^{-1}$, and the Sloan Extension for Galactic Understanding and Exploration (SEGUE; \citealt{yanny2009segue}), which extended radial velocity measurements across the Galactic disk and halo with  typical precision of $\sim$5 kms$^{-1}$, have enabled Milky Way dynamical studies at great details, advancing our understanding of the Galaxy’s formation and evolution.

With its wide field of view and ability to observe 4,000 targets simultaneously, the Large Sky Area Multi-Object Fiber Spectroscopic Telescope (LAMOST; \citealt{cui2012lamost,deng2012lamost,zhao2012lamost,Liu_2014}) has obtained RV measurements for millions of stars, building an unprecedented stellar kinematic database. LAMOST has two observing modes: low resolution ($R \approx 1800$) and medium resolution ($R \approx 7500$; \citealt{lamostMRS}). The low-resolution mode is the primary survey mode and has produced one of the largest RV datasets, providing a robust foundation for studies of stellar populations and Galactic dynamics.

\begin{figure}[htbp]
    \includegraphics[width=0.45\textwidth]{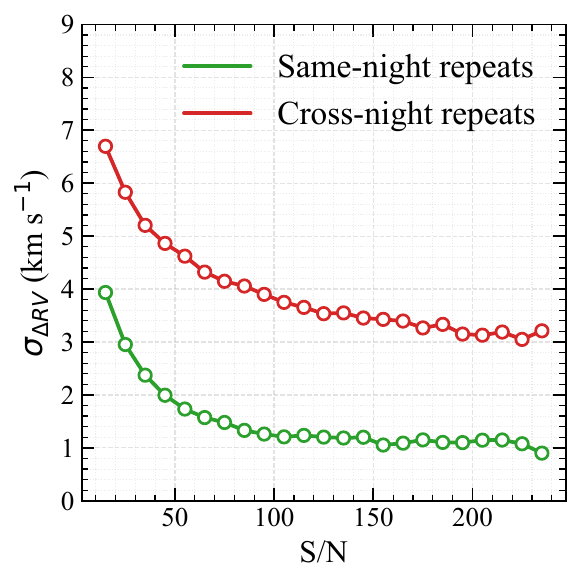}
    \caption{Consistency of radial velocity measurements from LAMOST repeat observations. The sample is divided into S/N bins of width 5 from 5 to 250. For each repeat pair, the lower S/N sets the bin. In each bin, the RV difference distribution is fitted with a Gaussian; its standard deviation ($\sigma$) is plotted as the vertical value. The green curve shows same-night repeats, and the red curve shows cross-night repeats.}
    \label{fig:rv_scatter_same_vs_cross}
\end{figure}

Due to the huge sample size of the LAMOST low-resolution survey, the scientific value of its radial velocities hinges on effective control of systematic errors. In this mode, RV accuracy is highly sensitive to the precision and stability of wavelength calibration. LAMOST wavelengths are first calibrated with arc-lamp spectra from calibration exposures, using a polynomial fit to map CCD pixels to theoretical wavelengths, and then refined with night-sky emission lines \citep{bai2017sky}. Temporal changes in observing conditions and instrument state can still cause calibration drifts between nights, introducing systematic errors that are hard to remove. As a result, radial velocities from observations taken on different nights are much less consistent than those from same-night repeats (see Fig.~\ref{fig:rv_scatter_same_vs_cross}).

Further analysis shows that the systematic errors causing the cross-night inconsistencies fall into two distinct categories:
(1) internal inconsistency of the dispersion solution, where the pixel–wavelength relation is inadequately modeled, leading to wavelength-dependent calibration errors and non-uniform velocity offsets across the spectrum; and
(2) a velocity zero-point offset, where the entire spectrum is uniformly shifted in velocity, creating a systematic difference between measured and true RVs (e.g., \citealt{zhangBO2021MRS,wangR2019MRS}).
In this work, we focus on these two types of systematics, systematically examine their behavior across spectrographs and fibers, and develop corresponding corrections to obtain more accurate radial velocities from the LAMOST low-resolution survey.

The paper is organized as follows. Section 2 describes the LAMOST LRS DR9 dataset, sample selection, and external data.
Section~3 explains how to derive correction curves for internal inconsistencies and determine RV zero points (RVZP) at different levels.
Section~4 analyzes the two systematic errors and summarizes their behavior, forming the basis of our hierarchical correction method.
Section~5 validates the corrected RVs using LAMOST repeat observations and cross-matches with external surveys (\textit{Gaia}, APOGEE, DESI).
Section~6 presents the final data product, and Section~7 summarizes the work.

\section{Data} \label{sec:Data}
\subsection{LAMOST DR9 LRS} \label{sec:dr9}
We utilized the low-resolution spectroscopic survey data from LAMOST DR9 v1.1, which covers both the Pilot Survey conducted from October 2011 to June 2021, and the first nine years of the Regular Survey starting in September 2012. In the DR9 release, the LAMOST Low-Resolution Spectroscopic Survey (LRS) provides a total of 11,226,252 spectra that have been wavelength calibrated, relatively flux calibrated, and sky-subtracted. These include 10,907,516 stellar spectra, 242,569 galaxy spectra, and 76,167 quasar spectra, offering a rich resource for studies of stellar populations, galaxy evolution, and quasar identification. The spectra span a wavelength range of 3690–9100 Å, with a resolving power of about $R \sim 1800$ at 5500 Å. Since the focus of this work is on stellar radial velocities, we restrict our analysis to stellar spectra only. In total, we selected 7,060,436 stellar spectra, each characterized by atmospheric parameters including effective temperature ($T_{\rm eff}$), surface gravity (log $g$), metallicity ([Fe/H]), and radial velocity. These spectra correspond to 5{,}153{,}361 unique targets. These parameters were derived using the LAMOST Stellar Parameter Pipeline (LASP; \citealt{wu2011LASP}).
\subsection{External Data} \label{sec:ext data}

The external radial velocity (RV) data used in this study are primarily drawn from two large-scale surveys: the Gaia mission  \citep{gaiadr3} and the Apache Point Observatory Galactic Evolution Experiment (APOGEE; \citealt{majewski2017apogee}) .

Gaia, launched by the European Space Agency (ESA), is a space-based astronomical mission designed to obtain precise astrometric and spectroscopic information for more than one billion stars \citep{gaia2016gaia}. It carries the Radial Velocity Spectrometer (RVS), which operates in the 847--874\,nm wavelength range and is capable of measuring line-of-sight velocities for stars brighter than $G \lesssim 14$ mag \citep{cropper2018gaia}. In Gaia DR3, the RVS provides radial velocities for about 33,812,183 stars. The key advantages of Gaia are its all-sky coverage and extremely large sample size. Its RV precision can reach better than 0.3\,km\,s$^{-1}$ for bright stars ($G < 12$ mag), but degrades rapidly to several km\,s$^{-1}$ for fainter targets ($G > 14$ mag). This makes Gaia data particularly suitable for large-sample statistical comparisons. In this work, we only retain Gaia sources with RV uncertainties smaller than 5\,km\,s$^{-1}$.

APOGEE is a sub-project of the Sloan Digital Sky Survey (SDSS), carried out with the 2.5\,m SDSS telescope using a high-resolution near-infrared H-band spectrograph ($R \sim 22,500$). APOGEE primarily targets red giant stars in the Milky Way. In APOGEE DR17, spectra for about 650,000 stars have been released, with typical RV precision better than 0.1\,km\,s$^{-1}$. The main strengths of APOGEE lie in its high spectral resolution and very high RV precision, but its sky coverage is much smaller than that of Gaia, focusing mainly on the Galactic disk and volume-limited fields.
%\begin{figure}
    %\centering
    %\includegraphics[width=0.4\textwidth]%{hist_gaia_apo_redlines.pdf}
    %\caption{Distribution of radial-velocity differences between common stars in \textit{Gaia}~DR3 and APOGEE~DR17, defined as $\Delta \mathrm{RV} = \mathrm{RV}_\mathrm{APOGEE} - \mathrm{RV}_\mathrm{Gaia}$. The solid curve shows a Gaussian fit to the histogram, with a median offset of $\mu = 0.21~\mathrm{km\,s^{-1}}$ and a dispersion of $\sigma = 0.76~\mathrm{km\,s^{-1}}$, as indicated in the upper-left corner. Vertical dashed lines mark $\mu \pm \sigma$.}
    %\label{fig:hist_gaia_apo}
%\end{figure}

It is worth noting that Gaia and APOGEE can be used as reliable external references. By combining the two surveys, we are able to achieve both high RV precision and a large sample size, thereby providing a more robust characterization of the systematic properties of LAMOST RVs.

\section{Method}
\subsection{Overview of the Correction Framework}

\begin{figure}[htbp]
    \centering
    \includegraphics[width=0.5\textwidth]{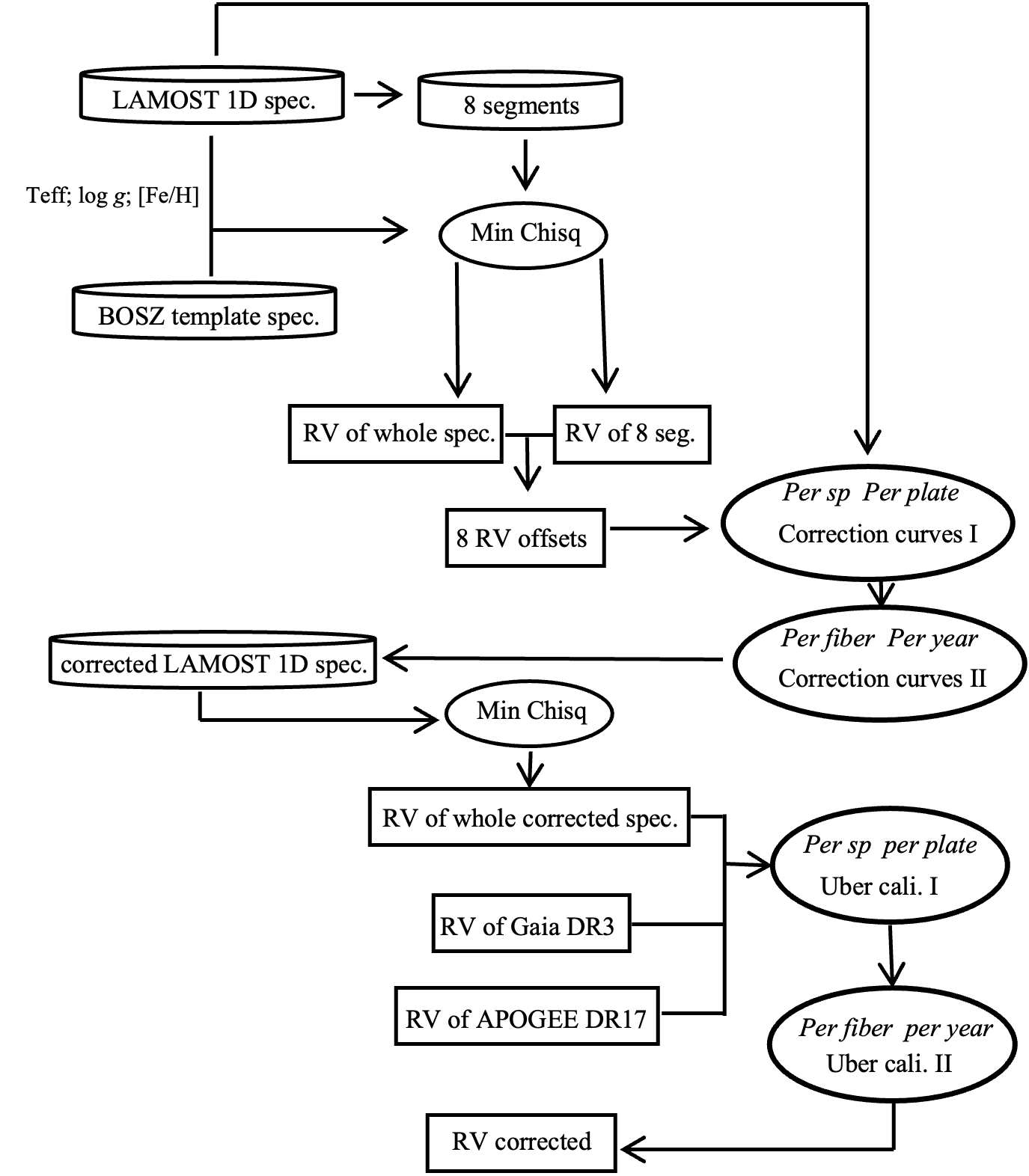}
    \caption{Flowchart of this work}
    \label{fig:flowchart}
\end{figure}

The precision of LAMOST low-resolution spectra is affected by multiple sources of systematic error, including instrumental drifts, environmental variations, and imperfections in wavelength calibration. As discussed in the introduction, these errors manifest in two distinct forms: 
(1) wavelength-dependent distortions, and 
(2) global zero-point shifts of the entire spectrum. 
To ensure both the accuracy and consistency of the LAMOST RVs, we constructed a hierarchical correction framework that identifies and eliminates different components of the systematic errors at multiple levels (see Fig.~\ref{fig:flowchart}).

The correction procedure consists of two major stages.  
In the first stage, we address the wavelength-dependent inconsistencies.  
For each spectrum, we measure relative velocity offsets across wavelength segments, compare these systematic offsets across instruments and observing timescales to identify common wavelength-dependent trends, and then define data groups to which a single correction function can be applied. 
We then fit wavelength-dependent polynomials to these segmental velocity offsets as functions of wavelength, separately for the blue and red arms. 
This yields wavelength calibration functions at the spectrograph and fiber levels, which are used to correct local calibration distortions.

In the second stage, we focus on the global zero-point shifts.  
We combine internal consistency constraints, derived from repeated LAMOST observations, with external constraints from APOGEE and \textit{Gaia} RVS, to perform multi-level zero-point calibration at both the spectrograph and fiber levels.  

This two-tier correction strategy -- addressing the wavelength-dependent distortions first and the global zero-point offsets afterwards -- ensures that the wavelength calibration achieves both local precision and global uniformity across the entire LAMOST observing system.  
The detailed methodology of each step is described in the following subsections.

\subsection{Internal inconsistency of LAMOST LRS wavelength calibration}

\begin{figure*}[htbp]
    \centering
    \includegraphics[width=0.95\textwidth]{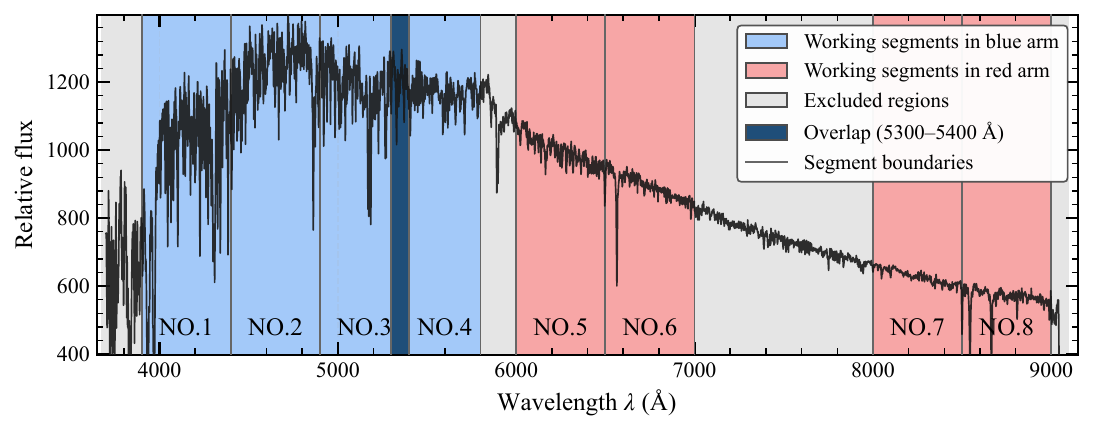}
    \caption{This figure illustrates the segmentation strategy applied to a representative low-resolution spectrum from LAMOST LRS DR9. The entire spectrum is divided into eight working segments, each used for independent radial velocity measurements. For clarity, the four segments in the red arm are highlighted with red shading, while the four segments in the blue arm are shaded in light blue. The overlapping region between the two arms (5300–5400\,\AA) is shown in dark blue. The gray regions mark the wavelength ranges excluded from the calculation. Each valid segment is labeled sequentially as NO.1–NO.8 according to its central wavelength from low to high.}
    \label{fig:lamost_segments}
\end{figure*}

In wavelength calibration, any deviation in the pixel–wavelength relation introduces a systematic error in the spectrum at the corresponding wavelength.  
According to the Doppler formula, a wavelength error $\Delta \lambda$ can be approximately expressed as a radial velocity error:  
\begin{equation}
\Delta RV \approx c \cdot \frac{\Delta \lambda}{\lambda},
\label{eq:Doppler relation}
\end{equation}
where $c$ is the speed of light, $\lambda$ is the wavelength, and $\Delta \lambda$ represents the calibration offset at that wavelength.  
Therefore, by measuring radial velocities in different wavelength segments, the local wavelength errors can be directly reflected by the deviations between the velocities from individual segments and that from the full spectrum.  
Specifically, following the approach of \cite{yuan2021star}, we divide each spectrum into several segments and independently measure the radial velocity of each segment using the minimum $\chi^2$ method.  
The velocity offset of each segment is then defined as the difference between the velocity measured in that segment and the average velocity derived from the entire spectrum:  

\begin{equation}
\Delta RV_{\text{seg}} = RV_{\text{seg}} - RV_{\text{all}}.
\end{equation}

In this way, wavelength calibration errors can be translated into relative velocity offsets among spectral segments, providing a more intuitive and physically medianingful representation. To investigate possible systematic patterns, we divide each spectrum into eight segments of about 500 Å (see Fig.~\ref{fig:lamost_segments}): 3900–4400 Å, 4400–4900 Å, 4900–5400 Å, 5300–5800 Å, 6000–6500 Å, 6500–7000 Å, 8000–8500 Å, and 8500–9000 Å. The radial velocity of each segment is measured independently. In this division we deliberately avoid certain problematic regions: 3670–3900 Å; 5800–6000 Å, the junction of the blue and red arms of LAMOST spectra, where independent calibration of the two arms may introduce discontinuities; 7000–8000 Å, where absorption lines are sparse and velocity determinations are unreliable; and 9000–9100 Å. These regions are therefore excluded from the velocity analysis.

The choice of eight segments is based on the following considerations: each segment must contain sufficient absorption lines to ensure reliable RV measurement; too few segments would fail to reveal systematic differences across wavelength, whereas too many would reduce the statistical robustness of individual segments. This segmentation thus strikes a balance between reliability and resolution, and is sensitive to wavelength-dependent calibration errors. Our results show that radial velocities among segments are not mutually consistent, and some segments even display significant discrepancies. We refer to this phenomenon as the internal inconsistency of spectra. Quantifying this inconsistency is crucial for improving the reliability of LAMOST radial velocities.

After obtaining the RV measurements in eight spectral segments for each spectrum, performing independent corrections for every individual spectrum. Therefore, it is necessary to seek a balance between computational efficiency and correction accuracy, and to design a strategy suitable for bulk spectral correction. To this end, we adopt a hierarchical evaluation framework, progressing from the smallest units (within a single spectrograph on a given plate) to larger scales (data across an entire observing year), in order to reveal potential systematic patterns. Specifically, we first examine the differences among segments within individual spectra, thereby assessing the wavelength-calibration error as a function of wavelength; second, we compare spectra from different fibers within the same spectrograph, in order to identify possible fiber-dependent systematic deviations; finally, we analyze the overall offsets among different spectrographs on the same plate, to investigate inter-spectrograph calibration differences as well as the influence of common observational conditions on systematic errors. Since the minimal correction unit cannot be predetermined, we proceed step by step, identifying stable patterns at each level of comparison. Using the statistical properties of these patterns and the minimal calibration unit, we fit the radial-velocity offsets of the eight spectral segments—described by the Gaussian-fitted mean value $\mu$—as a function of their central wavelength. We do not explicitly account for RV-variable objects (e.g., spectroscopic binaries or pulsators), because $\mu$ is dominated by the near-zero core and is insensitive to a few outliers. Thus, when RV-variable objects are rare (e.g., \citealt{tian2020catalog}), they do not significantly bias the final correction. For this purpose, we adopt separate quadratic polynomial fits for the blue and red parts of the spectra. This division is necessary because the LAMOST spectrographs employ a dichroic mirror to split the incoming light into blue and red channels, which are calibrated independently; the constraints and error sources differ between the two channels, and thus the wavelength calibration errors often exhibit distinct trends. Consequently, we obtain two functions, \(\Delta RV_{\mathrm{red}}(\lambda)\) and \(\Delta RV_{\mathrm{blue}}(\lambda)\), which are then converted through equation\eqref{eq:Doppler relation} into wavelength correction curves \(\Delta \lambda_{\mathrm{red}}(\lambda)\) and \(\Delta \lambda_{\mathrm{blue}}(\lambda)\). Finally, these correction curves are applied to bulk spectra to improve the consistency of radial velocity measurements.

\subsection{Zero point: spectrograph level}

The radial-velocity zero point denotes the global systematic offset between the wavelength scale of the observed spectra and the true physical wavelength scale, which manifests as a uniform shift of the entire spectrum in radial velocity. This type of error is distinct from the wavelength-dependent inconsistency discussed previously: the zero-point error corresponds to a global offset that is common to all spectral segments, whereas the wavelength-dependent error reflects relative discrepancies among different segments. Consequently, even after correcting for the wavelength-dependent inconsistency, it remains necessary to calibrate the zero point in order to ensure that the velocity scales of spectra obtained with different spectrographs, plates, fibers, and during different observing epochs are mutually consistent.

Analogous to the procedure adopted in the previous step, we first identify the most suitable minimal correction unit. This choice is guided by both physical and technical considerations. Because each spectrograph undergoes an independent wavelength calibration, it is natural to regard the spectrograph as the fundamental calibration unit. At the fiber level, individual fibers can exhibit distinct zero-point offsets under varying observing conditions and therefore can also be treated as independent correction units. Furthermore, zero-point drifts may arise between different observing years due to changes in the instrumental configuration or performance.

On the basis of these considerations, we define two hierarchical levels of correction in this study:  
(1) the \textit{spectrograph level}, at which each spectrograph on each plate is treated as an independent calibration unit, reflecting the fact that wavelength calibration is performed separately for each spectrograph; and  
(2) the \textit{fiber level}, at which all observations obtained with the same fiber within a single observing year (from September to July) are combined into one unit, thereby capturing long-term zero-point drifts induced by instrumental variations.

We then apply the so-called \textit{uber-calibration} method \citep{ubercali2008}, in which a global $\chi^2$ function is constructed to simultaneously satisfy two categories of physical constraints:  
(1) an \textit{internal constraint}, which minimizes the velocity differences among repeated LAMOST observations of the same targets; and  
(2) an \textit{external constraint}, which anchors the LAMOST velocity scale to external reference systems, such as APOGEE and \textit{Gaia} RVS, via sources observed in common.

For the zero-point correction at the \textit{spectrograph level} within each plate, we define the following $\chi^2$ function:
\begin{align}
\chi^2 &= \Lambda_1\frac{\sum_{j=1}^{m_1-1}\sum_{i=1}^{m_1}{n_{ij}\Big(\overline{(v_i - v_j)} + (\mu_i - \mu_j)\Big)^2}}{N_1} \nonumber \\
&\quad + \Lambda_2\frac{\sum_{i=1}^{m_2}{n_i^{\mathrm{apo}}\Big(\overline{(v_i - v_{\mathrm{apo}})} + \mu_i\Big)^2}}{N_2} \nonumber \\
&\quad + \Lambda_3\frac{\sum_{i=1}^{m_3}{n_i^{\mathrm{rvs}}\Big(\overline{(v_i - v_{\mathrm{rvs}})} + 0.2 + \mu_i\Big)^2}}{N_3},
\end{align}

where the first term encodes the constraint from repeated sources. For each pair of units $(i,j)$, we adopt the Gaussian-fitted central value $\mu$ as the velocity differences for their common sources $\overline{(v_i - v_j)}$, which is robust against a small number of RV-variable outliers, and weight this quantity by $n_{ij}$, the number of shared sources. This contribution drives $\Big(\overline{(v_i - v_j)} + (\mu_i - \mu_j)\Big) \to 0$. The second term represents the constraint from APOGEE: for each unit $i$, we evaluate the mean value of its velocity differences relative to APOGEE, $\overline{(v_i - v_{\mathrm{apo}})}$, weighted by $n_i^{\mathrm{apo}}$, the corresponding number of common sources. This enforces $\Big(\overline{(v_i - v_{\mathrm{apo}})} + \mu_i\Big)\to 0$. The third term incorporates the \textit{Gaia} RVS constraint, defined analogously using $\overline{(v_i - v_{\mathrm{rvs}})}$ and $n_i^{\mathrm{rvs}}$. A constant offset of $+0.2\,\mathrm{km\,s^{-1}}$ is applied to the \textit{Gaia} velocities to correct for the known systematic offset with respect to APOGEE. We apply a uniform correction of $+0.2\,{\rm km\,s^{-1}}$ to the \textit{Gaia} RVs. The \textit{Gaia} DR3 RVS zero-point offset depends on magnitude and color (\citealt{katz2023}; \citealt{blomme2023}), but the magnitude-dependent correction is typically $\sim 0$--$0.5\,{\rm km\,s^{-1}}$, comparable to our adopted constant. Using the literature magnitude-dependent prescription produces no significant change in the zero-point solution or precision metrics, so we adopt the constant correction for simplicity. The coefficients $\Lambda_1$, $\Lambda_2$, and $\Lambda_3$ regulate the relative weights of these three constraints. We assign different weights to the three constraints because internal repeat observations have many pairs but low single-measurement precision, whereas the Gaia/APOGEE cross-match has far fewer sources but much smaller uncertainties and a more stable radial-velocity zero point. With equal weights, the objective function would be dominated by the internal repeats, weakening the influence of the high-precision external anchors on the zero-point solution. Using weights prevents this dilution and ensures the final solution is both internally consistent and well tied to the external absolute zero point. In this analysis we empirically adopt $\Lambda_1=1$, $\Lambda_2=2$, and $\Lambda_3=2$.

By taking the derivative of $\chi^2$ with respect to each $\mu_i$ and setting it equal to zero, we obtain  
\begin{align}
\forall i = 1, \ldots, m_1 : \quad
\frac{\partial \chi^2}{\partial \mu_i}
&= 2 \Lambda_1 \sum_{j=1}^{m_1} \frac{n_{ij}}{N_1} 
   \big[ \overline{(v_i - v_j)} + (\mu_i - \mu_j) \big] \nonumber \\
&\quad + 2 \Lambda_2 \, \frac{n_i^{\mathrm{apo}}}{N_2} 
   \big[ \overline{(v_i - v_{\mathrm{apo}})} + \mu_i \big] \nonumber \\
&\quad + 2 \Lambda_3 \, \frac{n_i^{\mathrm{rvs}}}{N_3} 
   \big[ \overline{(v_i - v_{\mathrm{rvs}})} + 0.2 + \mu_i \big] \nonumber \\
&= 0 .
\end{align}

This set of equations can be compactly expressed in matrix notation as  
\begin{equation}
\frac{\partial \chi^2}{\partial \boldsymbol{\mu}} = 2A\boldsymbol{\mu} + \mathbf{b} = 0 ,
\end{equation}
which yields the linear system
\begin{equation}
A\boldsymbol{\mu} = -\tfrac{1}{2}\mathbf{b}.
\end{equation}

In this formulation, the diagonal elements of the coefficient matrix \(A\) are predominantly determined by weights (e.g., constraints from APOGEE, Gaia, or self-pair measurements), whereas the off-diagonal elements are induced by constraints involving repeated sources \((i,j)\). The right-hand-side vector \(\mathbf{b}\) is assembled from the weighted observed velocity offsets \(\bigl(\Delta v(i-j),\, \Delta v(i-\mathrm{apo}),\, \Delta v(i-\mathrm{rvs})\bigr)\). Aggregating all such contributions yields the full matrix \(A\) and vector \(\mathbf{b}\), from which the zero-point correction \(\mu_i\) for each unit can be inferred.

It should be emphasized that even if a correction unit $i$ has very few, or no, external common sources, it can still be constrained indirectly via its repeated sources in common with other units. Within our $\chi^2$ formalism, the $(\mu_i - \mu_j)$ term couples $\mu_i$ to any unit $\mu_j$ that is anchored by external references. Consequently, as long as unit $i$ shares at least one repeated source with a unit that is externally anchored, its zero point remains determinable. Conversely, if a unit has neither external matches nor repeated sources, its zero point cannot be solved within this framework. The resulting vector $\boldsymbol{\mu}$ therefore specifies the zero-point corrections for all solvable units. Subtracting the corresponding $\mu_i$ from each source (already corrected for wavelength-dependent systematics) completes the zero-point homogenization procedure.

\subsection{Zero point: fiber level}
Following the zero-point correction applied at the \textit{spectrograph level}, we identified small yet temporally stable zero-point offsets in the observations associated with individual fibers across different observing years. This behavior is likely attributable to the fact that different fibers exhibit intrinsic, fiber-specific zero-point deviations that remain approximately constant under varying observational conditions. If not corrected, such local biases can propagate and accumulate, ultimately giving rise to residual systematic errors in the final radial velocity (RV) measurements. The implementation of this additional correction, once identified, led to a substantial improvement in the accuracy of the RV determinations and thus represents a significant methodological contribution of this work. We therefore proceed to further calibrate the remaining systematic offsets at the fiber level.

In this step, all observations obtained with a given fiber in one observing year are treated as a single calibration unit. This allows us to capture both the long-term stability of each fiber and year-to-year instrumental changes. As in the previous step, we derive the calibration by minimizing a composite $\chi^2$ function combining \textit{internal} and \textit{external} constraints:
\begin{align}
\chi^2 &= 
\Lambda_1\frac{\sum_{j=1}^{m_1-1}\sum_{i=1}^{m_1}
{n_{ij}\Big(\overline{(v_i - v_j)} + (\mu_i - \mu_j)\Big)^2}}{N_1} \nonumber \\
&\quad + 
\Lambda_2\frac{\sum_{i=1}^{m_2}
{n_i^{\mathrm{ext}}\Big(\overline{(v_i - v_{\mathrm{ext}})} + \mu_i\Big)^2}}{N_2},
\end{align}
where the first term minimizes internal velocity differences among repeat observations of the same fiber, and the second enforces consistency between LAMOST radial velocities and external standards. Here, $m_1$ is the number of fiber-level calibration units entering the internal term (typically all units), and $m_2$ is the subset with enough external matches to enter the external term ($m_2 \le m_1$). The weights $n_{ij}$ and $n_i$ are the numbers of common sources for unit pair $(i,j)$ and of external matches for unit $i$, while $N_1=\sum_{1\le i<j\le m_1} n_{ij}$ and $N_2=\sum_{i=1}^{m_2} n_i$ are their normalization factors. The $\overline{(v_i-v_j)}$ is the mean of the RV differences of the common sources and the $\overline{(v_i-v_{\mathrm{ext}})}$ he RV differences relative to the external reference velocity. The coefficients $\Lambda_1$ and $\Lambda_2$ set the relative weights of the internal and external terms, we adopt  $\Lambda_1=1$, $\Lambda_2=2$ and $\mu_i$ is the zero-point correction for unit $i$.

The external reference velocity, $v_{\mathrm{ext}}$, is determined as a weighted average of the \textit{Gaia} and APOGEE radial velocities:
\begin{equation}
v_{\mathrm{ext}} = 
\frac{\bigl(v_{\mathrm{Gaia}} + 0.2\bigr)\, n_{\mathrm{Gaia}} + v_{\mathrm{APO}}\, n_{\mathrm{APO}}}
     {n_{\mathrm{Gaia}} + n_{\mathrm{APO}}},
\end{equation}
where a constant offset of $+0.2\,\mathrm{km\,s^{-1}}$ is applied to the \textit{Gaia} velocities to correct for the known systematic zero-point difference with respect to the APOGEE.

When the total number of external common sources is small, $(n_{\mathrm{Gaia}} + n_{\mathrm{APO}}) < 20$, we adopt a reference velocity defined at the fiber-bundle level. In LAMOST, each contiguous group of 25 fibers forms a \textit{fiber bundle}, within which all fibers share a similar observational environment and thus show similar RV zero-point offsets. In this case, the external reference velocity for fiber $i$ is the weighted mean of all common sources in its bundle:
\begin{equation}
v_{\mathrm{ext,\,bundle}} =
\frac{\sum_{\mathrm{bundle}} 
      \big[(v_{\mathrm{Gaia}} + 0.2)\, n_{\mathrm{Gaia}} 
          + v_{\mathrm{APO}}\, n_{\mathrm{APO}}\big]}
     {\sum_{\mathrm{bundle}} (n_{\mathrm{Gaia}} + n_{\mathrm{APO}})}.
\end{equation}

By determining optimal zero-point corrections $\mu_i$, we derived fiber-specific radial velocity offsets for each observing year, further homogenizing the LAMOST velocities and reducing small-scale residual systematics across all fibers.

\section{Result}

\subsection{Correction about Internal inconsistency}

To provide a more intuitive demonstration of the wavelength-dependent inconsistency, we selected a representative spectrograph from a single observing plate and visualized the radial-velocity (RV) deviations for all of its fibers. For each fiber, we calculated the RV difference between each of the eight spectral segments and the RV measured from the full spectrum, and displayed these differences in the form of a two-dimensional map. This representation simultaneously encodes the wavelength dimension (vertical axis) and the spatial arrangement of the fibers (horizontal axis), thereby exposing the internal spatial structure of the calibration errors within the spectrograph. In this manner, the relative offsets among the different spectral segments, as well as their variation across the spectrograph, can be more clearly identified and analyzed.

\begin{figure}[htbp]
    \centering
    \includegraphics[width=\columnwidth]{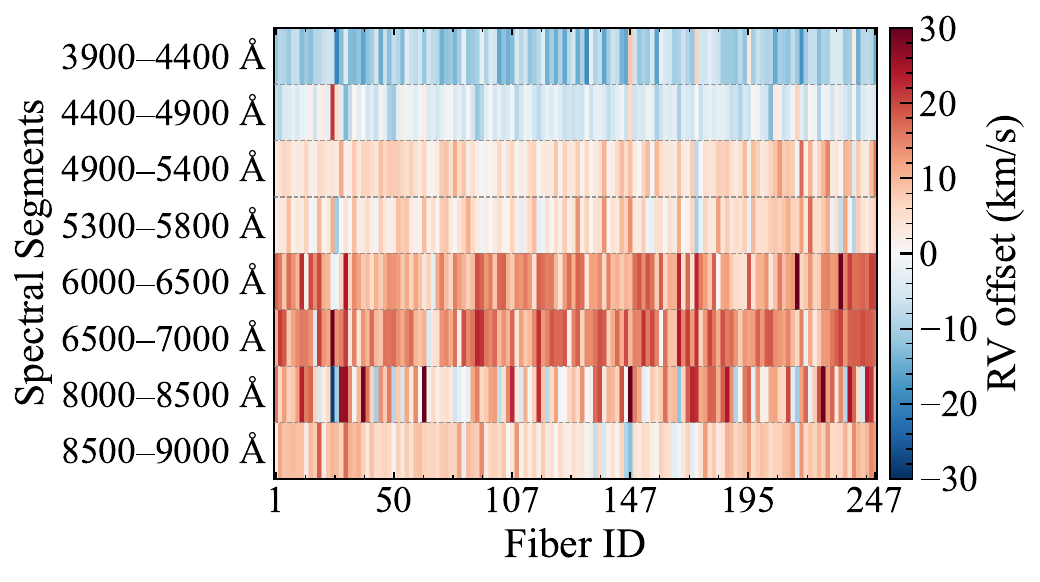}
    \caption{This figure shows segmented velocity offsets for 138 low-resolution spectra taken with spectrograph No.\,9 in LAMOST plate GACII008N38B1 on December 1, 2019. The horizontal axis is fiber ID, and the vertical axis shows eight spectral segments (3900--9000\,\AA). The color indicates the difference between the radial velocity from each segment and that from the full spectrum. Gray dashed lines separate the segments..}
    \label{fig:rv_offset}
\end{figure}

\begin{figure}[htbp]
    \centering
    \includegraphics[width=\columnwidth]{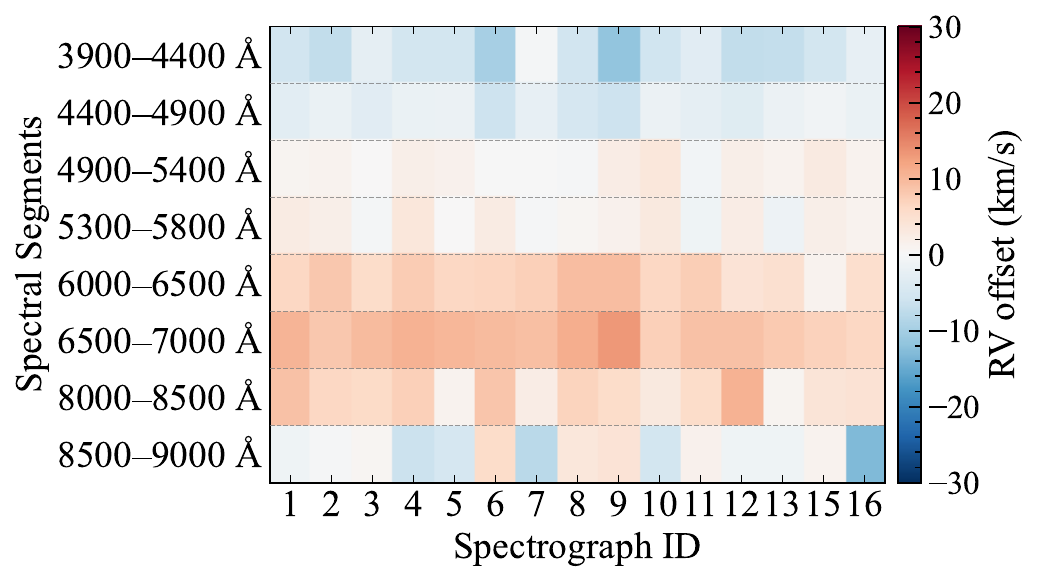}
    \caption{Segmented velocity offsets for all spectrographs in LAMOST plate GACII008N38B1 on December 1, 2019. The horizontal axis is spectrograph ID, and the vertical axis shows eight spectral segments (3900–9000\,\AA), separated by gray dashed lines. For each spectrograph and segment, the offset is the $\mu$ of the segmental radial velocity differences relative to the full-spectrum velocity, derived from Gaussian fits. Color indicates the value of these offsets, highlighting systematic velocity shifts across segments and spectrographs.}
    \label{fig:rv_offset_spid}
\end{figure}

%\begin{figure}[htbp]
    %\centering
    %\includegraphics[width=0.48\textwidth]{fig_sp9_plate_heatmap.pdf}
    %\caption{Heatmap of RV offsets for spectrograph 9 within one night, with spectral segments on the vertical axis, plates on the horizontal axis, and color indicating velocity offsets in km s$^{-1}$.}
    %\label{fig:sp9_plate_heatmap}
%\end{figure}

\begin{figure*}[htbp]
    \centering
    \includegraphics[width=0.85\textwidth]{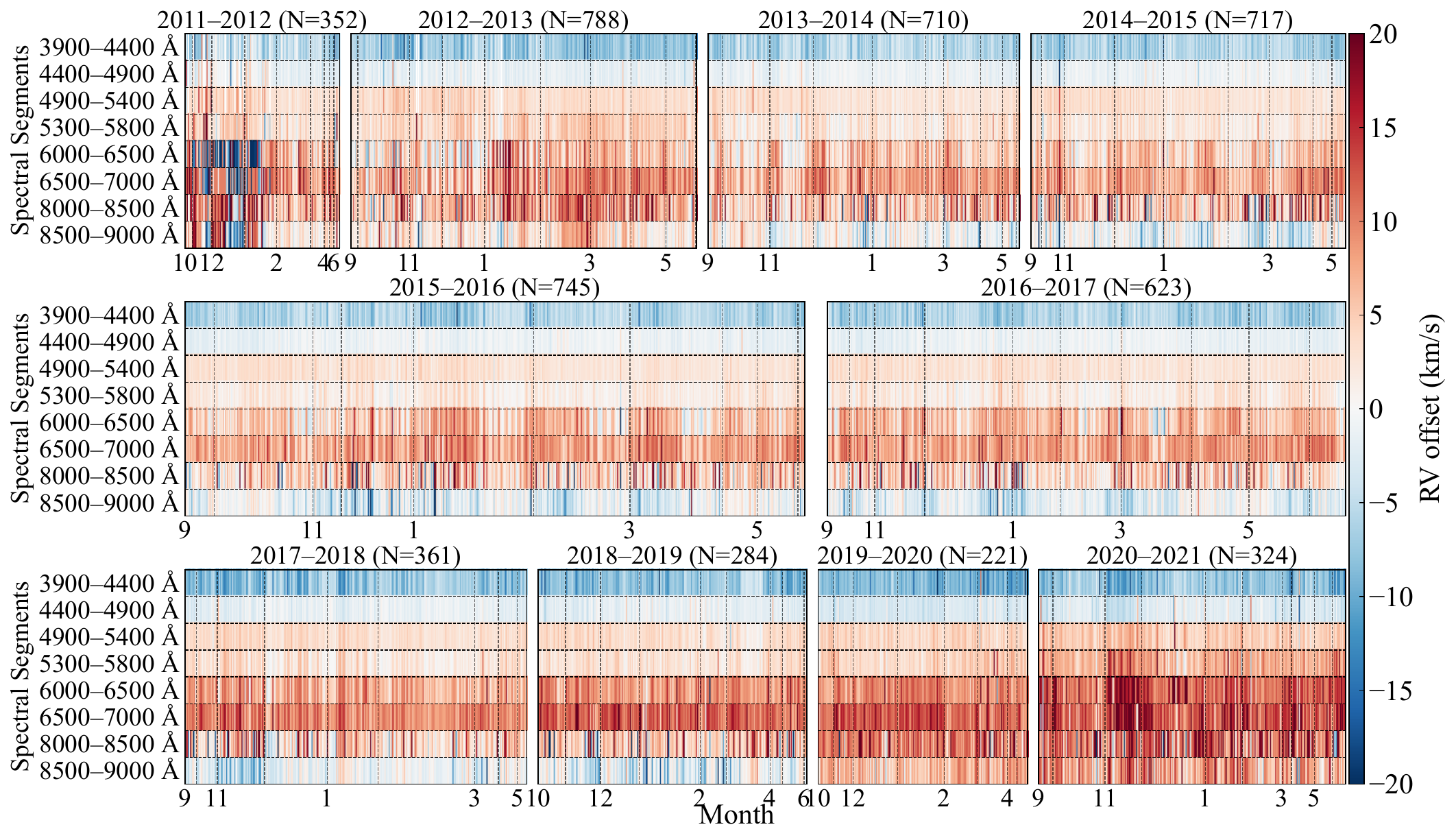}
    \caption{Segmental velocity offsets of Spectrograph\,9 by observing year. The offset is the difference between each segment’s radial velocity and that from the full spectrum, using the per-exposure $\mu$ offsets from a Gaussian fit to all spectra in that exposure. Each grid cell shows the offset for one segment in a single exposure. Panels show different years, with subtitles listing the year range and number of observing nights. The horizontal axis is time, labeled at the first day of each month; the vertical axis lists the eight spectral segments (3900--9000\,\AA). The color bar gives the velocity offset magnitude.}
    \label{fig:sp9_season_heatmaps}
\end{figure*}

Fig.~\ref{fig:rv_offset} shows velocity offsets across eight spectral segments for different fibers within a spectrograph on a plate. Most fibers share a similar pattern: negative offsets at the start of the blue arm (3900--4900\,\AA{}), positive offsets in the red arm (6000--9000\,\AA{}), and near-zero offsets in between. Fiber-to-fiber differences are mostly local and much smaller than this global trend, and are dominated by random S/N–related errors rather than stable systematics, so they average out statistically. Sources observed with the same spectrograph show highly consistent distributions, with no obvious systematic trends with stellar parameters, indicating shared calibration conditions. We therefore analyze velocity offsets of different sources within each segment and fit them with Gaussian functions, adopting the fitted $\mu$ as the representative velocity offset. Even when a distribution is not perfectly Gaussian, the fit yields a robust central value that suppresses noise and outliers and better reflects the true systematic offset. To ensure reliable Gaussian fitting, we require a minimum sample size of $N \ge 15$ for each calibration unit. The information in each spectral segment is limited, and template matching in some segments can fail to deliver stable RV measurements, which are then discarded, reducing the usable sample. For calibration units with too few samples, we do not fit a correction curve and mark them as unsolved; therefore, a small fraction of spectrograph (or fiber) units lack correction curves.

Fig.~\ref{fig:rv_offset_spid} shows that different spectrographs of the same plate have similar segment-wise velocity-offset trends but differ in absolute values, indicating real systematics from their independent wavelength calibrations. Spectrographs on the same plate share similar patterns, suggesting common environmental factors, such as that night’s observing conditions. Thus, the correction unit cannot be the plate because inter-spectrograph differences are significant, or the single fiber because fiber-to-fiber variations are dominated by random noise. We therefore adopt each spectrograph within a plate as the basic correction unit, capturing its independent systematics while maintaining statistical robustness.

\begin{figure}[H]
    \centering
    \includegraphics[width=0.48\textwidth]{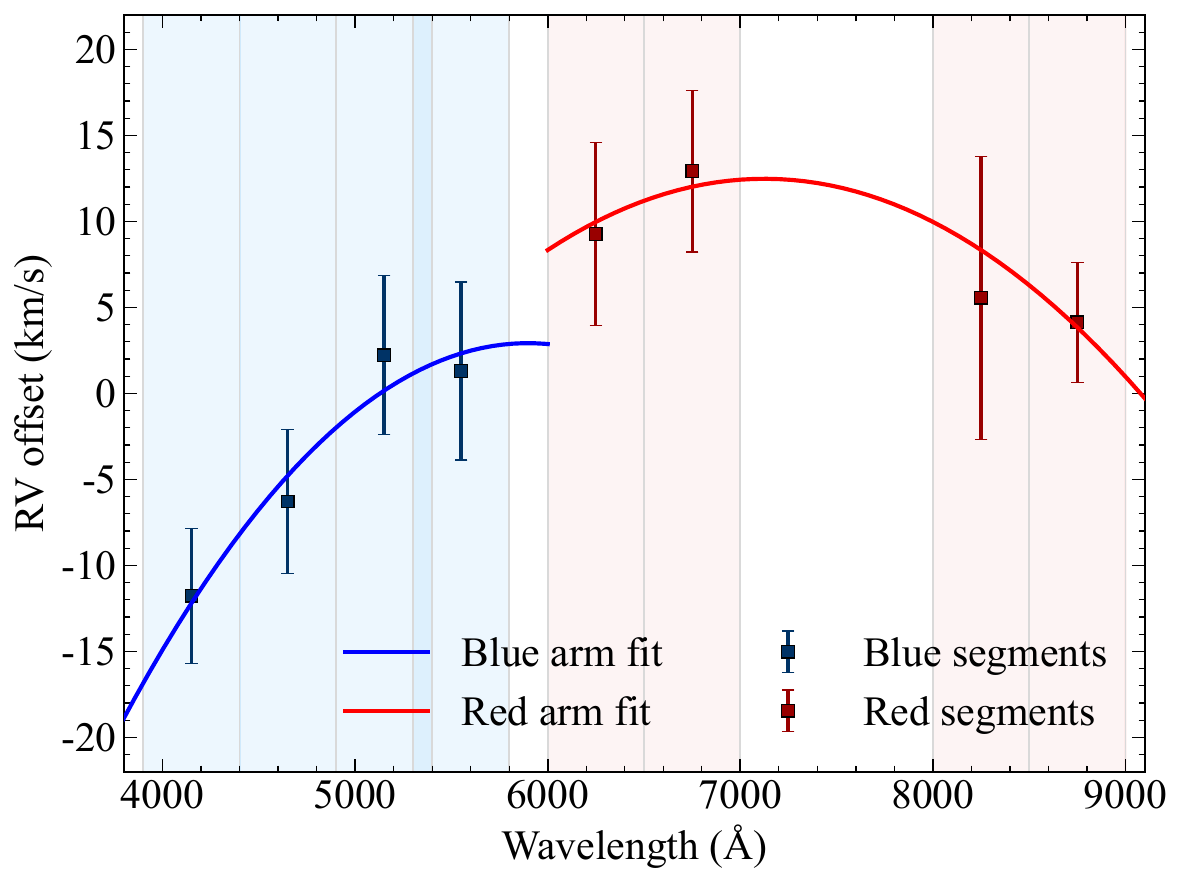}
    \caption{Example of velocity-offset correction curves for Spectrograph\,9 on LAMOST plate GACII008N38B1, observed on December 1, 2019. The horizontal axis is wavelength; the vertical axis is RV offset. Dark blue and dark red points show the offsets ($\mu$) of individual spectral segments in the blue and red arms, with error bars giving their deviations ($\sigma$). Each point is at the segment’s central wavelength. Blue and red curves are polynomial fits to the four blue-arm and four red-arm segments, showing the wavelength dependence of the RV offsets. Light blue and light red shading indicate the wavelength ranges of segments used for RV measurements.}
    \label{fig:rv_offset_sp9}
\end{figure}

\begin{figure*}[htbp]
    \centering
    \includegraphics[width=\textwidth]{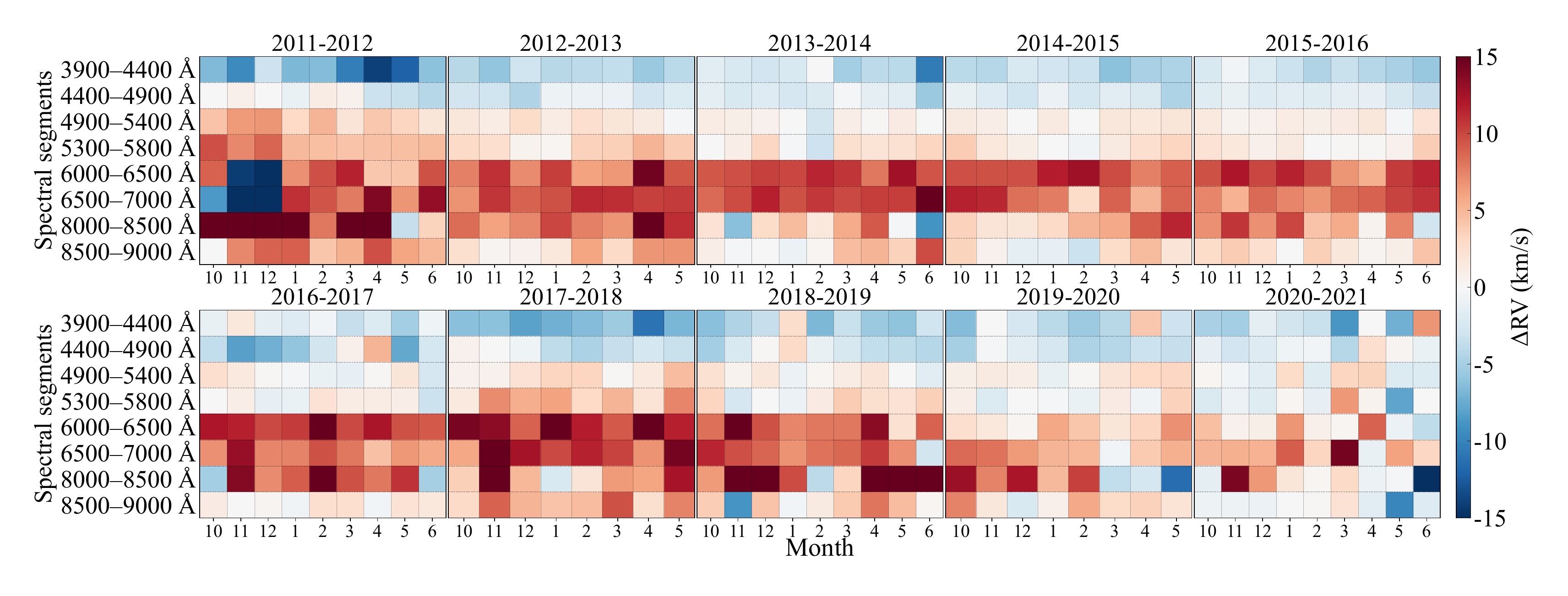}
    \caption{
    Monthly median residual velocity offsets for fiber 217 of spectrograph 3 after removing spectrograph-level offsets. Each observing year is shown in a separate panel, with months on the horizontal axis, spectral segments on the vertical axis, and color indicating residual velocity offset, enabling direct comparison of systematic deviations across years and spectral ranges.}
    \label{fig:fiber217_spectro3_heatmap}
\end{figure*}

\begin{figure*}[htbp]
    \centering
    \includegraphics[width=\textwidth]{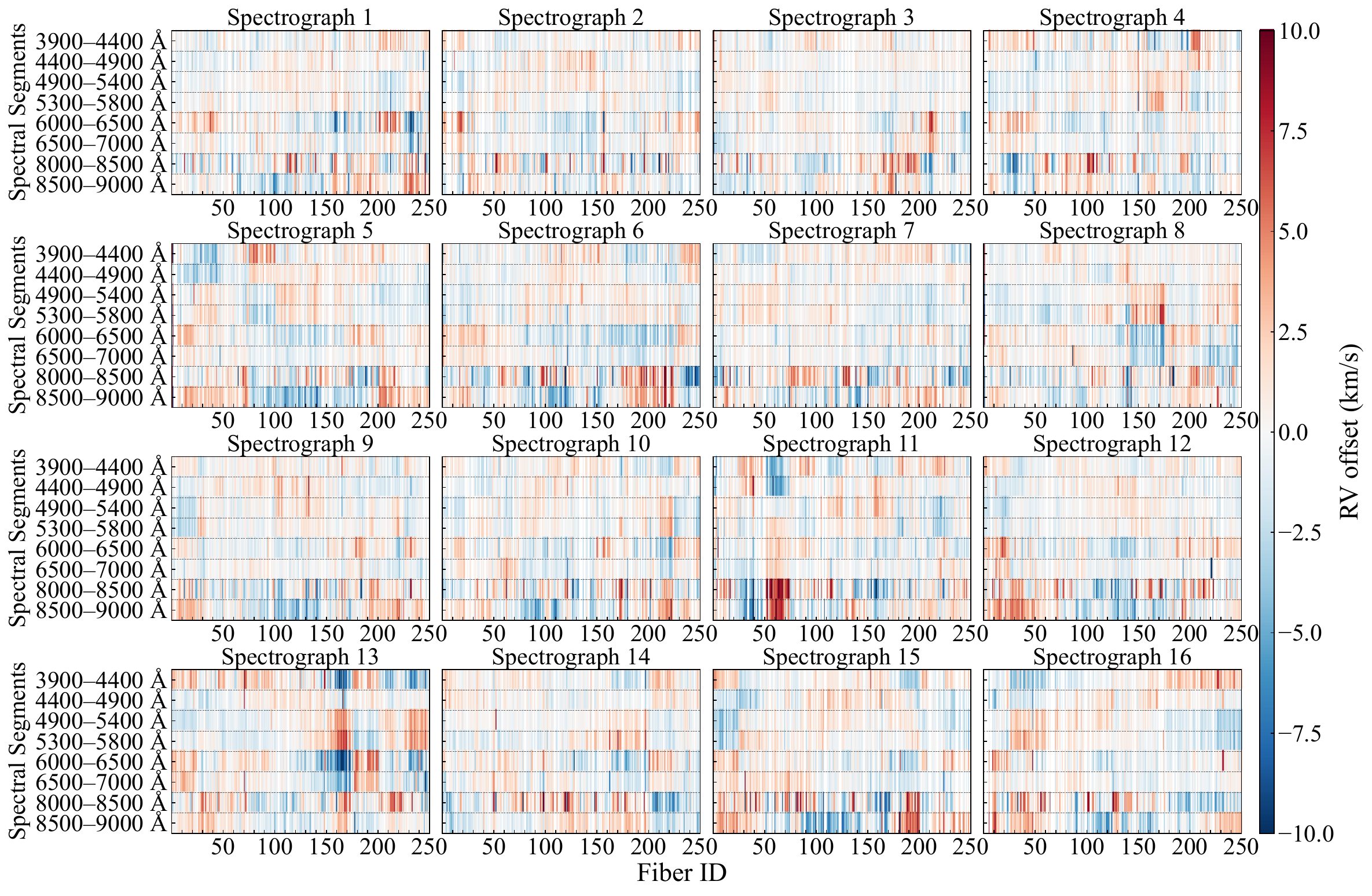}
    \caption{Radial velocity offsets per fiber for 2017–2018, after subtracting the overall spectrograph-level offset. For each fiber, RV offsets in eight spectral segments over the year were collected, and the offset was the $\mu$ obtained from a Gaussian fit. Each grid cell shows the segment-wise RV offset for one fiber. The horizontal axis is fiber ID; the vertical axis is the eight spectral segments. White vertical bands mark dead fibers (e.g., around Fiber 160 in Spectrograph 1). The color bar shows the RV offset magnitude.}
    \label{fig:rv_heatmap_sp1-16_2017-2018}
\end{figure*}

\begin{figure}[htbp]
    \centering
    \includegraphics[width=\linewidth]{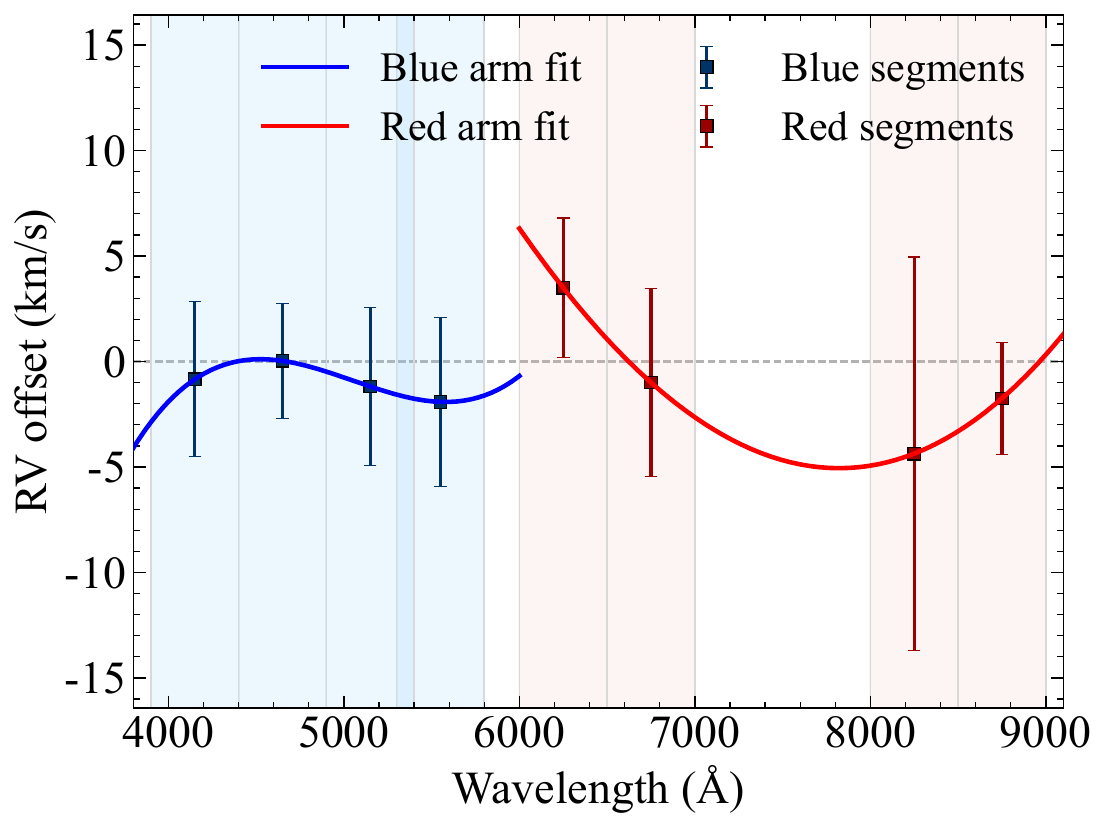}
    \caption{Example correction curve for Fiber\,3 of Spectrograph\,9 in the 2019--2020 observing year. The horizontal axis is wavelength and the vertical axis is RV offset. Blue and red points with error bars show the $\mu$ and $\sigma$ of RV offsets in the blue- and red-arm segments, plotted at each segment’s central wavelength. Solid blue and red curves are cubic polynomial fits to the four segments in each arm, showing how RV offset varies with wavelength. Light blue and light red shading mark the spectral ranges used for RV measurements.}
    \label{fig:sp9_fiber11_2019}
\end{figure}

%A detailed inspection shows that, within a single night, multiple exposures from the same spectrograph share similar velocity-offset patterns across all spectral segments (see Fig.~\ref{fig:sp9_plate_heatmap} as an example). Therefore, all exposures from the same night were averaged to study daily and yearly variations. 

Within a single observing year, the velocity-offset patterns of a given spectrograph remain relatively stable (see Spectrograph\,9 in Fig.~\ref{fig:sp9_season_heatmaps} as an illustrative example; for completeness, the annual RV offset heatmaps for the remaining spectrographs are presented in Appendix~B), but significant global shifts appear between different years. These “annual jumps” are likely caused by instrumental adjustments and calibration resets during the LAMOST summer shutdown and maintenance. %The figure also highlights typical cases: in 2012, the blue-end segments show strong negative offsets, whereas during 2017–2019 the red-end segments persistently exhibit positive offsets. Such systematic annual differences demonstrate that calibration errors are strongly influenced by long-term changes in instrumental and environmental conditions. 
%Accordingly, in subsequent corrections we treat different observing years separately and establish year-dependent correction curves, in order to remove long-term systematics more accurately.

Although the eight segments show time- and spectrograph-dependent variations, a clear wavelength trend emerges: the 3900--4400\,\AA{} and 4400--4900\,\AA{} segments are typically negative (about --10 km\,s$^{-1}$); the 4900--5400\,\AA{} and 5300--5800\,\AA{} segments are near zero; the 6000--6500\,\AA{}, 6500--7000\,\AA{}, and 8000--8500\,\AA{} segments are positive; and the 8500--9000\,\AA{} segment is again negative (about --10 km\,s$^{-1}$). Overall, the common offsets across segments are usually within $\pm 20$ km\,s$^{-1}$, revealing systematic wavelength-dependent behavior and providing a robust quantitative basis for constructing wavelength-dependent RV correction functions.

We fit the $\mu$ of radial velocities of eight spectral segments in each spectrograph of each plate against their central wavelengths and use the resulting curves to correct the wavelength scale (see Fig.~\ref{fig:rv_offset_sp9}). We adopt the $\mu$ because it is more robust to outliers. As the blue- and red-end velocities show parabolic trends, we fit the blue and red arms separately with quadratic polynomials. This is required because the LAMOST spectrographs split the spectrum with a dichroic and calibrate the two arms independently, so their wavelength calibration errors follow distinct trends.

Before analyzing fiber-level behavior, we first remove the spectrograph-level systematic offsets determined previously. For each spectrum, we subtract the corresponding $\mu$ velocity offset. The remaining differences are residuals, representing deviations that persist after the spectrograph-level, wavelength-dependent correction. Fig.~\ref{fig:fiber217_spectro3_heatmap} shows the monthly residual velocity offsets for a single fiber over all observing years. Within a given year, the residuals vary slowly. Between years, larger shifts appear, possibly caused by instrumental adjustments during LAMOST's summer maintenance.
Therefore, we define one observing year (from September to July) as the basic unit for fiber-level analysis. Fig.~\ref{fig:rv_heatmap_sp1-16_2017-2018} shows residual patterns of all fibers for the 2017--2018 year as an example. The patterns are different among fibers. 

%In the 3900--4400\,\AA{} and 4400--4900\,\AA{} segments, the typical residuals are around $-1.5\,\mathrm{km\,s^{-1}}$, reaching up to $5\,\mathrm{km\,s^{-1}}$ in certain observing years. 
%For the 4900--5400\,\AA{} and 5300--5800\,\AA{} segments, the typical residuals are about $1\,\mathrm{km\,s^{-1}}$. 
%The 6000--6500\,\AA{}, 6500--7000\,\AA{}, and 8000--8500\,\AA{} segments exhibit larger residuals, typically around $12\,\mathrm{km\,s^{-1}}$, with some months exceeding $15\,\mathrm{km\,s^{-1}}$. 
%The 8500--9000\,\AA{} segment shows larger fluctuations, but overall remains near $5\,\mathrm{km\,s^{-1}}$. 

In the 2017--2018 observing year, after subtracting the overall spectrograph-level zero-point, the RV offsets in the eight spectral segments fluctuate mildly around $0~\mathrm{km\,s^{-1}}$ with a slight negative bias. Typically, the $\mu$ of offsets in each segment are about $-0.5$ to $-1.0~\mathrm{km\,s^{-1}}$. For the six segments spanning 3900--7000~\AA, most fibers lie within $\sim[-3.5,\,+1.4]~\mathrm{km\,s^{-1}}$, whereas the 8500--9000~\AA\ segment shows the largest dispersion. The overall patterns are broadly similar among spectrographs, but the prevalence of outlier fibers and strong striping varies: Spectrograph~13 exhibits the most prominent large offsets, and Spectrographs~1 and 14--16 more frequently show extended vertical red/blue stripes, while Spectrograph~7 (as well as 3 and 6) appears relatively stable.

%To correct these residuals, we combine the distributions of the residual offsets from the previous step, adopt the median residual velocity of the eight spectral segments, and then perform further fitting for each fiber in each observing year (see Fig.~\ref{fig:sp9_fiber11_2019}). This approach ensures that both global (spectrograph-level) errors and local (fiber–level) deviations are addressed, forming a hierarchical correction scheme. According to the distribution patterns of the residuals, we again separate the red and blue arms and fit them individually with cubic polynomials. A cubic function is chosen because the residual trends often exhibit inflection-like features, which cannot be captured by lower-order polynomials, while higher-order fits risk overfitting. 

To correct these residuals, we further fit each fiber in each observing year (see Fig.~\ref{fig:sp9_fiber11_2019} for an illustrative example from the 2019 observing year; results for the other nine observing years are presented in Appendix~C). Based on the residual patterns, we again separate the red and blue arms and fit each with a cubic polynomial. We use a cubic because the residuals often show inflection-like behavior that lower-order polynomials cannot capture, while higher orders risk overfitting. This hierarchical correction procedure properly treats both global (spectrograph-level) systematics and local (fiber-level) residuals. After these corrections, all spectra share a consistent velocity reference frame, enabling subsequent zero-point drift corrections and construction of a high-precision radial velocity catalog; a total of 6,778,184 spectra successfully received the correction (96.00\%).

\subsection{Correction about zero point}

\begin{figure*}[htbp]
    \centering
    \includegraphics[width=0.8\linewidth]{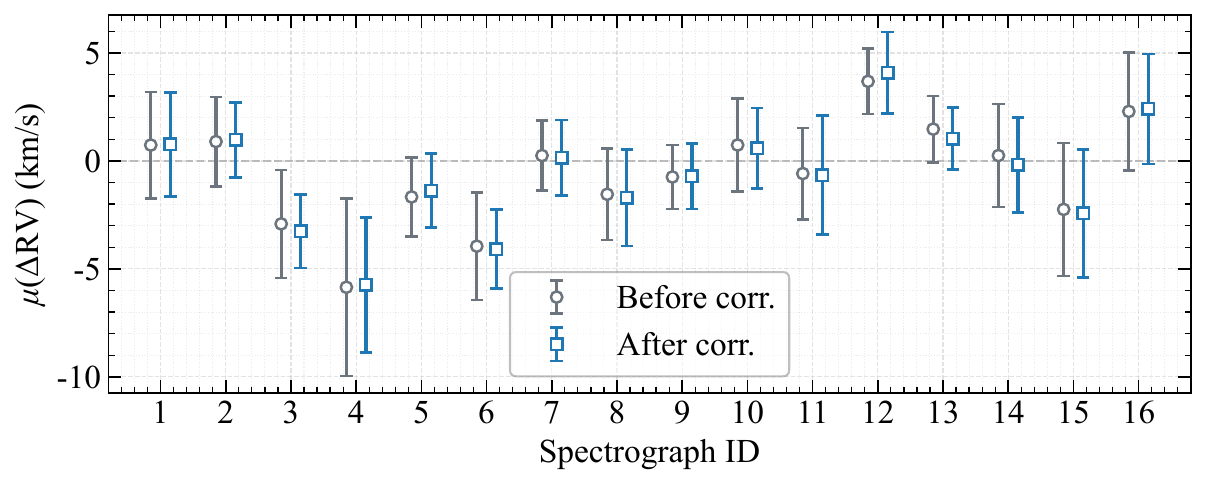}
    \caption{
    Comparison of radial velocity consistency between two plates with about 1700 common sources: plate \texttt{HIP29425K201} observed on 2016-11-23 and plate \texttt{HIP29425K2} on 2017-11-18. The horizontal axis shows spectrograph ID, and the vertical axis shows radial velocity differences ($\Delta RV$) of repeated sources. Each spectrograph has about 100 repeats, whose velocity differences are fitted with a Gaussian; the fitted $\mu$ and standard deviation $\sigma$ are plotted with error bars. Grey circles show results before dispersion-curve correction, blue squares after correction. For most spectrographs, the $\sigma$ decreases after correction, indicating better internal consistency. However, the $\mu$ still show large offsets, and for some spectrographs (e.g., No.~1, 2, 3, 6, 12, 15, and 16) the deviation from zero even increases, revealing persistent velocity zero-point offsets beyond wavelength-dependent errors.}
    \label{fig:deltarv_mu_spid}
\end{figure*}

\begin{figure*}[htbp]
    \centering
    \includegraphics[width=\textwidth]{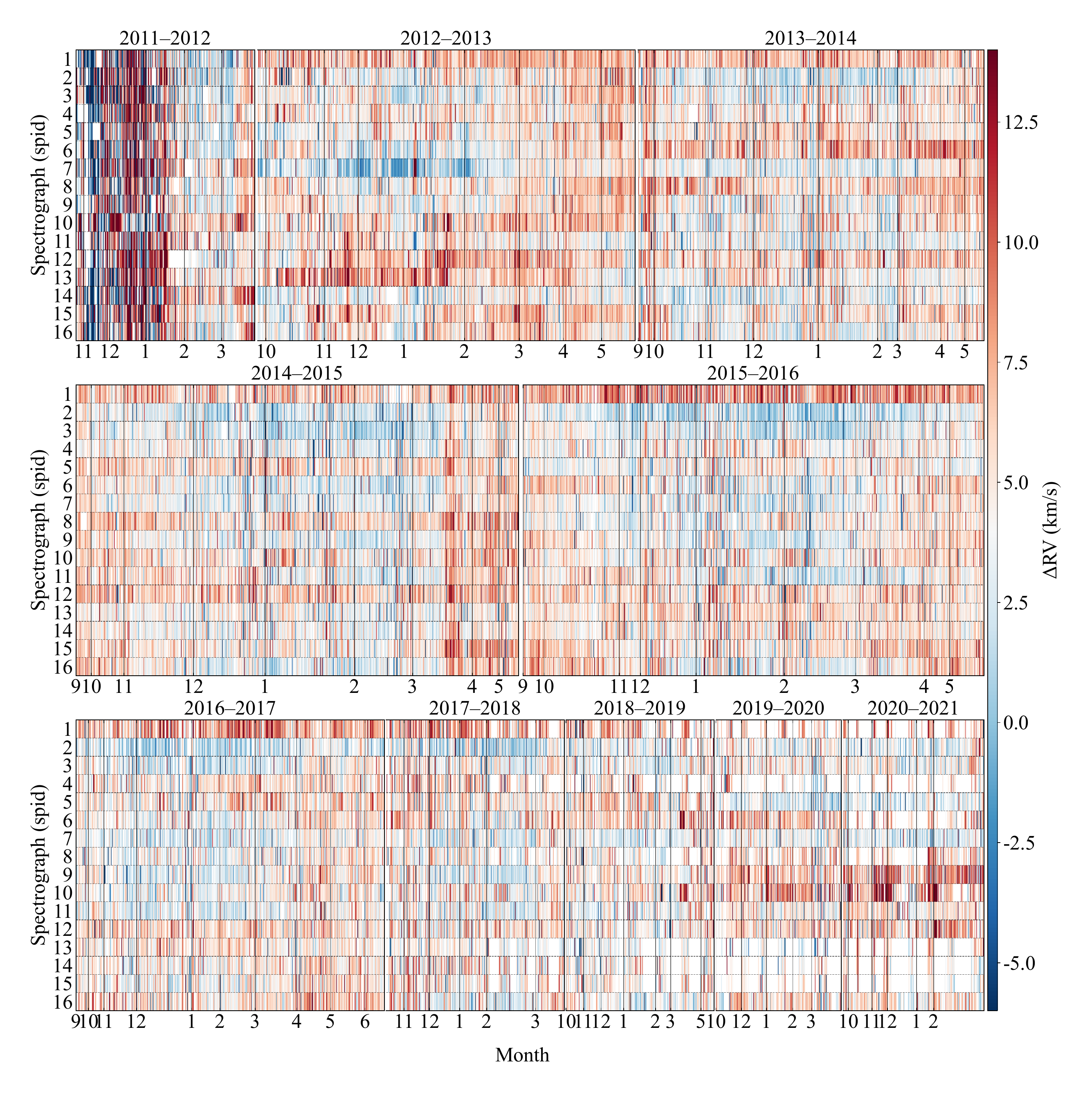}
    \caption{
    Zero-point corrections for individual spectrographs. % after spectrograph-level calibration. 
    Each vertical strip represents a plate, and the sixteen horizontal cells within each strip represent its sixteen spectrographs. Cell color indicates the zero-point correction for that spectrograph, and the white cells indicate spectrographs without correction. Plates are ordered by observing time along the horizontal axis; the axis is labeled by month, with tick marks marking the first plate of each month and gray vertical lines separating months. Horizontal lines separate different spectrographs. The color bar on the right shows the zero-point correction scale.}
    \label{fig:plate_spectrograph_offsets}
\end{figure*}

\begin{figure*}[htbp]
    \centering
    \includegraphics[width=\textwidth]{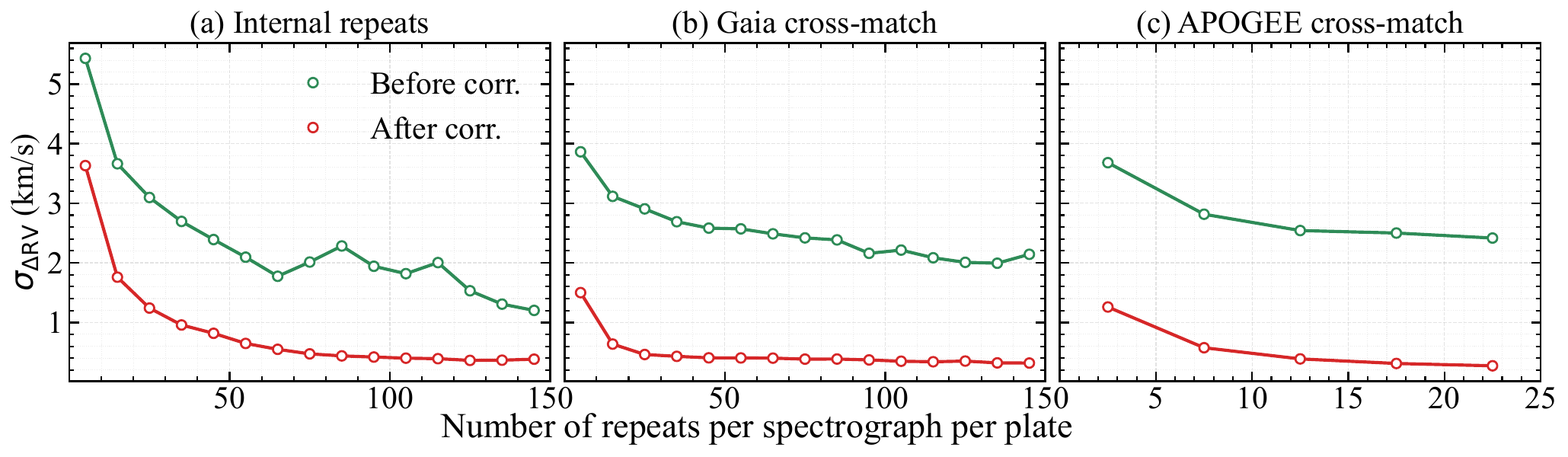}
    \caption{
    Comparison of radial velocity consistency before and after the zero-point correction. 
    Panel (a) shows the standard deviation of radial velocity differences ($\sigma_{\Delta RV}$) between repeated observations of the same sources in different LAMOST spectrographs, as a function of the number of repeated sources per spectrograph pair. Note that larger number of repeated sources  provides a more reliable standard-deviation estimate.
    Panel (b) presents the $\sigma_{\Delta RV}$ for common sources between individual LAMOST spectrographs and \textit{Gaia}, while Panel (c) shows the corresponding comparison with APOGEE. 
    The green symbols and lines denote the results before correction, and the red ones denote the results after correction. 
    In all cases, the post-correction $\sigma_{\Delta RV}$ values are significantly reduced, demonstrating that the applied zero-point correction improves the consistency of radial velocity measurements.}
    \label{fig:sigma_compare_panels}
\end{figure*}

\begin{figure*}[htbp]
    \centering
    \includegraphics[width=\linewidth]{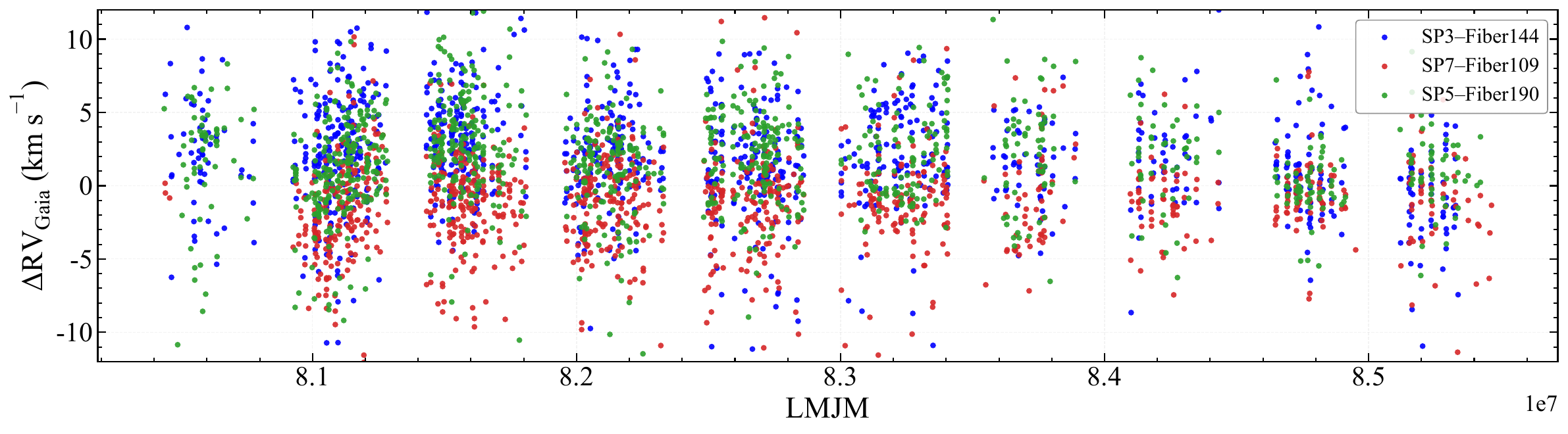}
    \caption{
        Distribution of velocity differences between LAMOST and \textit{Gaia} versus observing time for the three fibers with the most common sources: SP3–Fiber144 (blue; 1162 sources), SP7–Fiber109 (red; 1142), and SP5–Fiber190 (green; 1140). The vertical axis shows $\Delta RV_{\mathrm{Gaia}} = RV_{\mathrm{LAMOST}} - RV_{\mathrm{Gaia}}$ after applying the wavelength-dependent correction and spectrograph-level zero-point calibration. The horizontal axis gives the observing time in LAMOST Modified Julian Date (LMJM, in scientific notation). The figure shows that $\Delta RV_{\mathrm{Gaia}}$ varies smoothly with time, demonstrating that residual zero-point drifts remain at the fiber level over long timescales and motivating temporal modeling in the subsequent fiber-level zero-point correction.
    }
    \label{fig:fiber_gaia_deltarv_vs_time}
\end{figure*}

After applying wavelength-dependent corrections at the spectrograph and fiber levels, the velocity consistency of repeated observations within each spectrograph improves modestly, reducing relative scatter but not setting the absolute velocity scale.  Thus, each spectrograph becomes more internally consistent, yet its $\mu$ of velocity difference can still deviate from zero. For many spectrographs, the $\mu$ of velocity differences move even farther from zero after the wavelength-dependent correction (see Fig.~\ref{fig:deltarv_mu_spid}).  This residual offset is the radial-velocity zero point (RVZP) of each spectrograph, demonstrating the need for an additional spectrograph-level zero-point correction to place all spectrographs on a common velocity scale.

We adopt the uber-calibration method from Section~3.3, which incorporates both internal and external constraints.  
The internal term minimizes velocity differences among repeated LAMOST observations in different calibration units, while the external term aligns the velocity scale with high-precision standards such as APOGEE and \textit{Gaia}~RVS.  
The method yields a zero-point correction $\mu_i$ for each spectrograph, which is then subtracted from all corresponding spectra.
As shown in Fig.~\ref{fig:plate_spectrograph_offsets}, the derived zero-point corrections typically fall within $-$2.5 to 10.0 $\mathrm{km\,s^{-1}}$, and vary with time and spectrographs. The heatmap shows distinct RV zero-point patterns across spectrographs. Spectrographs~1 and~12 are consistently red, indicating persistently positive zero-point offsets, while spectrographs~2 and~3 contain more blue regions. The other spectrographs alternate between red and blue over time.

We quantify calibration improvement by comparing, before and after correction, the standard deviation of the velocity-difference distribution for each calibration unit. For internal repeats, a calibration unit is the set of spectra from a given spectrograph within a single exposure. For any pair of units sharing repeated sources, we compute the velocity differences $\Delta RV$ of their common sources and use the standard deviation of this $\Delta RV$ distribution, $\sigma_{\Delta RV}$, to quantify their relative consistency. An effective zero-point correction suppresses systematic differences between units, significantly reducing $\sigma_{\Delta RV}$; a large $\sigma_{\Delta RV}$ instead signals residual uncorrected offsets or other systematics.

For external validation, we cross-match each calibration unit with \textit{Gaia} RVS and APOGEE and compute the velocity differences for the common sources. We adopt the same metric, $\sigma_{\Delta RV}$, i.e., the standard deviation of the velocity-difference distribution, as the performance indicator. The simultaneous reduction of these dispersions after correction shows that our method improves the internal repeatability and also reduces the zero-point offsets with respect to high-precision external references, thereby achieving the goal of uber-calibration.

As shown in the left panel of Fig.~\ref{fig:sigma_compare_panels}, the dispersion of velocity differences drops markedly after the zero-point correction, demonstrating improved internal consistency among all spectrographs.  
Comparisons with external standards (APOGEE and \textit{Gaia}~RVS) show the same behavior: the scatter of velocity residuals relative to these references is significantly reduced after correction.  These trends confirm that our method reduces both internal and external components of the global velocity offset, yielding a more coherent and reliable calibration. This suggests that our weighting scheme is reasonable: after the zero-point correction, the dispersion decreases markedly and internal consistency improves substantially, while further fine-tuning of the weights yields only minor additional gains. Therefore, we do not perform a more exhaustive optimization or grid search over the weight values. 

%As a function of time, the zero points remain nearly constant within a %given observing month but change gradually between months. Although these %transitions do not always align with calendar months, their smoothness %indicates that each spectrograph’s instrumental state evolves %continuously rather than abruptly. This behavior shows that the RV zero %point responds to slow instrumental drifts and environmental changes yet %is stable on short timescales, providing a physically reasonable basis %for the subsequent fiber-level corrections.

This spectrograph-level correction removes large-scale systematics and unifies the velocity scale across spectrographs.  
Within a given spectrograph, however, small but stable residual offsets persist between fibers, reflecting subtle differences in local calibration.  
To remove these residuals and ensure consistency across fibers, we apply an additional zero-point correction at the fiber level.

For a given fiber, its zero-point offset relative to external datasets (APOGEE and \textit{Gaia}) varies smoothly within a single observing year but shows larger jumps between years (see Fig.~\ref{fig:fiber_gaia_deltarv_vs_time}).  
These inter-year differences are likely linked to the annual \textit{LAMOST} shutdown and maintenance.   
From repeated tests, we find that adopting one observing year (September to July) as the characteristic timescale for zero-point variation best balances physical realism, data volume, and statistical robustness.

\begin{figure*}[htbp]
    \centering
    \includegraphics[width=1\textwidth]{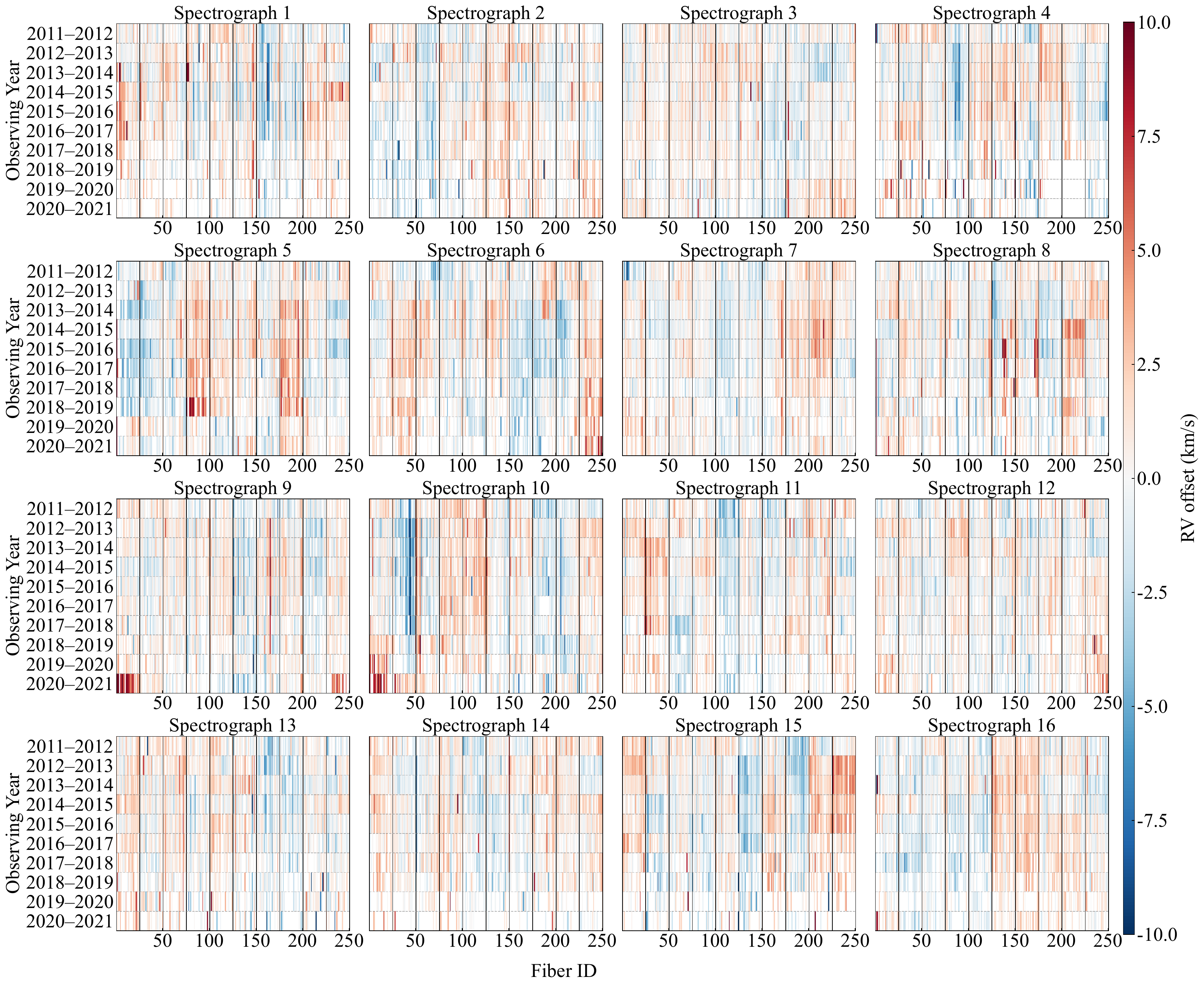}
    \caption{
    Radial velocity zero-points of individual fibers after spectrograph-level corrections. For each fiber and observing year, the zero-point is the weighted median RV difference between LAMOST and external references (Gaia, APOGEE), with weights proportional to the number of common sources. The horizontal axis shows fiber ID; the vertical axis lists observing years. Vertical black lines separate groups of 25 fibers, matching the LAMOST fixtures. Blank regions mark fibers and years without common Gaia or APOGEE sources. The color bar gives the RV zero-point offset in km s$^{-1}$.
    }
    \label{fig:fiber_heatmap}
\end{figure*}

With this timescale defined, each fiber shows a characteristic velocity offset relative to the external datasets, typically $\sim3\,\mathrm{km\,s^{-1}}$, with neighboring fibers having similar values (Fig.~\ref{fig:fiber_heatmap}).  
We attribute this to the \textit{LAMOST} design, where every 25 fibers share a common fixture and therefore similar instrumental and environmental conditions, producing correlated systematics.  
For fibers with too few external matches which are fewer than 20 common sources in a given year, we adopt the median offset of their 25-fiber group.  
The fiber-level zero point for each observing year is then given by the weighted mean of the offsets relative to \textit{Gaia} and APOGEE.

\begin{figure*}[htbp]
    \centering
    \includegraphics[width=\textwidth]{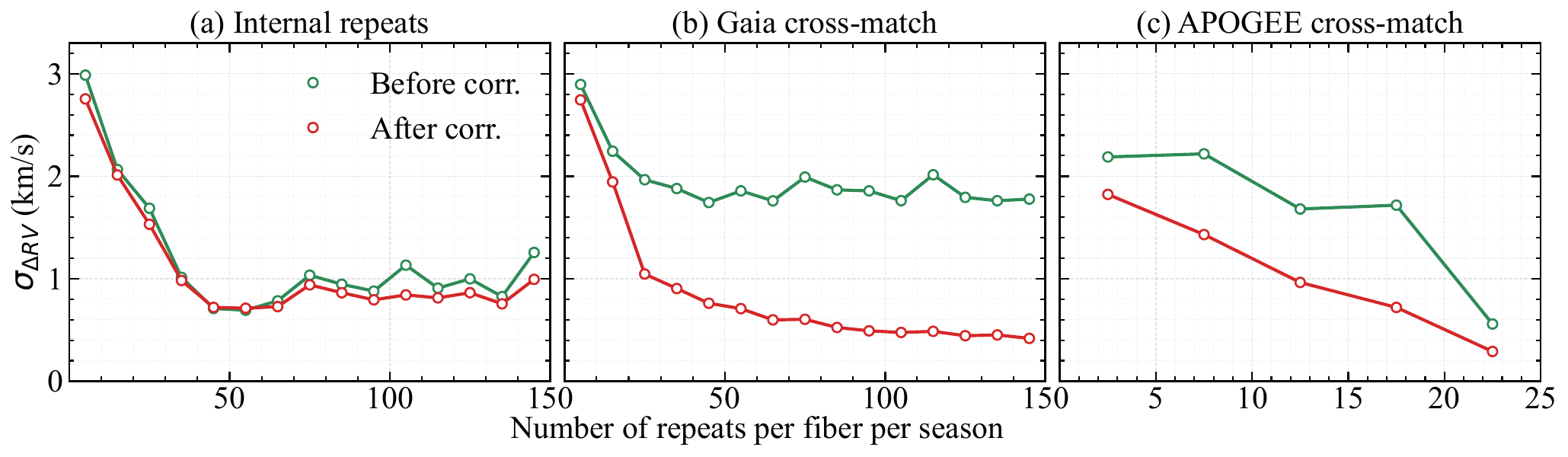}
    \caption{
        Comparison of radial-velocity consistency before and after the fiber level correction.
        Panel (a) shows the velocity scatter ($\sigma_{\Delta RV}$) of internal repeat observations within fiber-years as a function of the number of repeated pairs.
        Panels (b) and (c) show the velocity scatter of common sources between LAMOST and \textit{Gaia}, and between LAMOST and APOGEE, respectively, also as a function of the number of matched pairs per spectrograph.
        The green and red curves represent results before and after the zero-point corrections, respectively.
        A smaller $\sigma_{\Delta RV}$ after correction indicates improved internal and external RV consistency.
    }
    \label{fig:sigma_compare_panels_fiber}
\end{figure*}

\begin{figure*}[htbp]
    \centering
    \includegraphics[width=\textwidth]{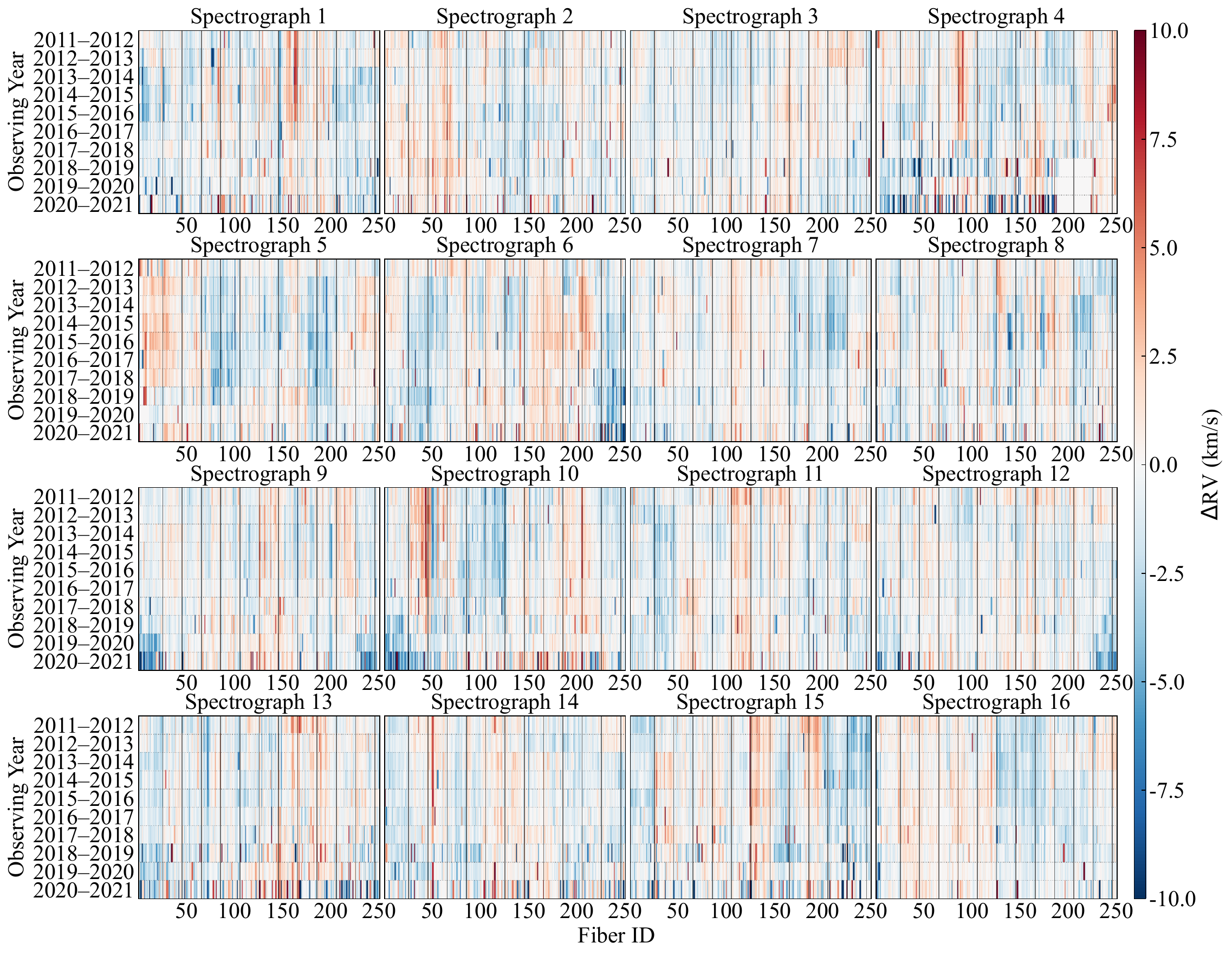}
    \caption{
        Fiber-level radial velocity zero-point corrections.
        Each panel corresponds to one spectrograph, with fiber ID (1–250) along the horizontal axis
        and observing years (2011–2012 to 2020–2021) arranged from top to bottom in chronological order.
        The color of each pixel represents the zero-point correction for the corresponding fiber
        in a given year (in km\,s$^{-1}$), as indicated by the color bar on the right.
        Gray dashed lines separate different observing years,
        while black solid lines mark intervals of 25 fibers.
    }
    \label{fig:spid_fiber_season_heatmaps_deltaRV}
\end{figure*}

%%HERE
As in the spectrograph-level zero-point correction, we quantify the calibration performance by comparing, before and after correction, the standard deviation of the velocity-difference distribution for each calibration unit, and we use as the horizontal axis the number of repeated sources contained in each unit, $N$.  
As shown in Fig.~\ref{fig:sigma_compare_panels_fiber}, the velocity-difference distribution becomes much narrower after correction, indicating improved internal consistency across all fiber-level units.  
Comparisons with external datasets (APOGEE and \textit{Gaia}~RVS) show similar improvements, confirming that the $\chi^2$ minimization reduces both internal and external zero-point offsets and yields a physically consistent velocity calibration. The specific fiber-level corrections are illustrated in Fig.~\ref{fig:spid_fiber_season_heatmaps_deltaRV}.  

This layered correction procedure removes zero-point systematics from the \textit{LAMOST} data, yielding a unified, self-consistent velocity reference frame across spectrographs, fibers, and observing years. Because the method links calibration units using RV differences from repeat observations and external observations, it cannot be applied to spectra that lie on isolated plates and lack sufficient external matches; for these cases, the zero point remains unsolved. We therefore exclude them from the final sample, resulting in corrected RVs for 5{,}785{,}300 spectra.

\section{Verification}

\begin{figure*}[t]
    \centering
    \includegraphics[width=0.8\textwidth]{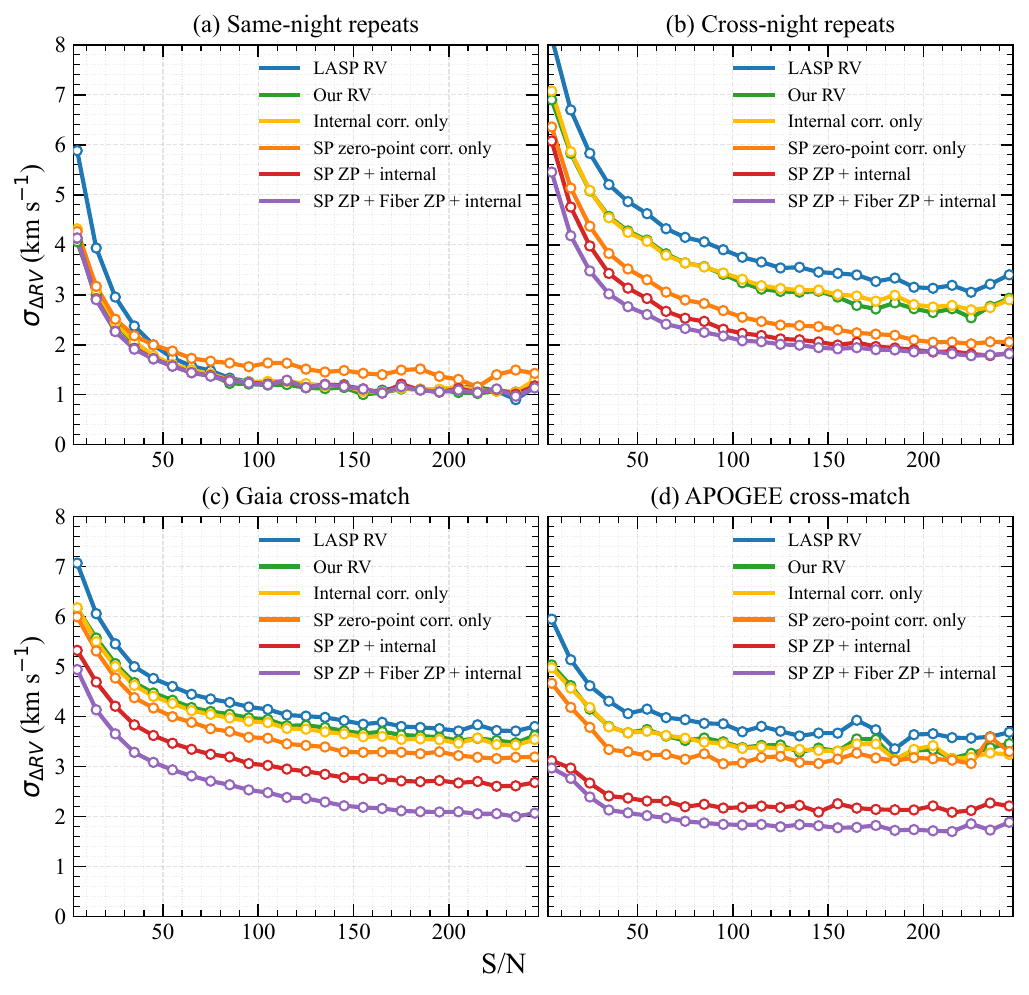}
    \caption{
    Comparison of RV precision and zero-point offsets before and after corrections. Panels (a) and (b) show internal consistency from LAMOST repeat observations: (a) same-night and (b) cross-night repeats. Each point is the Gaussian-fitted standard deviation $\sigma_{\Delta RV}$ of RV differences in S/N bins (bin size = 5). Panels (c) and (d) show external validation: (c) LAMOST–Gaia and (d) LAMOST–APOGEE common sources, with each point again the fitted $\sigma_{\Delta RV}$ in S/N bins. Colors indicate RV determinations: gray = original LASP pipeline, blue = our pipeline without correction, green = after internal consistency correction, orange = after spectrograph level zero-point correction, red = after spectrograph level zero-point correction and internal consistency correction, purple = after all. The purple curves have the lowest $\sigma$ and $\mu$ offsets closest to zero, indicating that the combined corrections substantially improve LAMOST RV precision and consistency.
    }
    \label{fig:rv_snr_sigma_mu}
\end{figure*}
\begin{figure*}[t]
    \centering
    \includegraphics[width=0.85\textwidth]{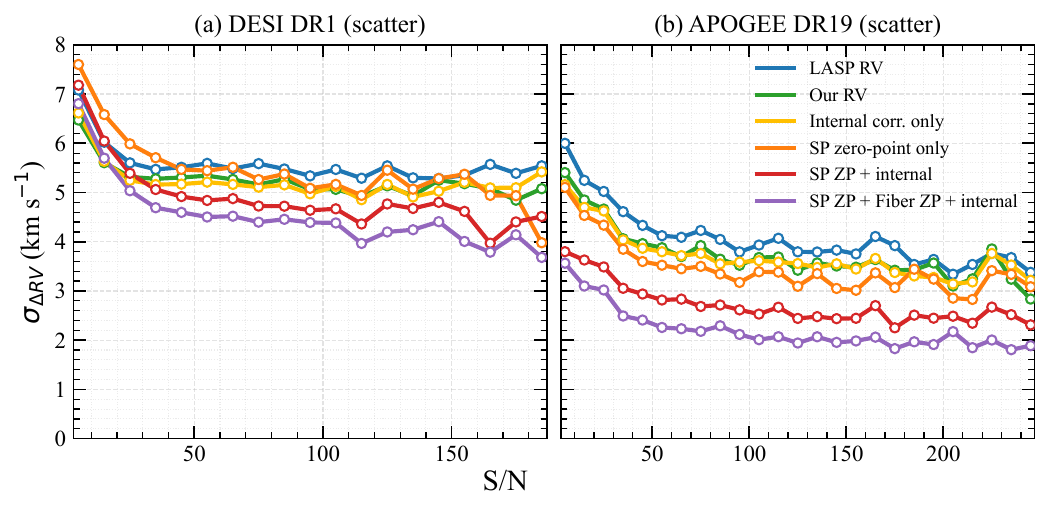}
    \caption{
    Comparison of RV precision and zero-point offsets between our measurements and those from DESI~DR1 and APOGEE~DR19 for common stars. Panels~(a) and~(b) show the standard deviation $\sigma_{\Delta RV}$ of RV differences versus S/N. For DESI~DR1 (panels~a), we divided stars into 20 S/N bins from 0 to 200 and fit a Gaussian to the RV differences in each bin to obtain $\sigma$ and $\mu$. For APOGEE~DR19 (panels~b), we used 25 S/N bins from 0 to 250 with the same procedure. The DESI comparison includes 239,355 stars and the APOGEE comparison 38,497 stars. All 38,497 APOGEE~DR19 stars are new relative to DR17, and both the DESI~DR1 and APOGEE~DR19 samples are completely independent of the calibration training set, providing an independent validation of our RV corrections.
    }
    \label{fig:rv_snr_2x2}
\end{figure*}

\begin{figure}[t]
    \centering
    \includegraphics[width=0.45\textwidth]{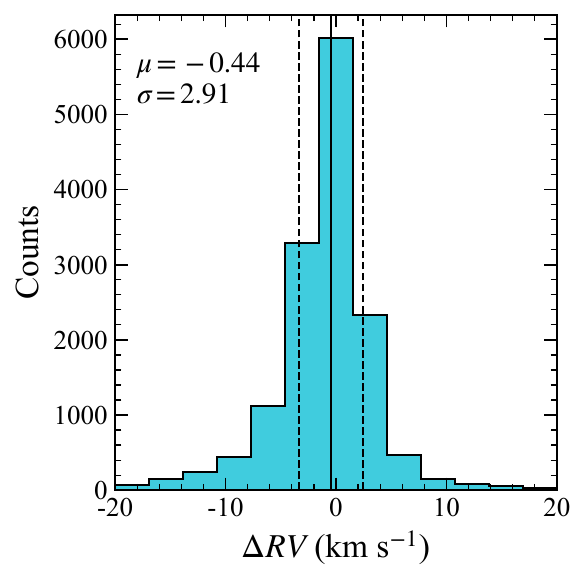}
    \caption{
    Distribution of RV differences between DESI~DR1 and APOGEE~DR17 for their 14,449 common stars, where $\Delta RV = v_{\mathrm{DESI}} - v_{\mathrm{APOGEE}}$. The solid black line shows the $\mu$ of offset, and the dashed black lines show the $\pm1\sigma$ dispersion.
    }
    \label{fig:hist_desi_apo}
\end{figure}

We assess the precision of the corrected radial velocities using repeated observations (see Fig.~\ref{fig:rv_snr_2x2}).  
For same-night observations, the scatter of velocity differences versus S/N is nearly unchanged by the correction, indicating that intra-night consistency is already at its limiting, baseline level.  
For observations on different nights, it is noteworthy that correcting wavelength-dependent internal inconsistencies alone does not significantly improve radial-velocity repeatability and can even slightly worsen it.  
The correction tightens the velocity-difference distribution within each spectrograph but amplifies their relative zero-point offsets.  
Thus, when spectra from all spectrographs are combined, the overall dispersion of velocity differences increases.  
However, once the internal inconsistencies are corrected, the subsequent zero-point calibration becomes more effective. In the orange curve we show the results of applying only the zero-point correction without correcting internal inconsistencies , and those of applying the zero-point correction after correcting internal inconsistencies with the red curve. For spectra with $\mathrm{S/N}>100$, correcting internal inconsistencies yields an additional improvement, reducing the dispersion from $\sim2.2~\mathrm{km\,s^{-1}}$ to $\sim2.0~\mathrm{km\,s^{-1}}$.

%We compare two approaches: (1) applying the spectrograph-level zero-point correction without internal-inconsistency correction, and (2) applying it after correcting internal inconsistencies.  
%The second approach yields higher consistency: the standard deviation of velocity differences among repeats decreases by $\sim0.2\,\mathrm{km\,s^{-1}}$.  
%The improvement is clearer against external datasets: the velocity dispersion relative to \textit{Gaia} decreases by $\sim0.6\,\mathrm{km\,s^{-1}}$, and relative to APOGEE by $\sim0.8\,\mathrm{km\,s^{-1}}$.  
%Thus, correcting internal inconsistencies significantly improves both internal and external precision. 

%Furthermore, we observe that after the fiber-level zero-point correction, the improvement relative to \textit{Gaia} is more significant than that relative to APOGEE.  
%This difference arises because the number of APOGEE common sources per fiber and observing year is typically only $\sim3$, whereas the corresponding number for \textit{Gaia} is about $\sim50$.  
%Considering this disparity, the external reference velocity ($v_{\mathrm{ext}}$) in our $\chi^2$ formulation is defined as a weighted combination of the two surveys, effectively making the final corrections closer to the \textit{Gaia} reference frame.  

After applying both spectrograph and fiber levels of zero-point correction, the standard deviation of velocity differences among repeated observations decreases from $\sim3.6\,\mathrm{km\,s^{-1}}$ to $1.8\,\mathrm{km\,s^{-1}}$.  
This corresponds to a single-measurement precision of $1.8/\sqrt{2} \approx 1.3\,\mathrm{km\,s^{-1}}$, indicating substantial improvement, although still slightly larger than the theoretical limit.  

For each spectrum with a successfully corrected radial velocity $rv_{\rm corr}$, we define a new single-visit uncertainty $\sigma_{rv_{\rm corr},i}$. Our approach uses error propagation to transfer the overall precision gain from cross-night repeats to individual spectra. From the repeat sample in Fig.~18, we measure the dispersion of velocity differences $\Delta v$ in SNR bins, both before (LASP) and after correction. In each SNR bin, the change in *single-measurement* variance is
\begin{equation}
\Delta\sigma_{\rm single}^2
=
\frac{\sigma_{\Delta v,{\rm LASP}}^2-\sigma_{\Delta v,{\rm corr}}^2}{2},
\end{equation}
where the factor of 2 reflects that $\Delta v$ is the difference of two independent, similarly precise measurements. We then take a robust average over SNR and obtain a constant variance-reduction term, 
$C=2.35~{\rm km\,s^{-1}}$ (i.e., a variance reduction of $C^2$). For spectrum $i$, we apply this reduction to the LASP uncertainty $\sigma_{rv_{\rm LASP},i}$ and define
\begin{equation}
\sigma_{rv_{\rm corr},i}
=
\sqrt{\sigma_{rv_{\rm LASP},i}^{2}-C^{2}},
\qquad
C=2.35~{\rm km\,s^{-1}} .
\end{equation}
To avoid unphysical values when $\sigma_{rv_{\rm LASP},i}$ is underestimated (yielding a negative argument of the square root), we impose an uncertainty floor
$\sigma_{\rm floor}=1.3~{\rm km\,s^{-1}}$, corresponding to the high-SNR plateau in Fig.~18.

To assess the absolute accuracy of our calibrated radial velocities, we validate them against two independent high-precision surveys: DESI~DR1 (\citealt{desidr1}) and APOGEE~DR19 (\citealt{apogeedr19}; after removing sources already present in APOGEE~DR17).  
Common sources between LAMOST and these surveys are used to estimate the velocity dispersion as a measure of precision.  
The results show that our calibration substantially improves the agreement with APOGEE~DR19: at high signal-to-noise ratio, the standard deviation of the RV differences with respect to APOGEE~DR19 decreases from $\sim3.6~\mathrm{km\,s^{-1}}$ (LASP) to $\sim2.0~\mathrm{km\,s^{-1}}$ for $\mathrm{S/N}>100$. 
The comparison with DESI also shows an improvement: for $\mathrm{S/N}>100$, the scatter decreases from $\sim5.6~\mathrm{km\,s^{-1}}$ (LASP) to $\sim4.2~\mathrm{km\,s^{-1}}$. 
These trends are illustrated in Fig.~\ref{fig:rv_snr_2x2}.

Fig.~\ref{fig:hist_desi_apo} shows that DESI~DR1 and APOGEE~DR19 themselves exhibit a mutual dispersion of about $2.91\,\mathrm{km\,s^{-1}}$.  
According to quadratic error propagation, if the dispersion between DESI~DR1 and APOGEE is $2.91\,\mathrm{km\,s^{-1}}$, and that between our velocities and APOGEE is $2.0\,\mathrm{km\,s^{-1}}$, the expected dispersion between our velocities and DESI~DR1 should be $\sim3.5\,\mathrm{km\,s^{-1}}$.  
The measured value deviates slightly from this expectation, suggesting possible systematic differences among the datasets that merit further investigation.  

For the APOGEE~DR19 comparison (these sources are new to DR17), the velocity dispersion drops from $3.5$ to $2.0\,\mathrm{km\,s^{-1}}$, showing that our hierarchical correction strategy greatly improves LAMOST RV precision.

\section{Data Product}

A primary product of this work is a value-added RV catalog with corrected velocities for $\sim5.7$ million spectra, corresponding to $4{,}235{,}600$ unique stars. For completeness and ease of use, the release retains all $7{,}060{,}436$ spectral entries from the initial sample; entries without a reliable correction are kept and identified via an added \texttt{flag} column that records the stage at which the calibration fails (e.g., the internal-consistency or zero-point step). The full value-added radial-velocity catalog is publicly available on Zenodo at DOI: 10.5281/zenodo.19039670.
Based on low-resolution ($R \approx 1800$) \textit{LAMOST}~DR9~v1.1 spectra, it applies the calibration procedures developed here, including the wavelength-dependent inconsistency correction and RV zero-point adjustment.  
Relative to the original LASP measurements, the calibrated RVs show substantially improved precision:  
Using internal cross-night repeats as a reference, for spectra with $\mathrm{S/N}>100$ the dispersion decreases from $\sim3.6~\mathrm{km\,s^{-1}}$ to $\sim1.8~\mathrm{km\,s^{-1}}$, implying a single-measurement uncertainty of $\sim1.3~\mathrm{km\,s^{-1}}$; 
using APOGEE DR17 as an external reference, for spectra with $\mathrm{S/N}>100$ the dispersion likewise decreases from $\sim3.6~\mathrm{km\,s^{-1}}$ to $\sim1.8~\mathrm{km\,s^{-1}}$.
The catalog also provides observational metadata and stellar parameters, enabling flexible subsample selection for diverse scientific applications.  
Its main columns are summarized in Table~\ref{tab:rv_columns}.  
This catalog offers a robust dataset for studies of Galactic kinematics, stellar populations, and the chemo-dynamical structure of the Milky Way.

\begin{table*}[htbp]
\centering
\caption{Description of the value-added radial velocity catalog.}
\label{tab:rv_columns}
\begin{tabular}{llll}
\hline
\textbf{Column} & \textbf{Name} & \textbf{Unit} & \textbf{Description} \\
\hline
1  & obsid            & --     & Unique observation identifier \\
2  & uid              & --     & Unique star identifier \\
3  & gp\_id           & --     & Survey identifier of the source (Pan-STARRS, Gaia, or LAMOST) \\
4  & designation      & --     & LAMOST target designation \\
5  & obsdate          & --     & Observation date (yyyy-mm-dd) \\
6  & lmjd             & --     & Local modified Julian day \\
7  & mjd              & --     & Modified Julian day \\
8  & planid           & --     & Plan name \\
9  & spid             & --     & Spectrograph ID \\
10 & fiberid          & --     & Fiber ID \\
11 & ra\_obs          & deg    & Fiber pointing right ascension \\
12 & dec\_obs         & deg    & Fiber pointing declination \\
13 & snru             & --     & Signal-to-noise ratio in $u$ band \\
14 & snrg             & --     & Signal-to-noise ratio in $g$ band \\
15 & snrr             & --     & Signal-to-noise ratio in $r$ band \\
16 & snri             & --     & Signal-to-noise ratio in $i$ band \\
17 & snrz             & --     & Signal-to-noise ratio in $z$ band \\

18 & subclass         & --     & Spectral subclass \\
29 & gaia\_source\_id & --     & Gaia DR3 source identifier \\
20 & gaia\_g\_median\_mag & mag  & Gaia $G$-band median magnitude \\

21 & fibertype        & --     & Fiber type (star, sky, etc.) \\
22 & ra               & deg    & Right ascension from input catalog \\
23 & dec              & deg    & Declination from input catalog \\
24 & teff             & K      & Effective temperature obtained by the LASP \\
25 & teff\_err        & K      & Effective temperature uncertainty obtained by the LASP\\
26 & logg             & dex    & Surface gravity obtained by the LASP \\
27 & logg\_err        & dex    & Surface gravity uncertainty obtained by the LASP \\
28 & feh              & dex    & Metallicity obtained by the LASP \\
29 & feh\_err         & dex    & Metallicity uncertainty obtained by the LASP\\
30 & rv\_LASP               & km s$^{-1}$ & Heliocentric radial velocity obtained by the LASP \\
31 & rv\_LASP\_err          & km s$^{-1}$ & Uncertainty of heliocentric radial velocity obtained by the LASP \\
32 & rv\_corr         & km s$^{-1}$ & Corrected radial 
velocity (this work) \\
 33 &  rv\_corr\_err        &  km s$^{-1}$ &  Uncertainty of corrected radial velocity (this work) \\
34 &  flag        &  -- &  Calibration status flag: 0 =rv\_corr available; 1 =rv\_corr missing (stopped at  \\
&       & &  internal-consistency step); 2 = rv\_corr missing (stopped at zero-point step). \\
\hline
\end{tabular}
\end{table*}

%This catalog will serve as a robust dataset for studies of Galactic kinematics, stellar populations, and the chemo-dynamical structure of the Milky Way, providing one of the largest homogeneous RV samples currently available.

\section{Summary}

We investigate two dominant systematic errors in RV measurements from LAMOST low-resolution spectra: wavelength-dependent inconsistencies and RV zero-point offsets.  
We present a hierarchical correction scheme and release a value-added catalog  with corrected velocities for $\sim5.7$ million spectra. 
Using large samples of repeat observations and external high-precision surveys (\textit{Gaia} and APOGEE), we develop a calibration framework that is both accurate and computationally efficient.  
The main conclusions are:

This work examines two main systematic errors in RV measurements from LAMOST low-resolution spectra: wavelength-dependent inconsistency and the RV zero-point offset.  
We then propose a hierarchical correction scheme and release a value-added catalog  with corrected velocities for $\sim5.7$ million spectra. 
Using large samples of repeat observations and external high-precision surveys (\textit{Gaia} and APOGEE), we develop a calibration framework that is both accurate and computationally efficient.  
The main conclusions are:

\begin{enumerate}
\item \textbf{Correction of wavelength-dependent inconsistencies.}  
LAMOST spectra show strong wavelength-dependent velocity systematics, with negative offsets in the blue and positive offsets in the red.  
We split each spectrum into eight wavelength segments, measured their local velocities, and mapped the distribution of these systematic errors, finding spectrograph-dependent patterns that evolve smoothly over time.  
Using these trends, we derived wavelength-dependent correction functions at both spectrograph and fiber levels, fitting the blue and red arms separately.  
This procedure substantially improved internal consistency among spectral segments.

\item \textbf{Zero-point correction.}  
We removed global velocity offsets using a hierarchical calibration that combines repeated observations (internal constraints) with external high-precision data (external constraints).  
A unified $\chi^2$ optimization solves for the zero-point of each spectrograph, mitigating large-scale systematics.  
Spectrograph zero-points range from $-5$ to $+15\,\mathrm{km\,s^{-1}}$, varying smoothly on short timescales with occasional long-term jumps.  
Fiber zero-points evolve smoothly within a given observing year but show discontinuities between years, likely due to LAMOST summer maintenance adjustments.  
We therefore define a zero-point correction function for each fiber in each observing year (September–July), achieving a three-level calibration in spectrograph, fiber, and time.

\item \textbf{Correction performance and precision evaluation.}  
After calibration, the RV precision of LAMOST low-resolution spectra improves markedly.  
The dispersion of cross-night velocity differences drops from $\sim3.6$ to $\sim1.8\,\mathrm{km\,s^{-1}}$, implying a single-measurement precision of $\sim1.3\,\mathrm{km\,s^{-1}}$.  
Cross-matches with \textit{Gaia} and APOGEE confirm that the systematic zero-point bias is effectively removed, with external dispersions reduced from $\sim3.8$ to $\sim2.0\,\mathrm{km\,s^{-1}}$ and from $\sim3.6$ to $\sim1.8\,\mathrm{km\,s^{-1}}$, respectively.  
Comparison with DESI~DR1 and APOGEE~DR19 shows consistent results: the dispersion relative to APOGEE decreases to $\sim2.0\,\mathrm{km\,s^{-1}}$, and agreement with DESI improves moderately but clearly over LASP, indicating a physically coherent and statistically reliable calibrated velocity system.

\item \textbf{Value-added catalog and scientific significance.}  
Based on the above corrections, we release 
a value-added RV catalog with corrected velocities for $\sim5.7$ million spectra, providing calibrated RVs with relevant observational metadata and stellar parameters. This catalog greatly improves the kinematic utility of LAMOST low-resolution spectra, offering a precise velocity reference for Galactic kinematics, stellar populations, and chemo-dynamical studies.
\end{enumerate}

%This study shows that systematic errors in LAMOST low-resolution spectra have a clear hierarchy:  
%the wavelength-dependent term mainly reflects local calibration deviations, while the zero-point drift traces long-term instrumental changes and calibration instability.  
%Using hierarchical modeling with multi-source constraints, we suppress these systematics.

Several limitations remain:  
(1) external reference coverage, especially from APOGEE, is uneven, giving spectrographs different external constraint strengths;  
(2) the fiber-level correction assumes a non-varying zero point within each observing year, but unresolved short-term variations likely exist, which may help explain why the achieved precision falls short of the theoretical limit;  
(3) a $\sim0.5\,\mathrm{km\,s^{-1}}$ discrepancy persists between internal and external precision, and its physical origin %-- possibly atmospheric effects, template mismatch, or residual calibration drift—
requires further study.

This work greatly improves RV precision for LAMOST low-resolution spectra and provides a general calibration framework for other large spectroscopic surveys, laying the groundwork for future high-precision studies of Galactic dynamics and stellar populations.

\vspace{7mm} \noindent {\bf Acknowledgments}
%The authors thank the anonymous referee for his/her suggestions that improved the clarity of our presentation. 
This work is supported by the National Key R\&D Program of China via 2024YFA1611901 and 2024YFA1611601, and the National Natural Science Foundation of China through the projects 12222301, and 12173007.

The Guoshoujing Telescope (the Large Sky Area Multi-Object Fiber Spectroscopic Telescope LAMOST) is a National Major Scientific Project built by the Chinese Academy of Sciences. Funding for the project has been provided by the National Development and Reform Commission. LAMOST is operated and managed by the National Astronomical Observatories, Chinese Academy of Sciences. 

This work has made use of data from the European Space Agency (ESA) mission {\it Gaia} (\url{https://www.cosmos.esa.int/gaia}), processed by the Gaia Data Processing and Analysis Consortium (DPAC, \url{https:// www.cosmos.esa.int/web/gaia/dpac/ consortium}). Funding for the DPAC has been provided by national institutions, in particular the institutions participating in the Gaia Multilateral Agreement. 

\bibliography{ref}
\bibliographystyle{aasjournal}

\appendix

\renewcommand{\thefigure}{A\arabic{figure}}
\setcounter{figure}{0}

\section{Details of Radial Velocity Measurement} \label{app:rv_method}

\subsection{Spectral Normalization}
\begin{figure*}[htbp]
    \centering
    \includegraphics[width=0.95\textwidth]{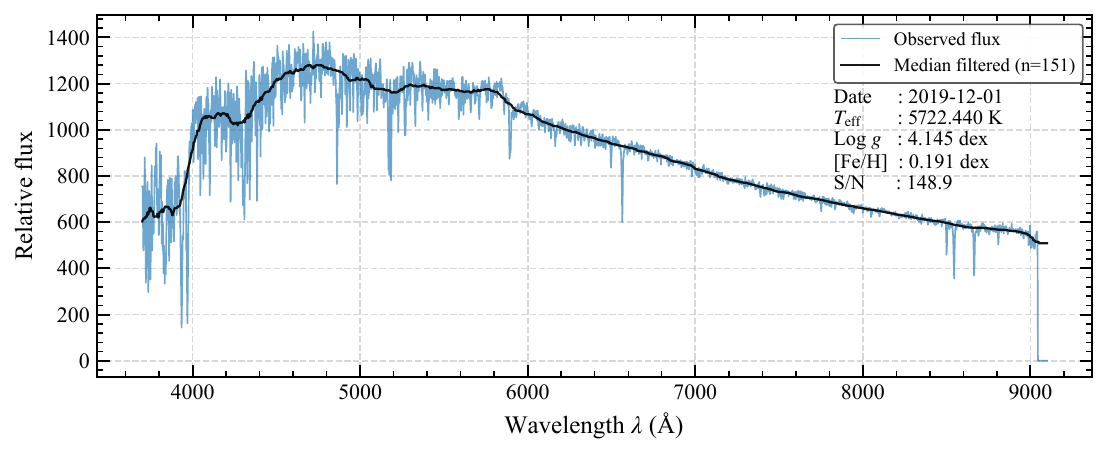}
    \caption{An example low-resolution stellar spectrum from LAMOST LRS DR9, illustrating continuum estimation via median filtering. The blue curve is the original noisy spectrum with absorption lines; the black curve is the continuum estimate obtained with a median filter using a 151-pixel window. Basic spectral information (observation date, effective temperature $T_{\rm eff}$, surface gravity $\log g$, metallicity [Fe/H], and signal-to-noise ratio S/N) is shown in the upper-right corner. Here, $T_{\rm eff}$, $\log g$, and [Fe/H] are derived from LASP \citep{Luo_2015}, and S/N is the median signal-to-noise ratio per pixel.}
    \label{fig:lamost_spectrum}
\end{figure*}

Both observed and template spectra are normalized to a common scale, improving the accuracy of subsequent analysis and comparisons. Because defining clean continuum regions is difficult, especially at low spectral resolution, we estimate the continuum with a median filter using a 151-pixel window (see Fig.~\ref{fig:lamost_spectrum}). This approach yields a robust continuum estimate and reduces noise effects. We apply no special treatment to boundary pixels, since the full spectral range of neither the observations nor the templates is used. The normalized spectrum is obtained by dividing the original spectrum by the median-filtered continuum.

\subsection{Template Spectra: BOSZ Library}
In this study, we utilize the BOSZ stellar atmosphere model for template matching. The BOSZ model, built on ATLAS9 \citep{bohlin2017new}, spans 1000--30,000 Å and covers a wide parameter space, including $T_{\rm eff}$, log $g$, [Fe/H], [C/M], and [$\alpha$/M]. For template selection, we perform a nearest-neighbor match in the three atmospheric parameters. Specifically, we discretize $T_{\rm eff}$ onto the BOSZ grid by rounding to the nearest grid point: for $T_{\rm eff}<12124\,\mathrm{K}$ we use 250~K steps, and for $T_{\rm eff}\ge 12124\,\mathrm{K}$ we use 500~K steps. Likewise, $\log g$ and [Fe/H] are rounded to the nearest nodes with step sizes of 0.5~dex and 0.25~dex, respectively. We then adopt the BOSZ spectrum at $(T_{\rm eff}^{\rm grid}, \log g^{\rm grid}, [{\rm Fe/H}]^{\rm grid})$ as the template, fixing [C/M] and [$\alpha$/M] to zero. %Since the resolution of LAMOST low-resolution spectra is $R \sim 1800$, we select templates broadened to $R=2000$ and then interpolate them to match the resolution of the observed spectra.

\subsection{Radial Velocity Measurement}
We employ the \textbf{minimum-$\chi^2_{red}$} method to measure the radial velocity (RV). The normalized template spectrum is shifted to generate 120 velocity-shifted templates, covering $\pm 300 \,\mathrm{km\,s^{-1}}$ around the LASP velocity with a step size of $5 \,\mathrm{km\,s^{-1}}$. Because the template and LAMOST wavelength grids are misaligned, we use cubic-spline interpolation to resample the templates onto the LAMOST grid for the subsequent $\chi^2_{red}$ calculation. %This range encompasses the velocity distribution of Galactic stars, while the chosen step size matches the sampling scale of the LAMOST spectra.  

The $\chi^2_{red}$ is defined as:
\begin{equation}
\chi^2_{red} = \frac{\chi^2}{N} = \frac{1}{N} \sum_{i=1}^{N} \left( \frac{f^{\mathrm{obs}}_i - f^{\mathrm{temp}}_i}{\sigma_i} \right)^2 ,
\end{equation}
where $N$ is the number of pixels, $f^{\mathrm{obs}}_i$ and $f^{\mathrm{temp}}_i$ are the normalized fluxes of observed and template spectra, and $\sigma_i$ is the noise derived from the S/N. Under an adequate model with correctly estimated uncertainties, $\chi^2_{\rm red}$ should be close to 1; values much larger than 1 indicate underestimated uncertainties or model mismatch. Unlike the raw $\chi^2$, the reduced chi-square is normalized by the number of pixels $N$, so it remains comparable across spectra with different wavelength coverage or numbers of valid pixels. In contrast, $\chi^2$ grows roughly linearly with $N$ and is less suitable for cross-spectral comparison. Since $\chi^2_{\rm red}$ approximates the mean squared, inverse-variance-weighted residual per pixel, it more intuitively reflects the average pixel-level mismatch.

The template velocity is refined by fitting a quadratic to the $\chi^2$ values of the best point and its nine nearest neighbors on each side; the polynomial minimum gives the final RV estimate.

\subsection{Validation}
\begin{figure}[htbp]
    \centering
    \includegraphics[width=0.48\textwidth]{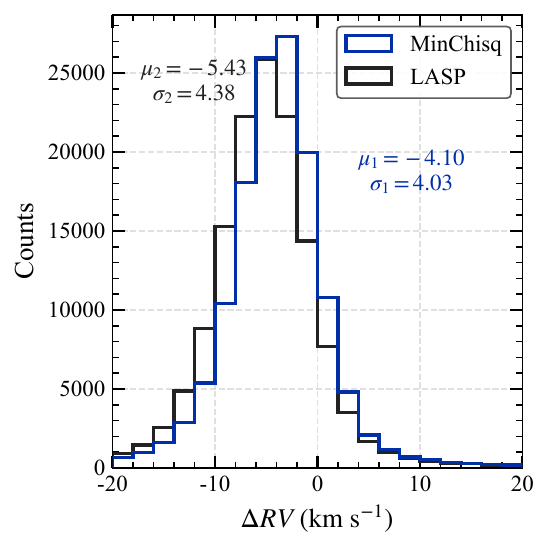}
    \caption{Comparison with APOGEE for the RV zero point and scatter. The cyan histogram shows our minimum-$\chi^2$ (MinChisq) pipeline, and the gray histogram shows LASP \citep{Luo_2015}. }
    \label{fig:deltaV_hist}
\end{figure}

We validate our method by cross-matching with APOGEE DR17 and comparing with LASP RVs \citep{Luo_2015}. Figure~\ref{fig:deltaV_hist} shows the distributions of $\Delta RV = RV_{\rm LAMOST} - RV_{\rm APOGEE}$ for 139,239 common stars. Our pipeline (cyan) exhibits a smaller zero-point offset ($\mu=-4.10$ km s$^{-1}$) and reduced scatter ($\sigma=4.03$ km s$^{-1}$) compared with LASP (gray, $\mu=-5.43$ km s$^{-1}$, $\sigma=4.38$ km s$^{-1}$).

\section{RV Offset Heatmaps for Each Spectrograph}
The main text uses Spectrograph~9 as an example to illustrate fiber-dependent radial velocity offset patterns. The LAMOST LRS instrument has 16 spectrographs in total. For completeness, this appendix shows annual RV offset heatmaps for the other 15 spectrographs. All figures are generated with the same methodology and use the same layout, color scale, and visualization scheme as Fig.~\ref{fig:sp9_season_heatmaps}, allowing direct comparison across spectrographs.

\begin{figure}[p]
  \centering
  \includegraphics[width=0.9\textwidth]{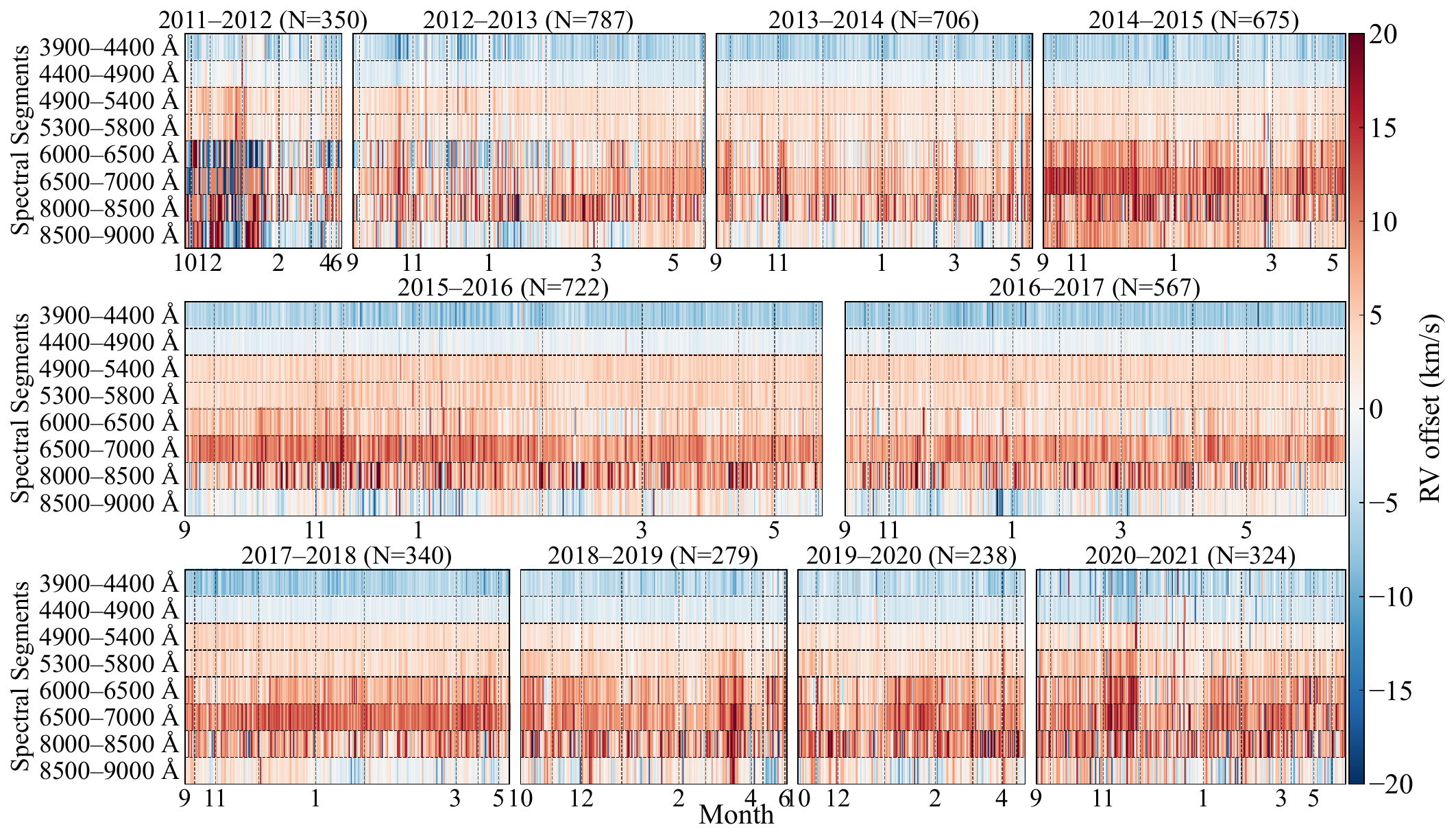}
  \caption{
    Annual RV offset heatmaps for Spectrograph~1 from 2011--2021.
  }
  \label{fig:sp1_heatmaps}
\end{figure}

\begin{figure}[p]
  \centering
  \includegraphics[width=0.9\textwidth]{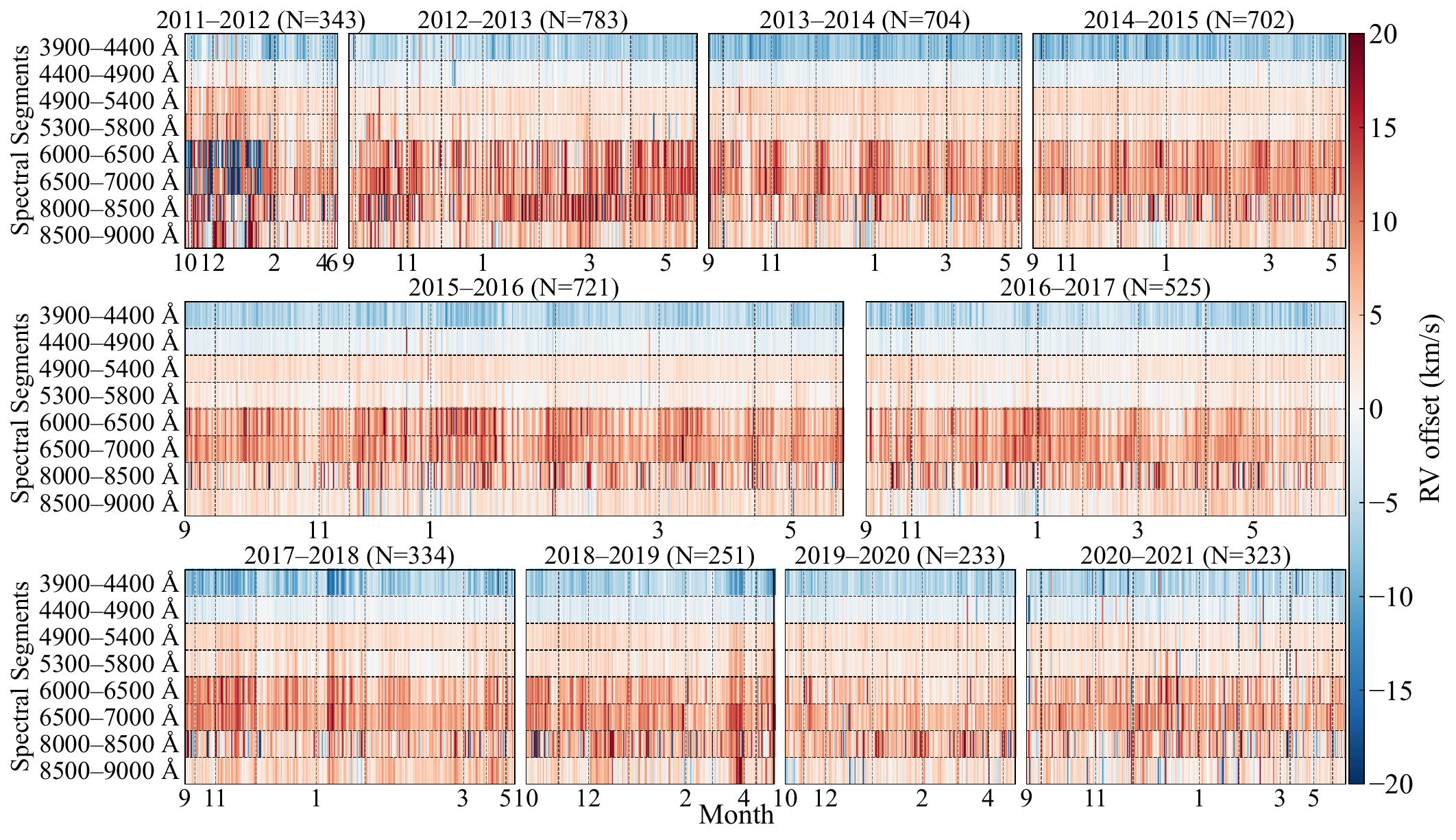}
  \caption{
    Same as Fig.~\ref{fig:sp1_heatmaps}, but for Spectrograph~2.
  }
  \label{fig:sp2_heatmaps}
\end{figure}

\begin{figure}[p]
  \centering
  \includegraphics[width=0.9\textwidth]{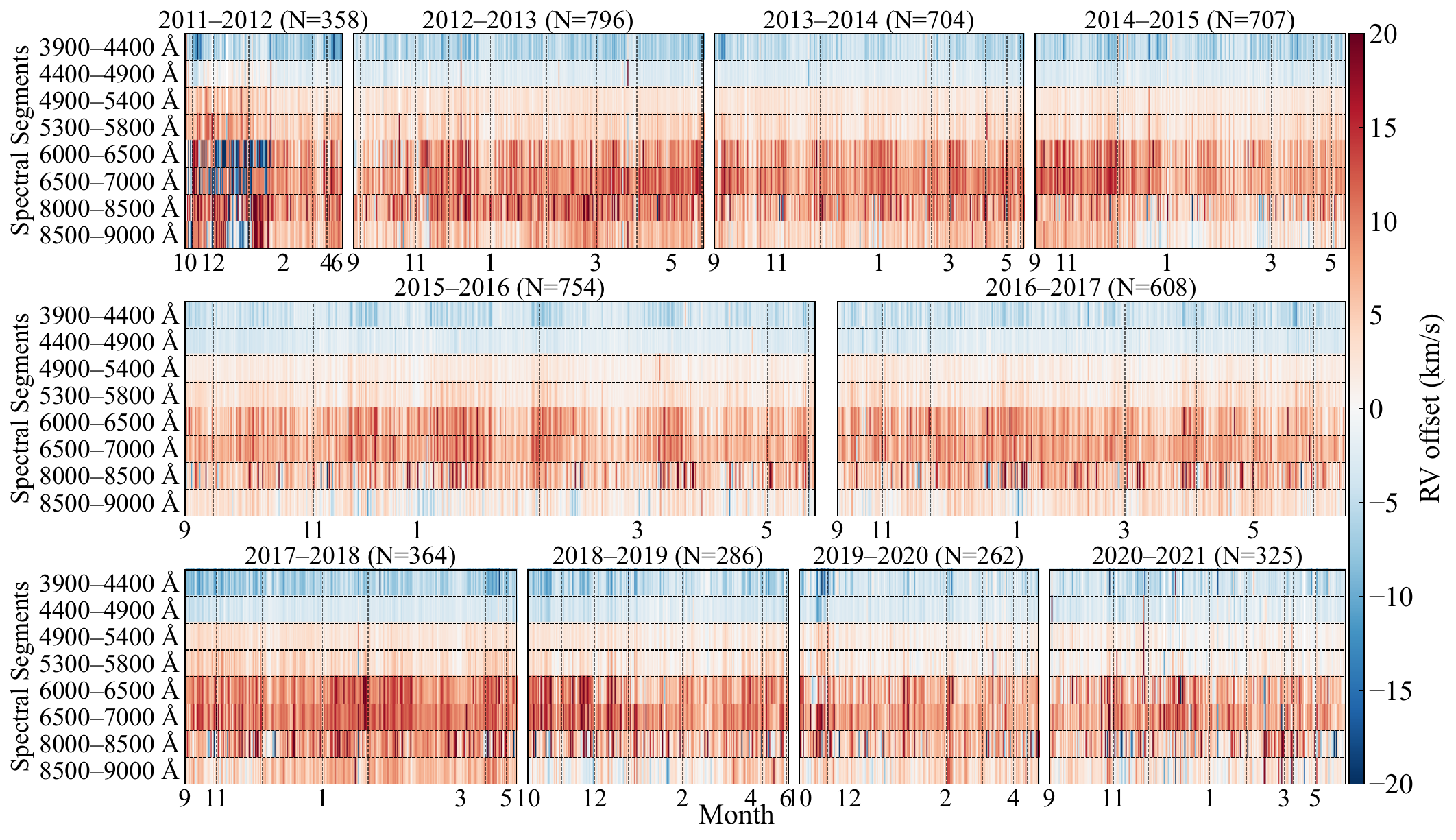}
  \caption{
    Same as Fig.~\ref{fig:sp1_heatmaps}, but for Spectrograph~3.
  }
  \label{fig:sp3_heatmaps}
\end{figure}

\begin{figure}[p]
  \centering
  \includegraphics[width=0.9\textwidth]{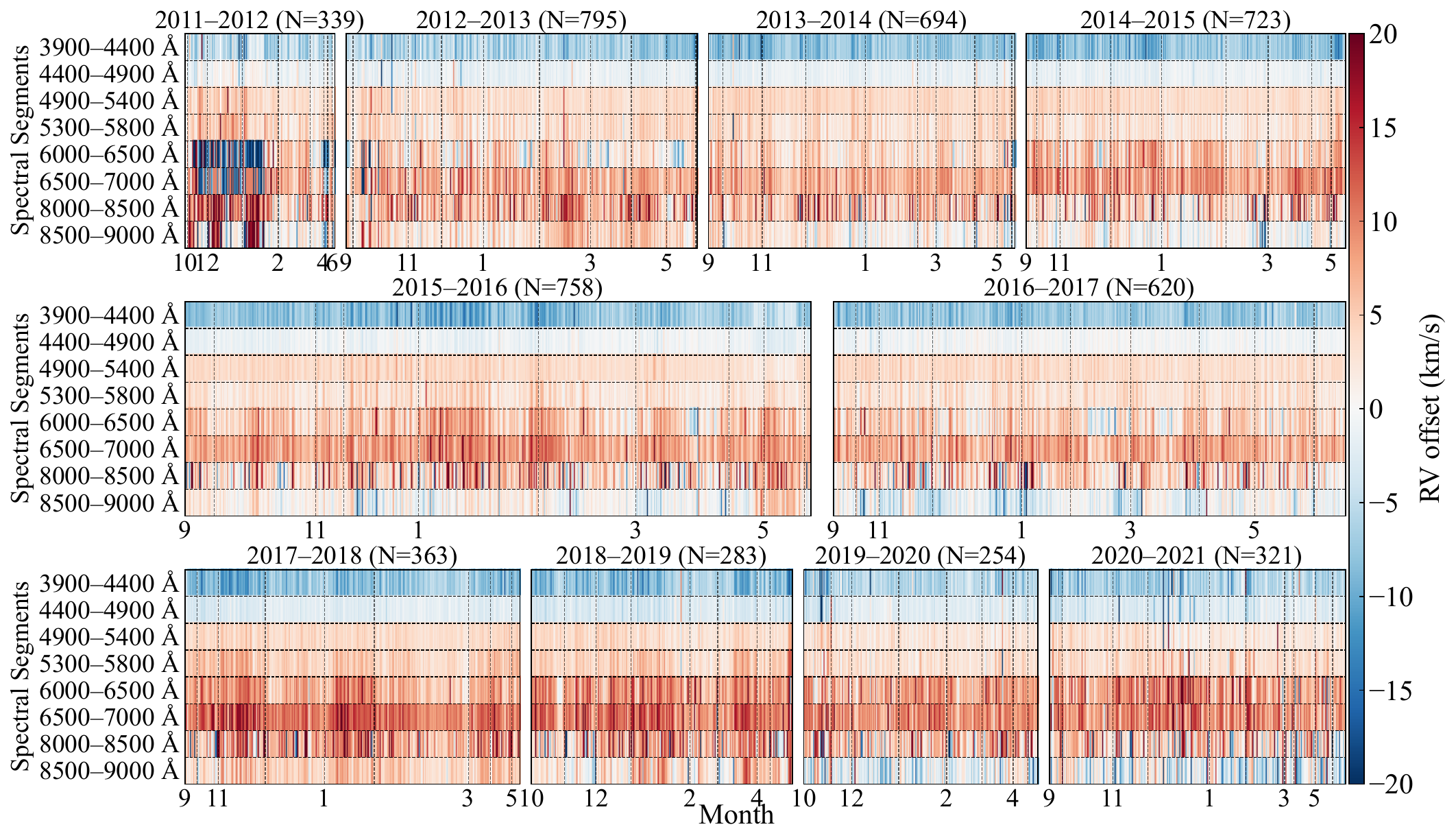}
  \caption{
    Same as Fig.~\ref{fig:sp1_heatmaps}, but for Spectrograph~4.
  }
  \label{fig:sp4_heatmaps}
\end{figure}

\begin{figure}[p]
  \centering
  \includegraphics[width=0.9\textwidth]{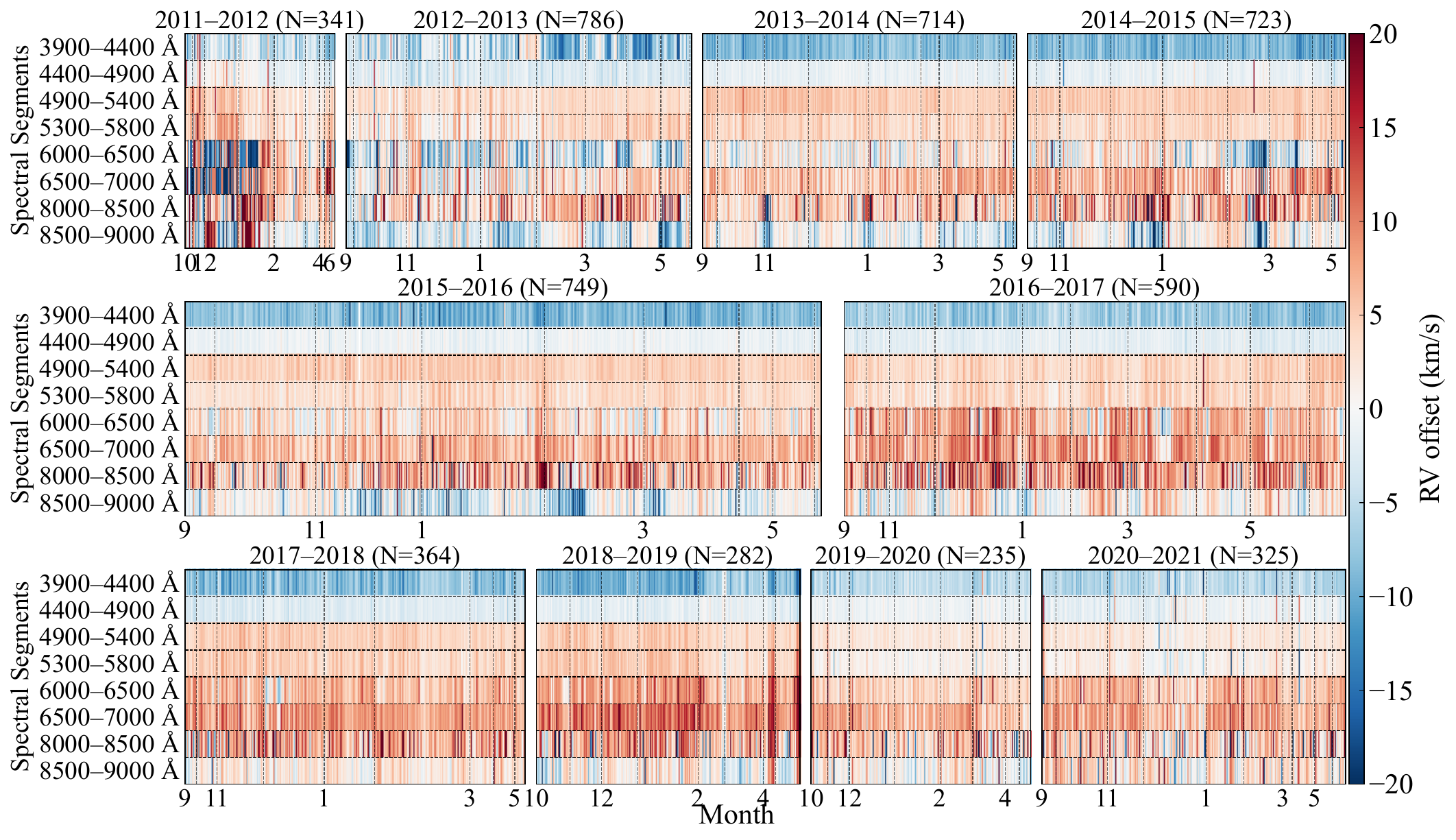}
  \caption{
    Same as Fig.~\ref{fig:sp1_heatmaps}, but for Spectrograph~5.
  }
  \label{fig:sp5_heatmaps}
\end{figure}

\begin{figure}[p]
  \centering
  \includegraphics[width=0.9\textwidth]{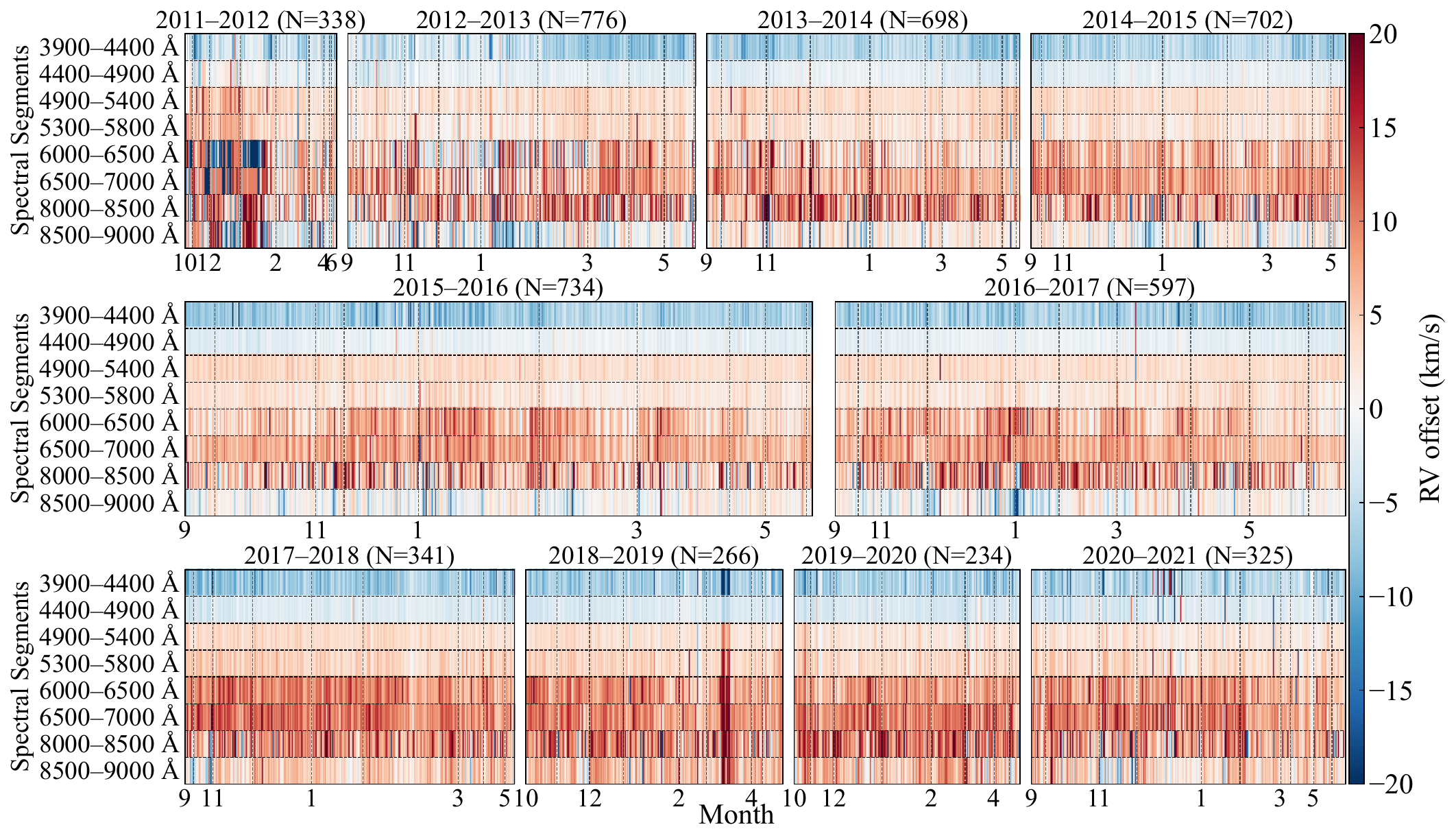}
  \caption{
    Same as Fig.~\ref{fig:sp1_heatmaps}, but for Spectrograph~6.
  }
  \label{fig:sp6_heatmaps}
\end{figure}

\begin{figure}[p]
  \centering
  \includegraphics[width=0.9\textwidth]{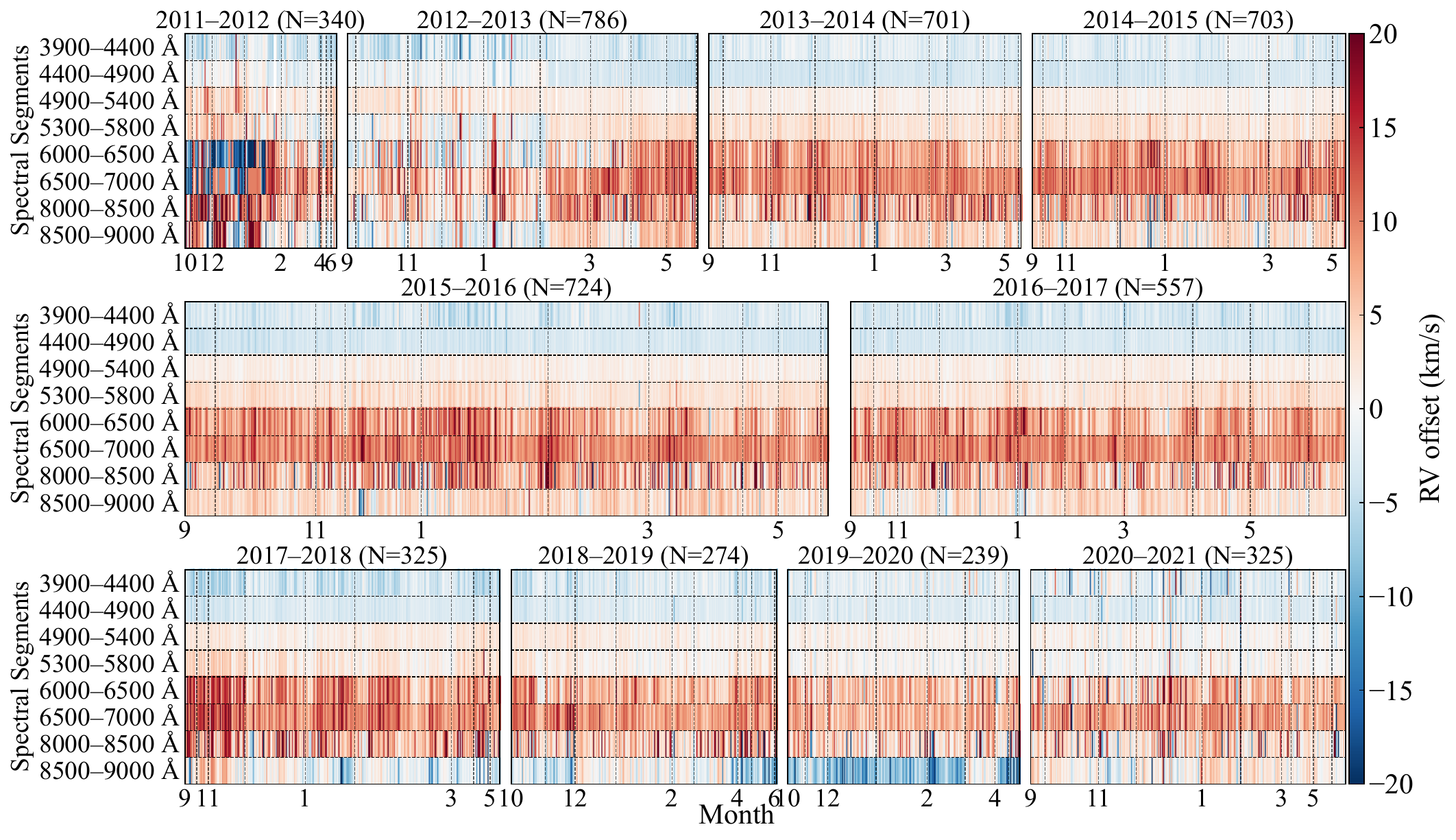}
  \caption{
    Same as Fig.~\ref{fig:sp1_heatmaps}, but for Spectrograph~7.
  }
  \label{fig:sp7_heatmaps}
\end{figure}

\begin{figure}[p]
  \centering
  \includegraphics[width=0.9\textwidth]{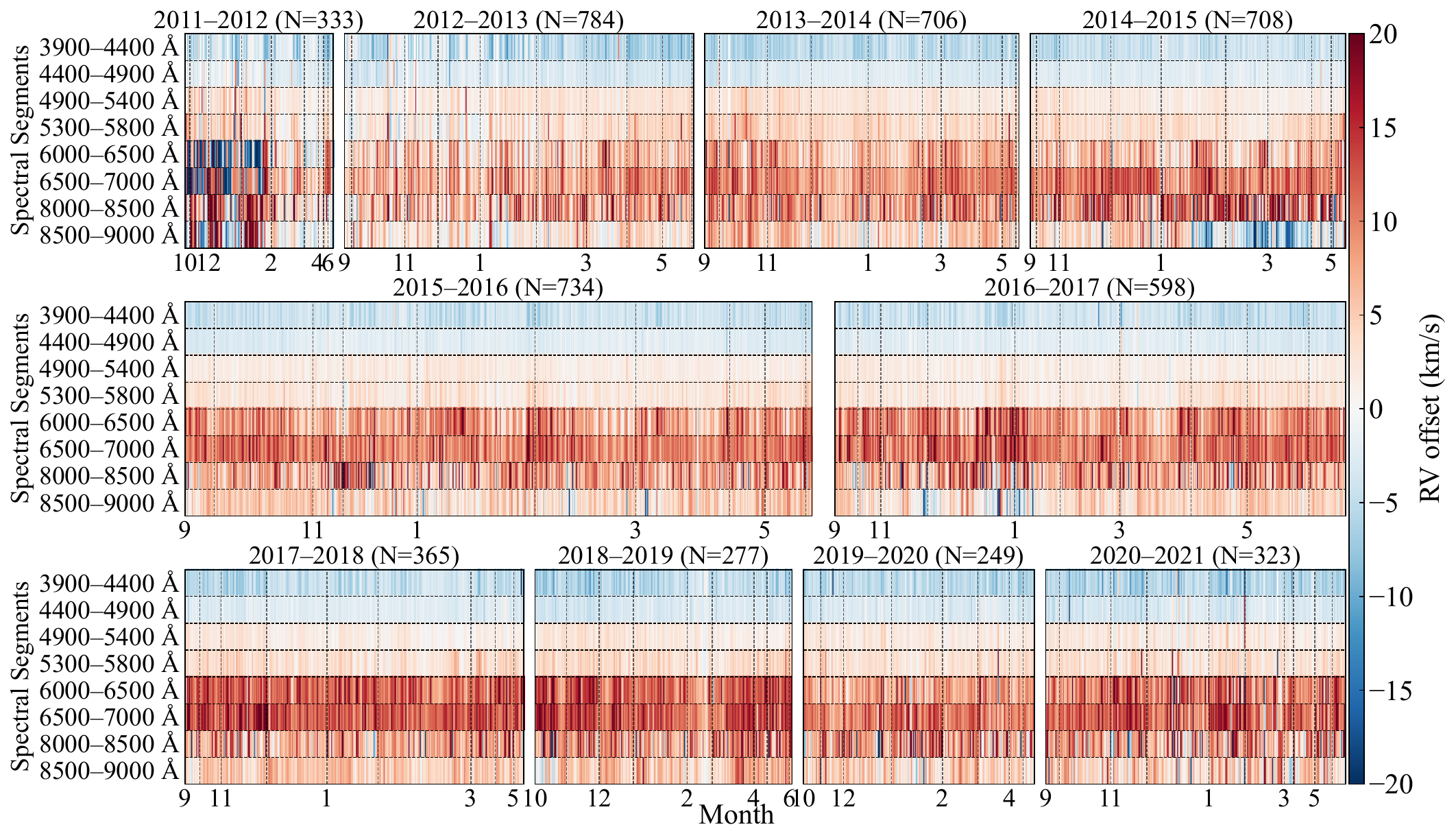}
  \caption{
    Same as Fig.~\ref{fig:sp1_heatmaps}, but for Spectrograph~8.
  }
  \label{fig:sp8_heatmaps}
\end{figure}

% sp9 skipped

\begin{figure}[p]
  \centering
  \includegraphics[width=0.9\textwidth]{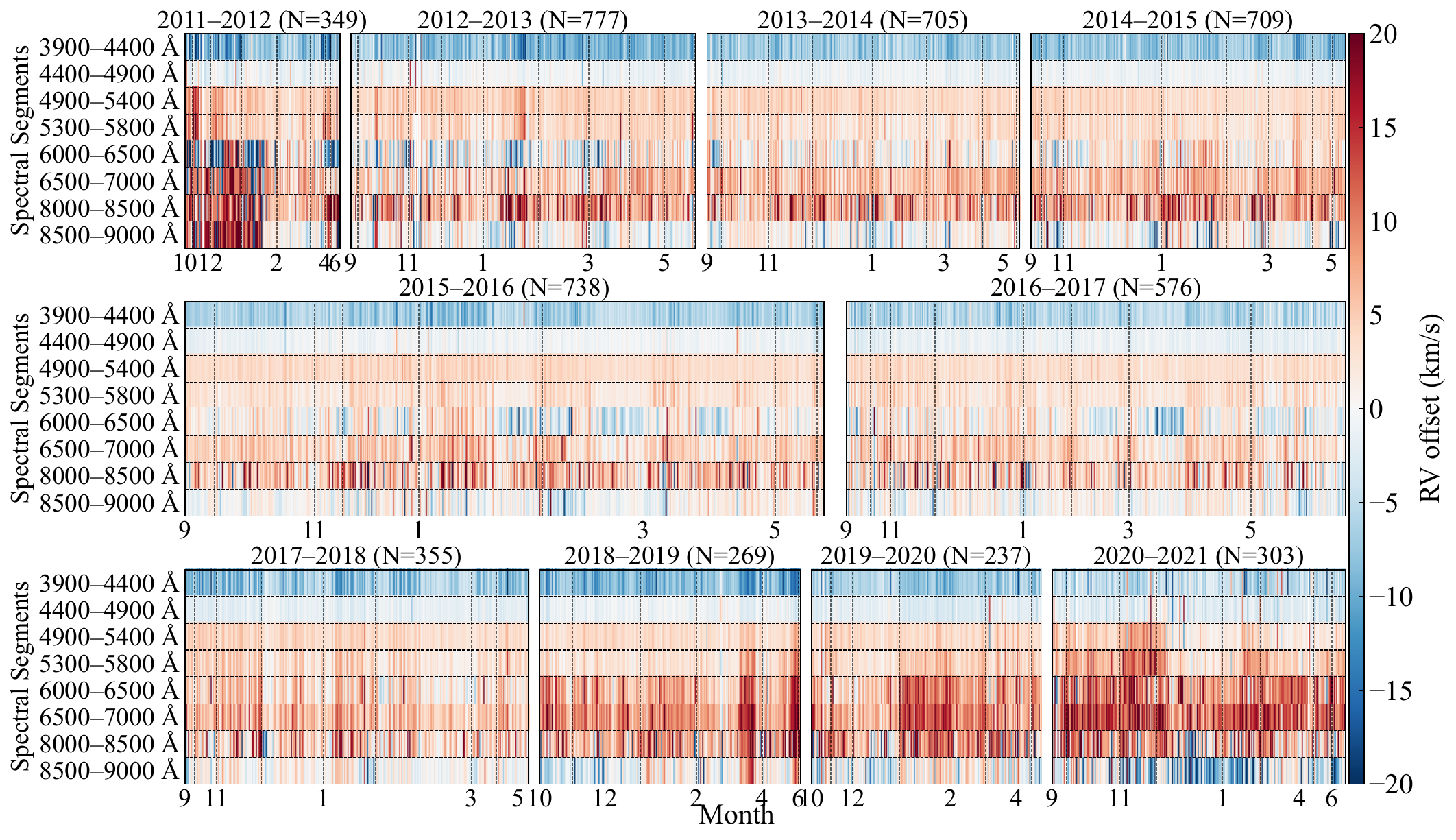}
  \caption{
    Same as Fig.~\ref{fig:sp1_heatmaps}, but for Spectrograph~10.
  }
  \label{fig:sp10_heatmaps}
\end{figure}

\begin{figure}[p]
  \centering
  \includegraphics[width=0.9\textwidth]{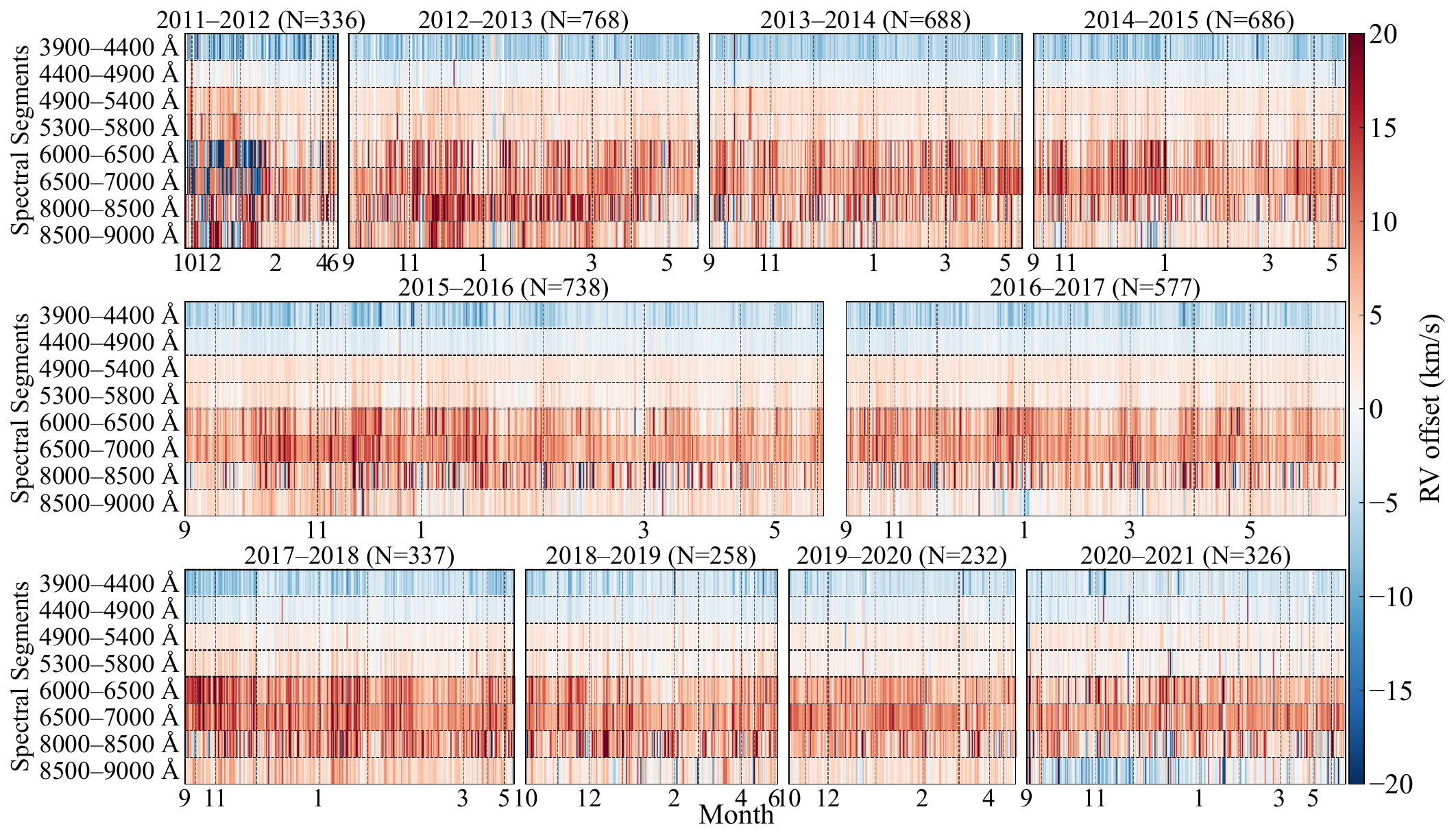}
  \caption{
    Same as Fig.~\ref{fig:sp1_heatmaps}, but for Spectrograph~11.
  }
  \label{fig:sp11_heatmaps}
\end{figure}

\begin{figure}[p]
  \centering
  \includegraphics[width=0.9\textwidth]{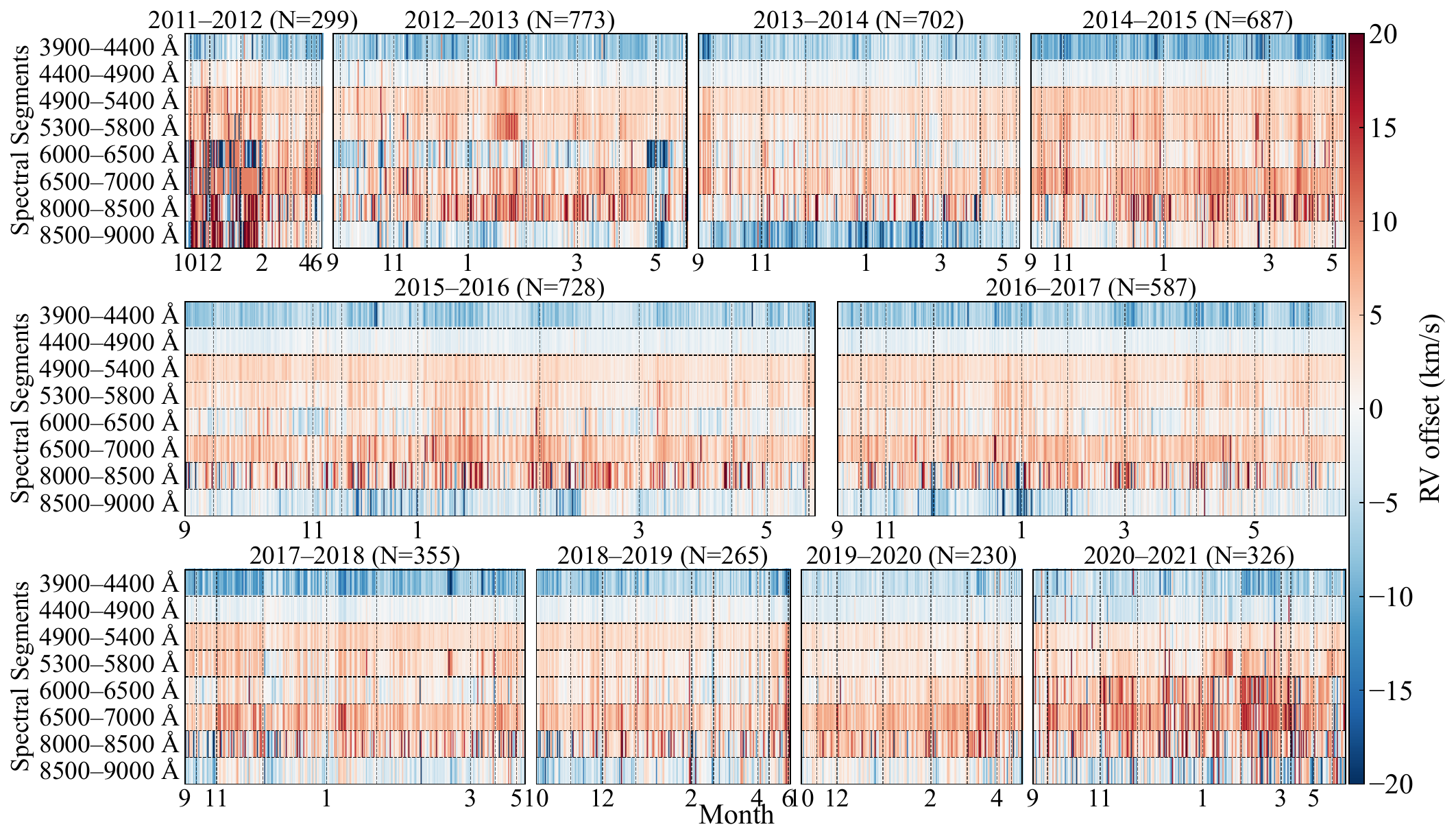}
  \caption{
    Same as Fig.~\ref{fig:sp1_heatmaps}, but for Spectrograph~12.
  }
  \label{fig:sp12_heatmaps}
\end{figure}

\begin{figure}[p]
  \centering
  \includegraphics[width=0.9\textwidth]{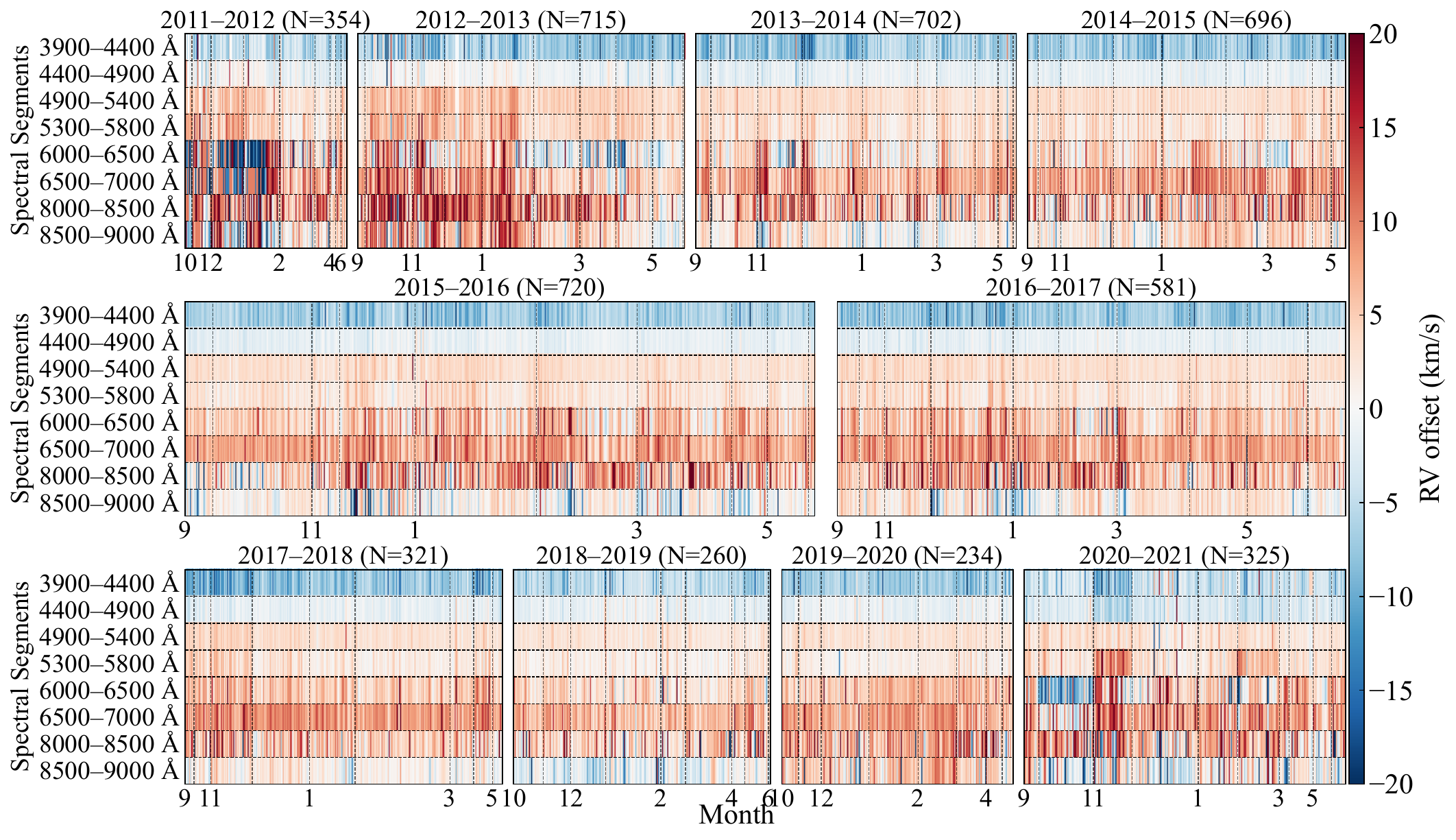}
  \caption{
    Same as Fig.~\ref{fig:sp1_heatmaps}, but for Spectrograph~13.
  }
  \label{fig:sp13_heatmaps}
\end{figure}

\begin{figure}[p]
  \centering
  \includegraphics[width=0.9\textwidth]{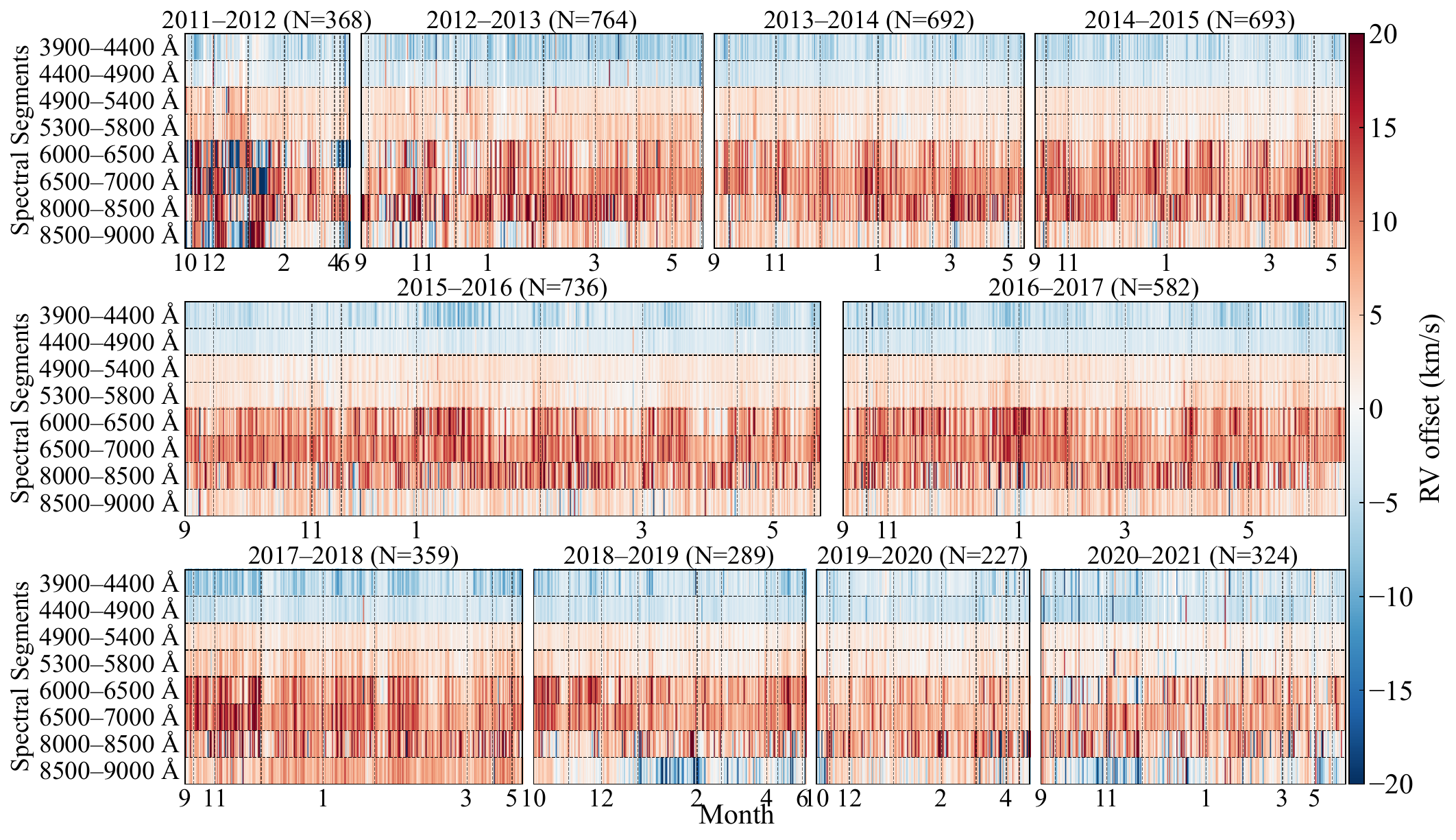}
  \caption{
    Same as Fig.~\ref{fig:sp1_heatmaps}, but for Spectrograph~14.
  }
  \label{fig:sp14_heatmaps}
\end{figure}

\begin{figure}[p]
  \centering
  \includegraphics[width=0.9\textwidth]{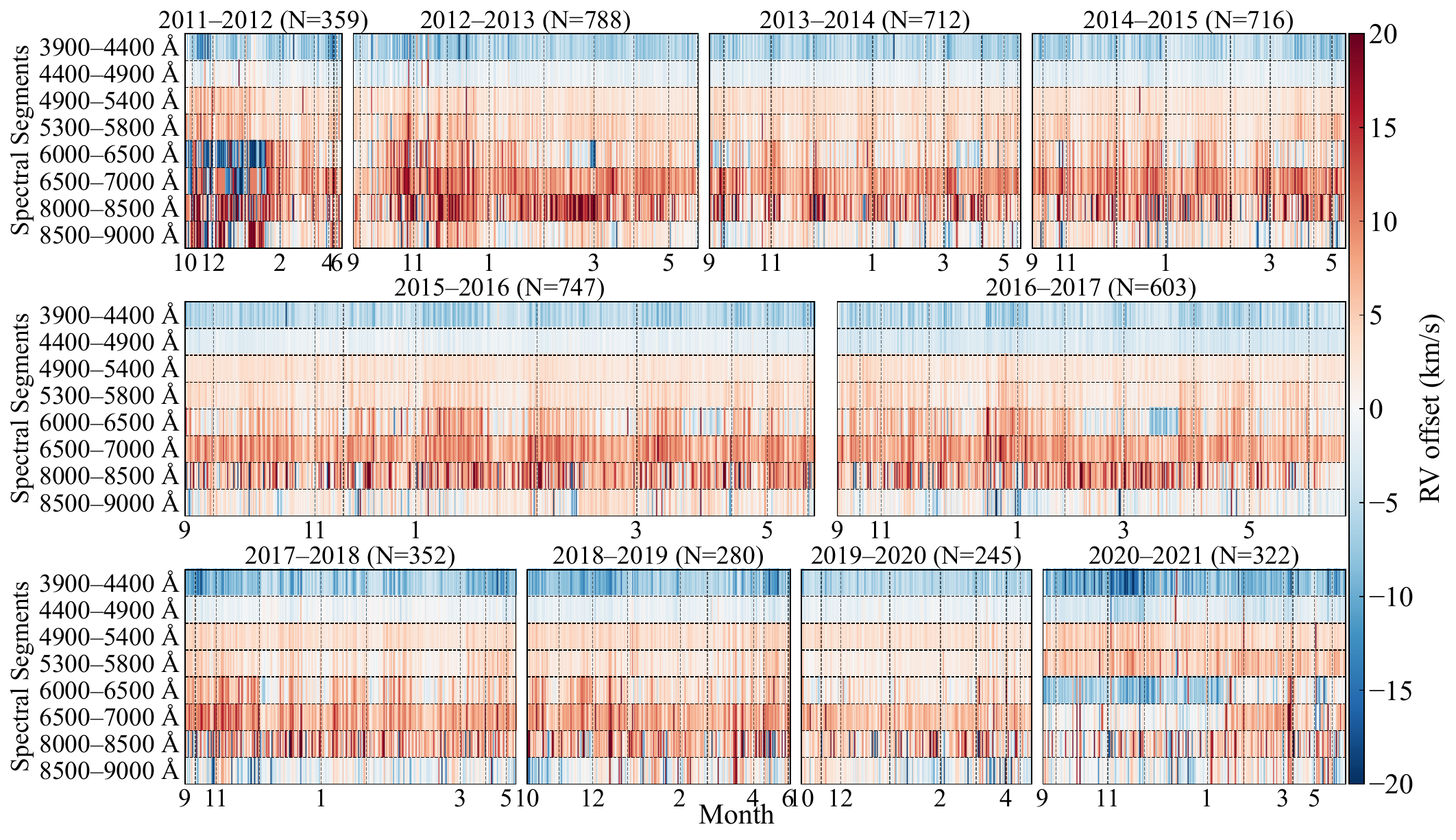}
  \caption{
    Same as Fig.~\ref{fig:sp1_heatmaps}, but for Spectrograph~15.
  }
  \label{fig:sp15_heatmaps}
\end{figure}

\begin{figure}[p]
  \centering
  \includegraphics[width=0.9\textwidth]{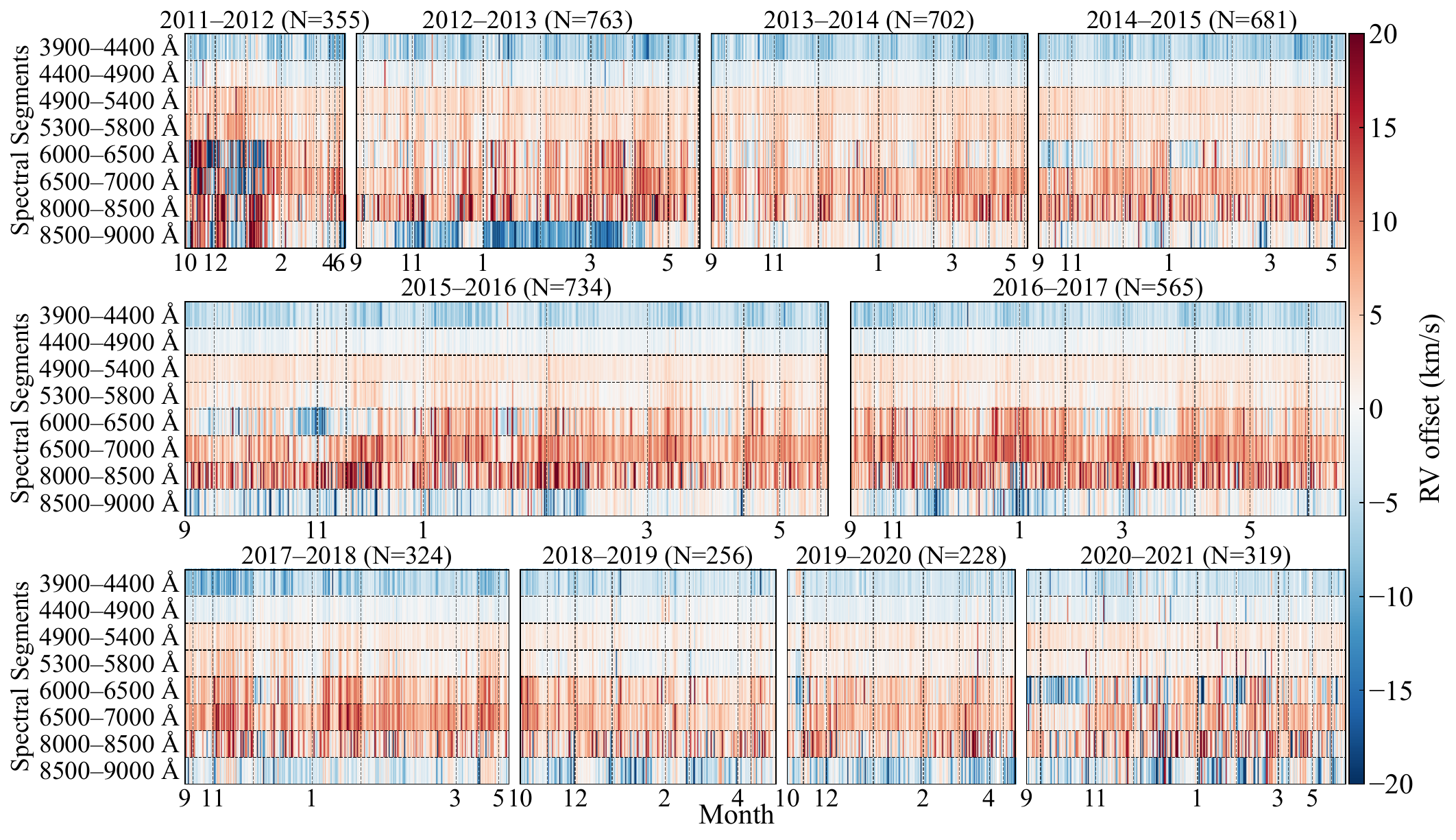}
  \caption{
    Same as Fig.~\ref{fig:sp1_heatmaps}, but for Spectrograph~16.
  }
  \label{fig:sp16_heatmaps}
\end{figure}
\section{RV Offset Heatmaps for Each Fiber in Each Observing Year (Spectrograph-corrected)}

In the main text, only the 2017--2018 observing year is shown as an example (Fig.~\ref{fig:rv_heatmap_sp1-16_2017-2018}). This appendix provides the radial velocity offsets per fiber for the other nine LAMOST observing years. All figures use the same methodology, color scale, and layout as Fig.~\ref{fig:rv_heatmap_sp1-16_2017-2018} to allow direct comparison across years.

\begin{figure}[p]
  \centering
  \includegraphics[width=0.87\textwidth]{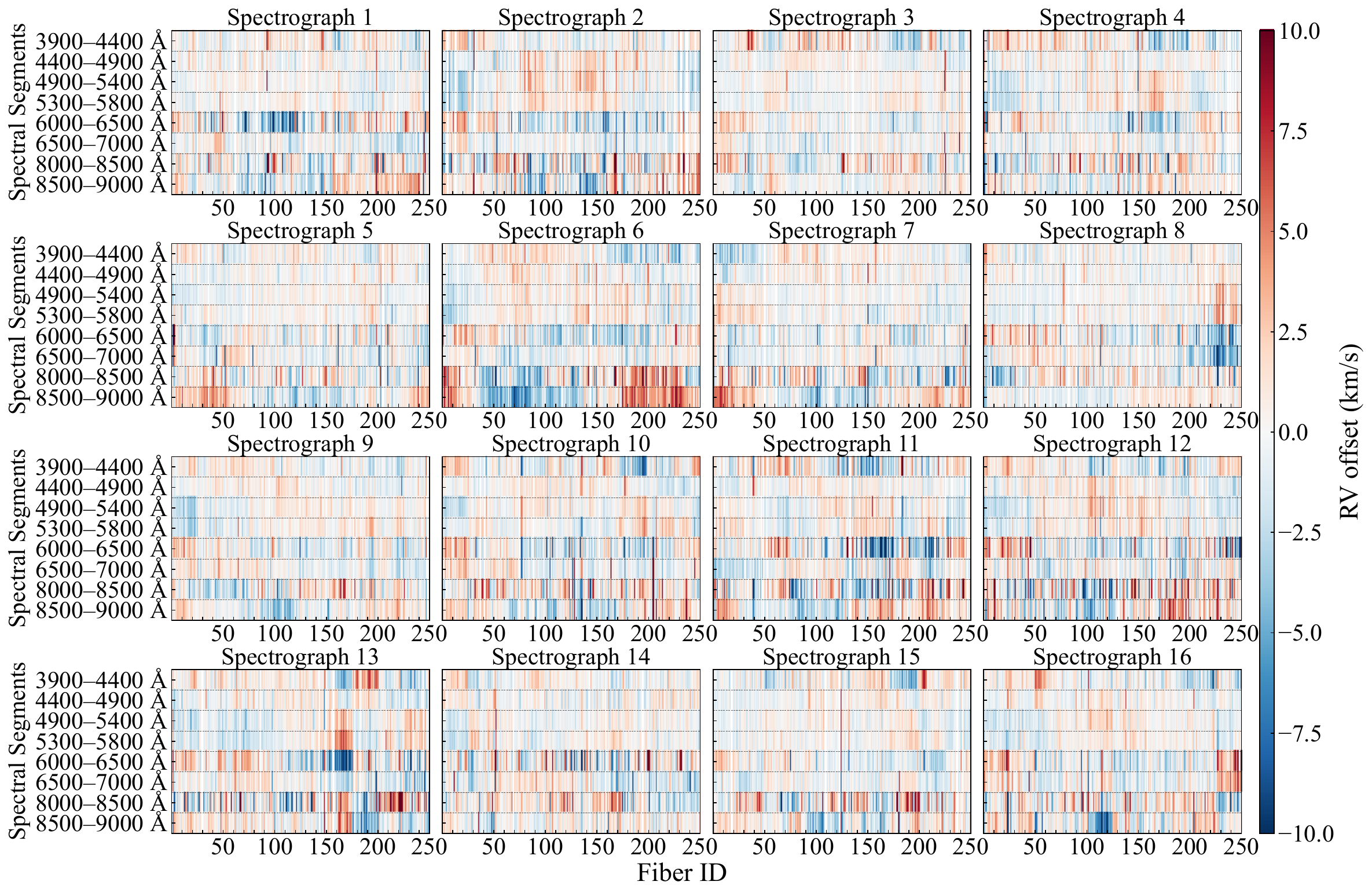}
  \caption{
    Radial velocity offsets per fiber for observing year 2011-2012,
    after subtracting the overall spectrograph-level offset.
  }
  \label{fig:heatmap-2011-2012}
\end{figure}

\begin{figure}[p]
  \centering
  \includegraphics[width=0.87\textwidth]{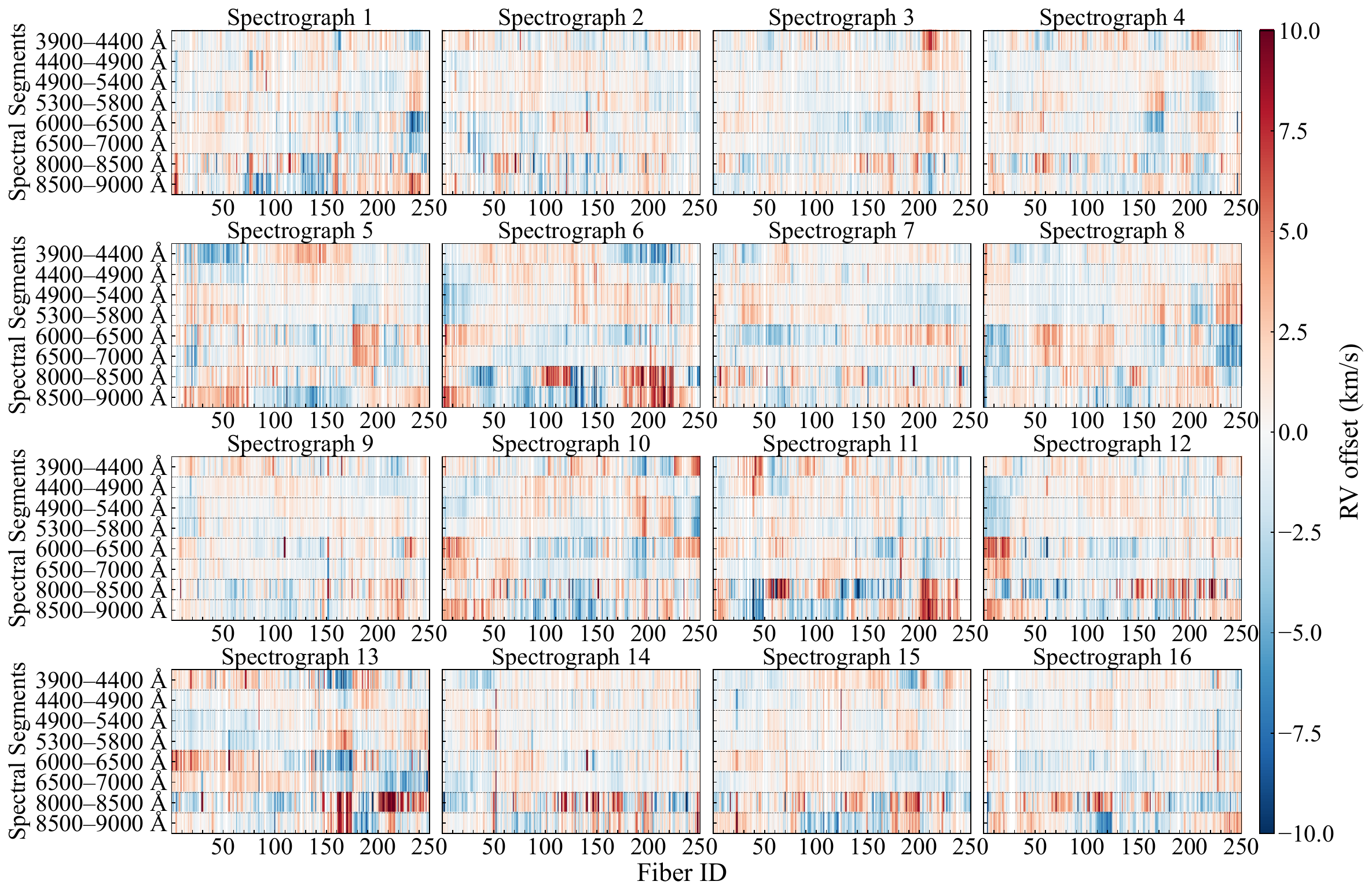}
  \caption{
    Same as Fig.~\ref{fig:heatmap-2011-2012},
    but for observing year 2012-2013.
  }
  \label{fig:heatmap-2012-2013}
\end{figure}

\begin{figure}[p]
  \centering
  \includegraphics[width=0.87\textwidth]{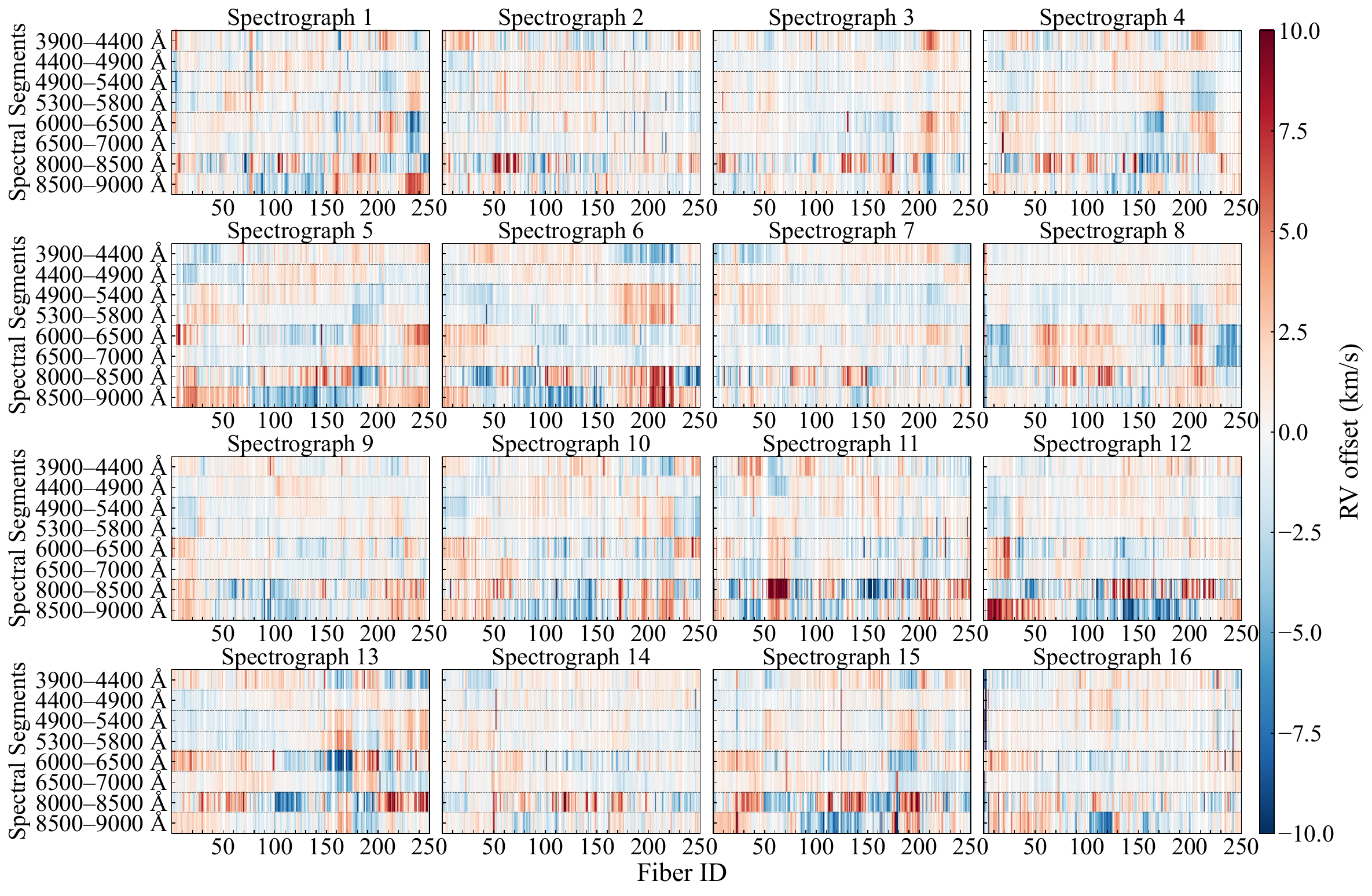}
  \caption{
    Same as Fig.~\ref{fig:heatmap-2011-2012},
    but for observing year 2013-2014.
  }
  \label{fig:heatmap-2013-2014}
\end{figure}

\begin{figure}[p]
  \centering
  \includegraphics[width=0.87\textwidth]{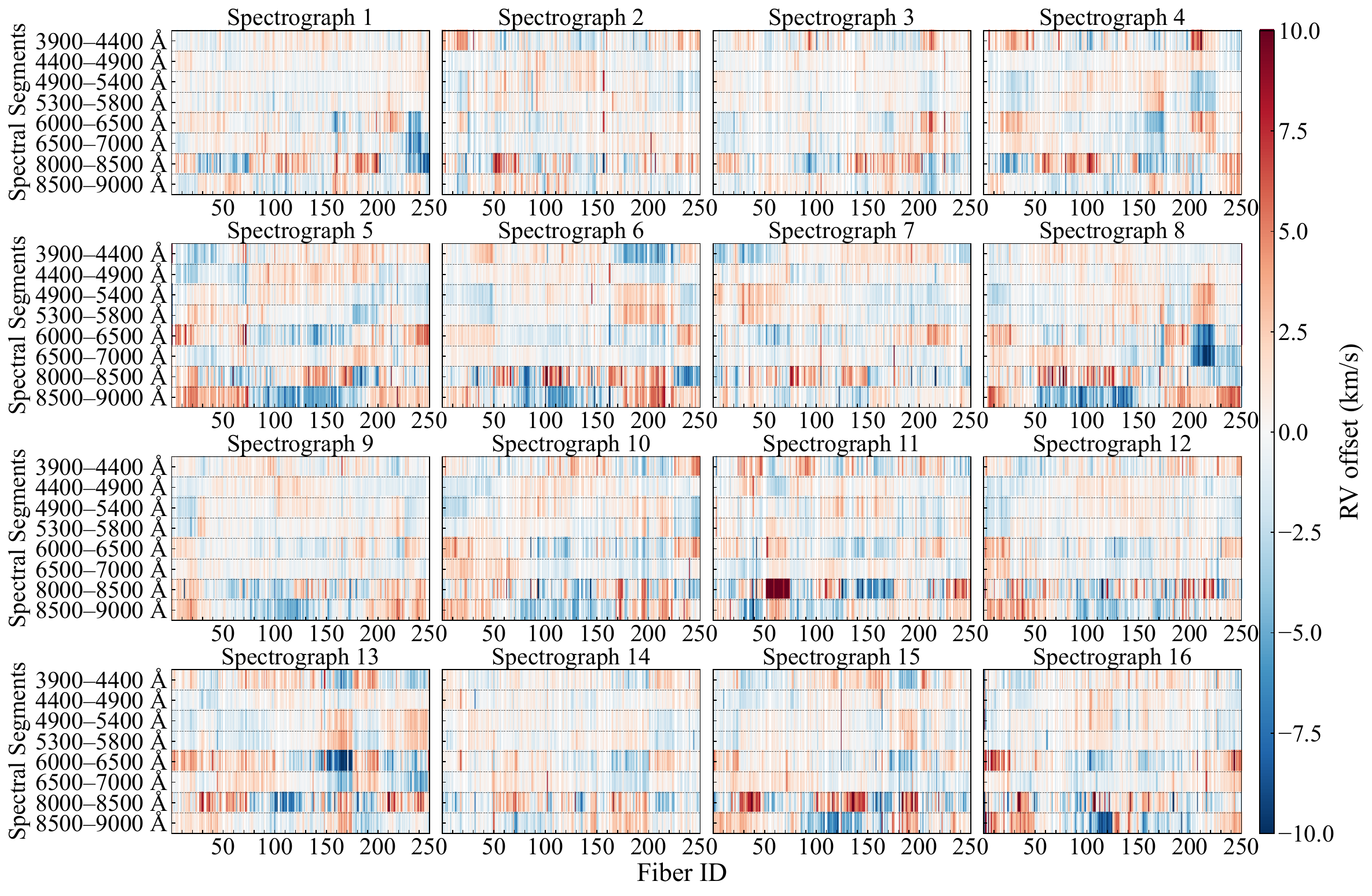}
  \caption{
    Same as Fig.~\ref{fig:heatmap-2011-2012},
    but for observing year 2014-2015.
  }
  \label{fig:heatmap-2014-2015}
\end{figure}

\begin{figure}[p]
  \centering
  \includegraphics[width=0.87\textwidth]{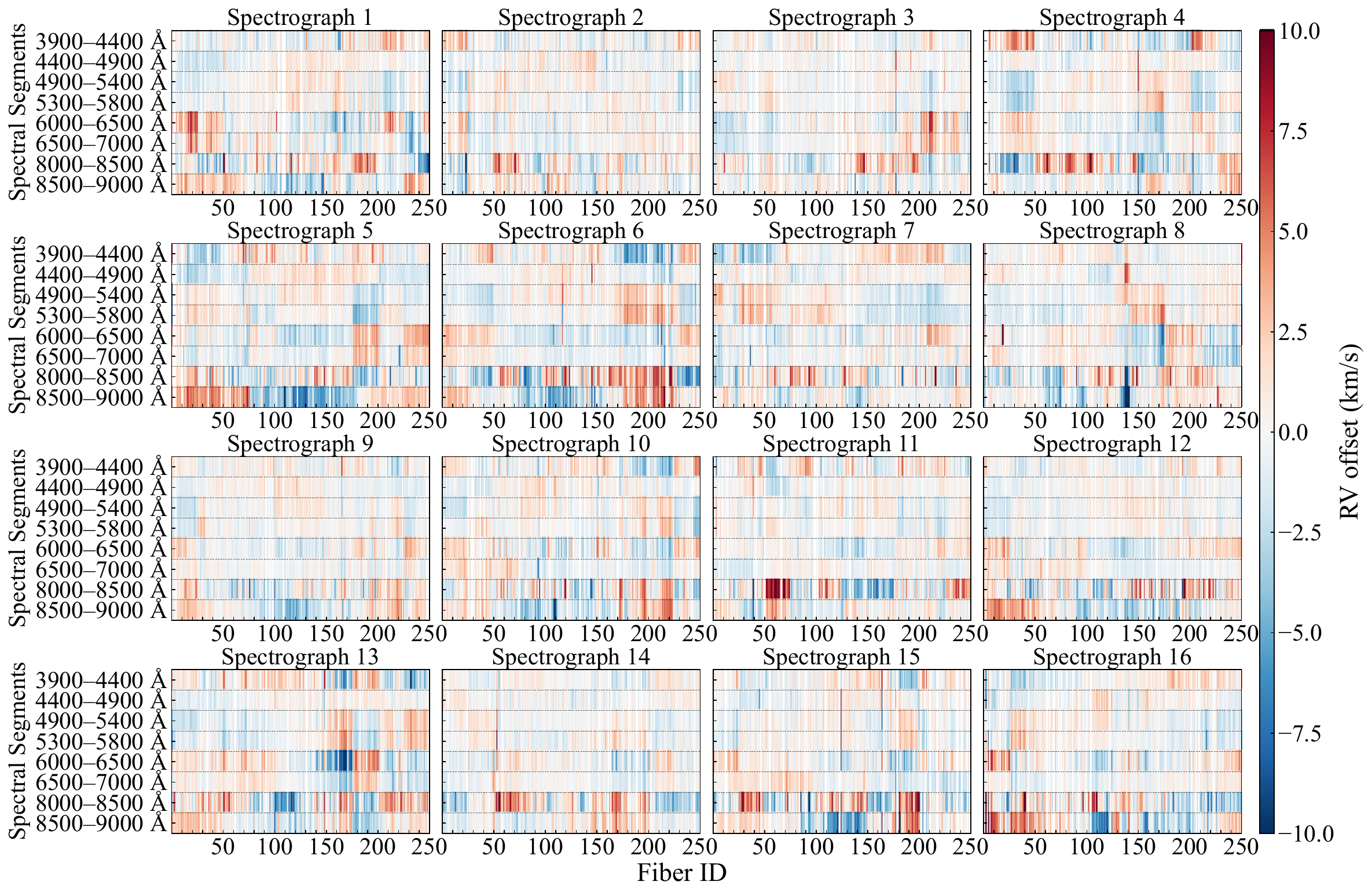}
  \caption{
    Same as Fig.~\ref{fig:heatmap-2011-2012},
    but for observing year 2015-2016.
  }
  \label{fig:heatmap-2015-2016}
\end{figure}

\begin{figure}[p]
  \centering
  \includegraphics[width=0.87\textwidth]{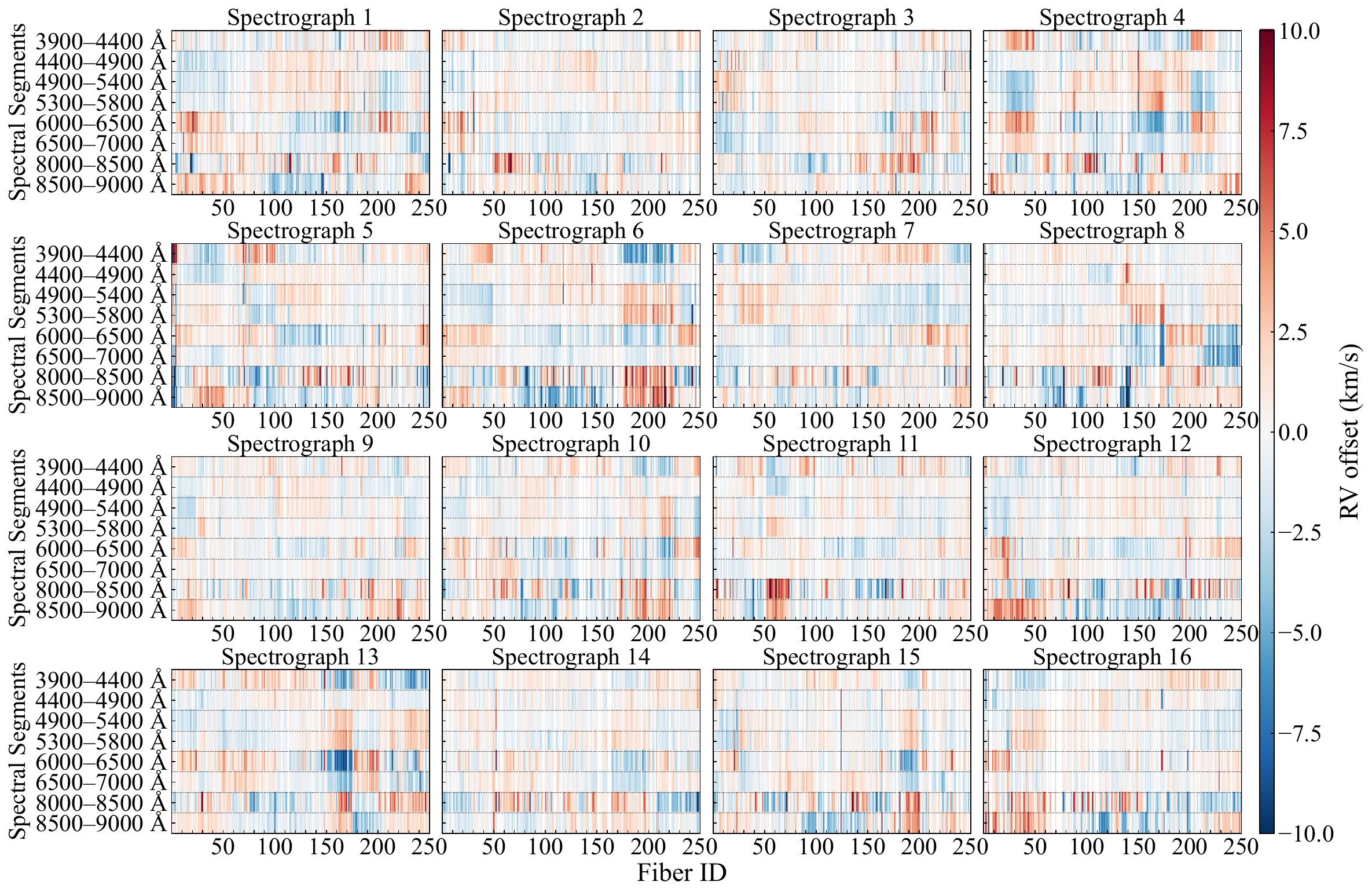}
  \caption{
    Same as Fig.~\ref{fig:heatmap-2011-2012},
    but for observing year 2016-2017.
  }
  \label{fig:heatmap-2016-2017}
\end{figure}

% 2017-2018 skipped

\begin{figure}[p]
  \centering
  \includegraphics[width=0.87\textwidth]{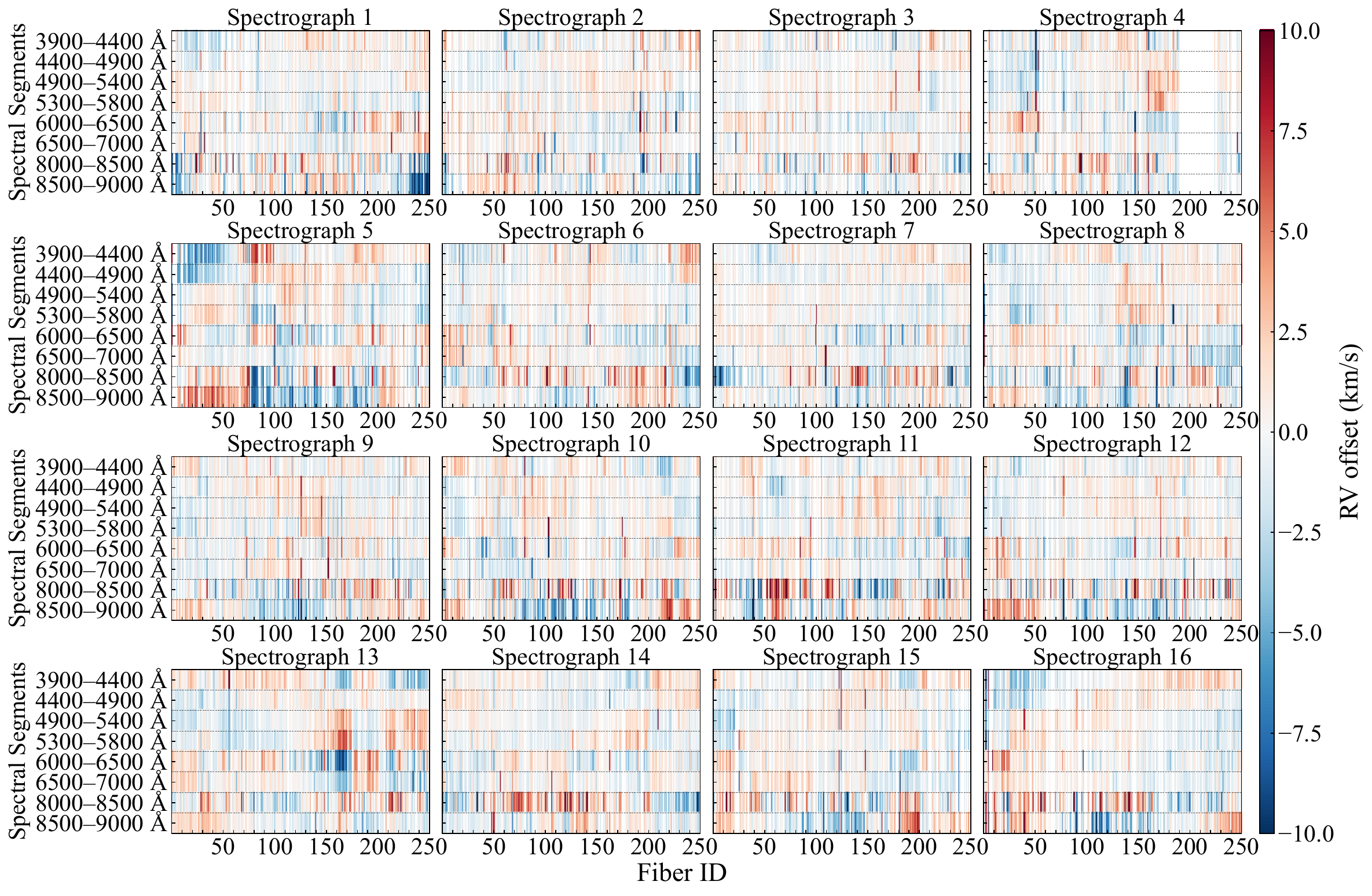}
  \caption{
    Same as Fig.~\ref{fig:heatmap-2011-2012},
    but for observing year 2018-2019.
  }
  \label{fig:heatmap-2018-2019}
\end{figure}

\begin{figure}[p]
  \centering
  \includegraphics[width=0.87\textwidth]{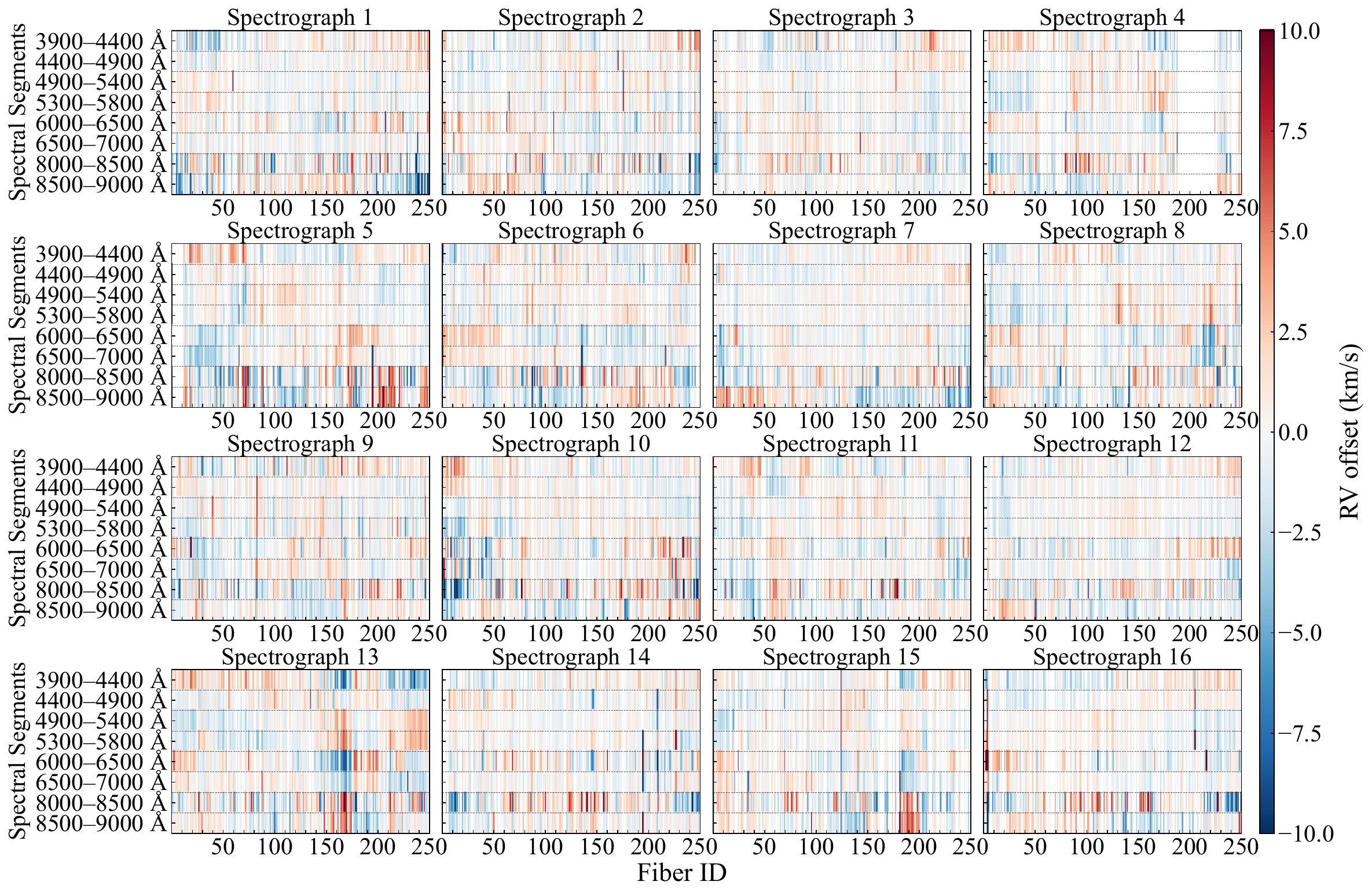}
  \caption{
    Same as Fig.~\ref{fig:heatmap-2011-2012},
    but for observing year 2019-2020.
  }
  \label{fig:heatmap-2019-2020}
\end{figure}

\begin{figure}[p]
  \centering
  \includegraphics[width=0.87\textwidth]{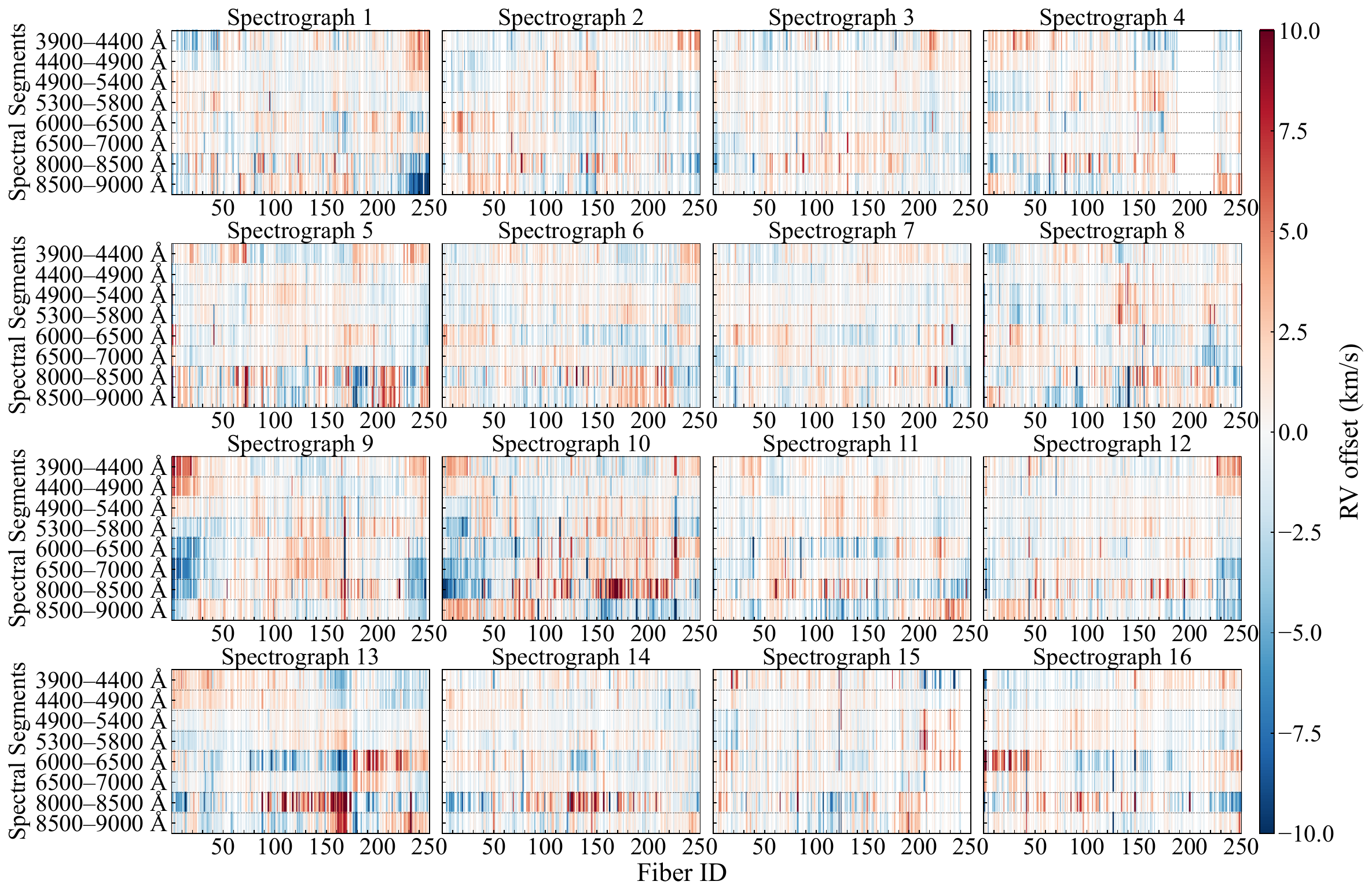}
  \caption{
    Same as Fig.~\ref{fig:heatmap-2011-2012},
    but for observing year 2020-2021.
  }
  \label{fig:heatmap-2020-2021}
\end{figure}

\clearpage

\end{document}